\documentclass[twocolumn,eqsecnum]{revtex4-1}
\usepackage{amsthm,setspace,afterpage}
\usepackage[usenames,dvipsnames]{xcolor}
\usepackage{amsmath,amssymb,graphicx,bm,subfigure, tikz}
\usepackage{hyperref}
\usepackage{float}
\usepackage{ulem}
\usepgflibrary{fpu}
\usepackage{pgf}
\usepackage{pgffor}
\usepackage{pgfplots}
\usepackage{dsfont}
\usepgfmodule{shapes}
\usepgfmodule{plot}
\usetikzlibrary{decorations}
\usetikzlibrary{arrows}
\usetikzlibrary{snakes}
\usetikzlibrary{calc,fadings,decorations.pathreplacing,spy,3d}
\graphicspath{ {./figures/} }
\pgfplotsset{compat=newest}

\begin{document}
\title{Invariant Theory and Orientational Phase Transitions.}

\author{Joseph Rudnick$^{1}$, and Robijn Bruinsma$^{1,2}$}
\affiliation{$^{1}$Department of Physics and Astronomy, University of California, Los Angeles, CA 90095, USA}
\affiliation{$^{2}$Department of Chemistry and Biochemistry, University of California, Los Angeles, CA 90095, USA}

\begin{abstract}
The Landau theory of phase transitions has been productively applied to phase transitions that involve rotational symmetry breaking, such as the transition from an isotropic fluid to a nematic liquid crystal. It even can be applied to the orientational symmetry breaking of simple atomic or molecular clusters that are not true phase transitions. In this paper we address fundamental problems that arise with the Landau theory when it is applied to rotational symmetry breaking transitions of more complex particle clusters that involve order parameters characterized by larger values of the $l$ index of the dominant spherical harmonic that describes the broken symmetry state. The problems are twofold. First, one may encounter a thermodynamic instability of the expected ground state with respect to states with lower symmetry. A second problem concerns the proliferation of quartic invariants that may or may not be physical. We show that the combination of a geometrical method based on the analysis of the space of invariants, developed by Kim \cite{Kim3} to study symmetry breaking of the Higgs potential, with modern visualization tools provides a resolution to these problems. The approach is applied to the outcome of numerical simulations of particle ordering on a spherical surface and to the ordering of protein shells.
\end{abstract}
\maketitle

\section{Introduction}

The freezing of fluids has fascinated generations of scientists. When temperature is lowered, interacting atoms and molecules can transform spontaneously from a shapeless, entropy-dominated fluid into an ordered crystal that has a well-defined, faceted shape. The ordering transition involves a loss of symmetry: an extended fluid is symmetric with respect to any translation or rotation but as a crystal, this same system is symmetric only with respect to a discrete set of translations and rotations. Spontaneous symmetry breaking of this type is not restricted to the phase transitions of extended or bulk systems. When a nanometer-sized cluster of atoms freezes, it also can adopt an ordered state with reduced symmetry~\cite{honeycutt}. An important difference between the freezing of bulk liquids and that of particle clusters is that the freezing of a particle cluster can not be a true phase transition because it only involves a small, finite system. Nevertheless, an extended system of interacting atomic clusters---a possible model of a glass---still can exhibit a genuine phase transition of this type~\cite{Steinhardt}. 

Another important difference is that a cluster of atoms in the liquid state has full rotational symmetry but no translational symmetry. The symmetry group of a cluster in the liquid state typically is  $O(3)$, the group of all rotations and reflections, or the group of all rotations $SO(3)$ in the case of chiral molecules. Neither the ordered nor the disordered cluster has any form of translational symmetry. Rotational symmetry breaking without translational symmetry breaking is encountered as well in extended systems, such as the transition from an isotropic fluid to a nematic liquid crystal with broken orientational symmetry \cite{CL}. In an important paper, Steinhardt, Nelson and Ronchetti~\cite{Steinhardt} (SNR) proposed in 1983 that a version of the Landau theory for orientational phase transitions of liquid crystals could be applied to the freezing of particle clusters. The order parameter was the radially-averaged angle-dependent density $\rho(\Omega)$ of the cluster, with $\Omega$ a solid angle measured from the center of a cluster of atoms or molecules. This density is then expanded in series of spherical harmonics:
\begin{equation}
\rho(\Omega)=\sum_{l=0}^\infty\sum_{m=-l}^l c_{l,m}Y_l^m(\Omega)
\label{OP}
\end{equation}
Under the symmetry operations of $O(3)$, each set of $2l+1$ expansion coefficients $c_{l,m}$ in this series transforms as an irreducible representation of $O(3)$. One of the principles of the Landau theory of phase transitions states that continuous or near-continuous symmetry breaking transitions should be associated with just one irreducible representation of the symmetry group of the high-symmetry phase, so just one particular value of $l$ should characterize spontaneous orientational symmetry breaking. The set of $2l+1$ expansion coefficients $c_{l,m}$ associated with that $l$ value is then the \textit{primary order-parameter} of the transition. For example, the onset of orientational order in nematic and cholesteric liquid crystals are associated by $l=2$, with various combinations of the azimuthal quantum number $m$.  Irreducible representations with different $l$ values may well be ``entrained'' by the primary order parameter through non-linear terms in the free energy but these play only a limited role, so the associated $c_{l,m}$ are known as secondary order-parameters. SNR proposed that the ordering of small particle clusters is dominated by an $l=6$ state with icosahedral symmetry. This was based on the construction of a variational free energy in the form of a functional $F([\rho(\Omega)])$ expressed in terms of the $c_{l,m}$. Such a variational free energy has to transform as a scalar under the symmetry operations of $O(3)$ of the disordered state. This is achieved by constructing $F([\rho(\Omega)])$ from sums of combinations of $c_{l,m}$ that transform individually as invariants under $O(3)$ or $SO(3)$. 

\begin{figure}
	\centering
	\includegraphics[width=1.5in]{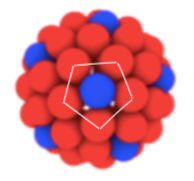}
	\includegraphics[width=1.5in]{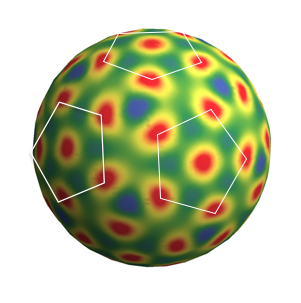}	
	\caption{Left: Icosahedral state of N=72 Lennard-Jones particles on a sphere (from Ref. \cite{paquay}). Right: the l=16 Icosahedral Spherical Harmonic with 72 maxima (from Ref. \cite{sanjay2}).}
	\label{16}
\end{figure}

\begin{figure}
	\centering
	\includegraphics[width=3.5in]{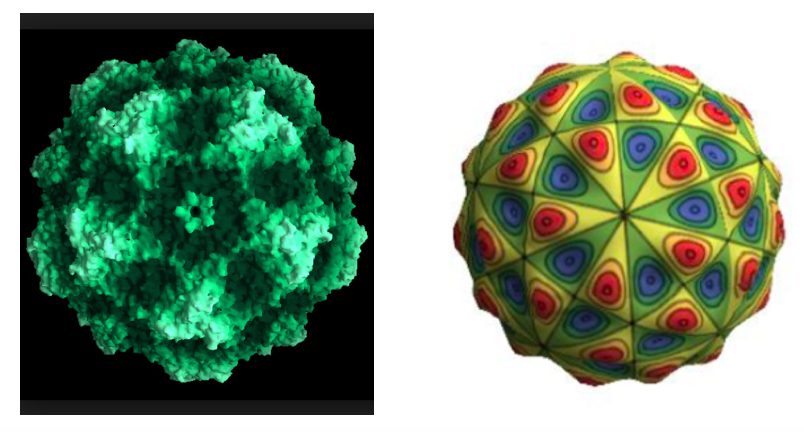}
	\caption{Left: Capsid of the Canine Parvovirus, as reconstructed by X-ray diffraction (from Ref. \cite{tsao}). It is composed of 60 identical proteins placed in an icosahedral pattern. Right: The l=15 icosahedral spherical harmonic with 60 maxima (from Ref. \cite{sanjay2}).}.
	\label{CP}
\end{figure}

The focus of the present article is on orientational ordering transitions with $l$ values larger than 6. Numerical simulations of 72 particles on a spherical surface interacting via a Lennard-Jones potential reported that there were various competing forms of orientational ordering~\cite{paquay}. The example shown in Fig. \ref{16} (left), has icosahedral symmetry. Icosahedral states can be constructed from certain linear combinations of spherical harmonics known as \textit{icosahedral spherical harmonics}, though only for certain values of $l$~\cite{golub}. The $l=16$ icosahedral spherical harmonic has 72 maxima, as shown in Fig. \ref{16}, right, so it could serve as the primary order parameter for the icosahedral ordering of 72 particles on a spherical surface.  
A second case is provided in the work of Lorman and Rochal~\cite{lorman2} who systematically compared the surface densities of icosahedral viral capsids with the icosahedral spherical harmonics. An example is shown in Fig.~\ref{CP} where the capsid of the Canine Parvovirus, which is composed of 60 identical proteins, is compared with the $l=15$ icosahedral spherical harmonic, which also has 60 maxima.  
The Parvovirus belongs to the smallest class of icosahedral viruses. Larger viral capsids correspond to icosahedral spherical harmonics values of $l$ that are larger than 15~\cite{lorman1, lorman2}. Since transitions from disordered to ordered viral capsids have been observed experimentally~\cite{garmann}, one could ask whether such transitions can be described by SNR-type Landau theories. 

Fundamental problems are encountered if one attempts to directly extend SNR to larger values of $l$. The first problem concerns thermodynamic stability. The simulations for 72 particles on a spherical surface and the observations on the Parvovirus seem to indicate that icosahedral shells that have a primary order parameter that transforms either as an $l=16$ or as an $ l=15$ icosahedral spherical harmonic should be thermodynamically stable for some appropriate choice of thermodynamic system parameters. However, when the $l=15$ and $l=16$ cases were investigated, states with icosahedral symmetry turned to be thermodynamically unstable~\cite{sanjay,sanjay2}. Separately, Matthews~\cite{Matthews} found that rotational symmetry breaking in the $l=16$ sector produces states with \textit{tetrahedral} symmetry. Strangely, the thermodynamic stability of the $l=15$ icosahedral state could be restored by mixing in small amounts of $l=16$~\cite{sanjay,sanjay2}. The fact that icosahedral symmetry appears to be associated with a \textit{mixed} $l=15+16$ state is unsatisfactory from the viewpoint of Landau theory because it seems to associate rotational symmetry breaking with \textit{two} irreducible representations of the symmetry group of the uniform state. Note that the $l=16$ contribution could not be viewed in this case as a secondary order parameter since secondary order parameters should not determiner the stability of the primary order parameter. 

A second issue concerns the number of invariants that are to be included in the Landau variational free energy. SNR effectively included two invariants for $l=6$ (one cubic and one quartic) but Jar\'{i}c~\cite{jaric, jaric2} showed that there are actually three independent quartic invariants for $l=6$. Depending on the coefficients of these three invariants, the $l=6$ icosahedral state may or may not be the ground state. We will see that the number of independent quartic invariants increases in a step-wise linear fashion with $l$, and that these new invariants in general are \textit{non-local}.  Do all these non-local invariants really have to be included even if the underlying physical system itself only involves short-range interactions? We will show that these two issues are actually intimately connected

\begin{figure}
	\centering
	\includegraphics[width=3in]{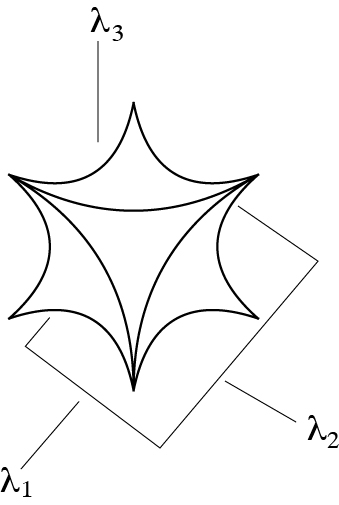}
	\caption{Three-dimensional space of invariants. The surface encloses the set of points generated when the order-parameter set $Q_{L,M}$ varies over its allowed values. The plane represents a surface of constant free energy. Intersection of the plane with the surface corresponds to a state with broken symmetry ( from \cite{Kim3}).}
	\label{Kim}
\end{figure}

To analyze this confusing state of affairs, we apply in this article a geometrical method that was developed by Kim~\cite{Kim3} in the context of symmetry breaking of the Higgs potential. This method starts from a vector space spanned by a set of linearly  independent invariants constructed from the order parameter.  By letting the order parameter range over all possible values, a volume is generated in the invariant space. This ``Kim volume" is a purely mathematical construct that is independent of the parameters of the physical system. For the present case, the invariants are polynomial expressions of the $c_{l,m}$ parameters in Eq. (\ref{OP}). A schematic example of  a Kim plot is shown in Fig. \ref{Kim} for the case of three invariants $\lambda_{1-3}$. Next, families of constant free energy hypersurfaces are constructed by allowing the physical parameters to vary. A broken symmetry state is associated with a point where such a constant free energy surface touches the Kim volume~\cite{Kimgeneralization}. 

In Section II we show how these invariants can be constructed systematically in terms of the $c_{l,m}$ parameters. Next, we practice with the Kim method for the simpler cases of $l=2$, $l=6$, and $l=7$.  In Section III we apply this method to orientational ordering in the $l=15$ sector, the $l=16$ sector and the combined $l=15+16$ sector. The fact that the icosahedral state is unstable in the pure $l=15$ sector and largely unstable in the pure $l=16$ sector is confirmed in full generality for variational free energies with local invariants. We also confirm that stable icosahedral states appear in the mixed $l=15+16$ sector. Finally, we show that the principles of Landau theory can be ``saved" if one includes the non-local invariants, at least for the case of the $l=15$ sector. Using diagrammatic perturbation theory, we show that the non-local invariants are generated \textit{from a purely local variational free energy} when one integrates out the $l\neq15$ sectors. At least formally, an orientational symmetry-breaking transition from an isotropic to an $l=15$ icosahedral state can be constructed with the $l=15$ sector---including the thin sliver of intervening tetrahedral states that was noted in the earlier numerical work---using a Kim construction with non-local invariants. More generally, a mixed $l=15+16$ representation provides an economical description of such transitions.

\section{Invariants and the Kim Method.}
 
In this section we lay the mathematical groundwork. We first demonstrate a systematic construction method to obtain the independent invariants for given $l$ based on the \textit{Wigner 3-j} symbols. The method generalizes that of Ref. \cite{jaric} for the $l=6$ case. Next, we construct Landau free energies, in the form of sums of invariants, up to quartic order for the $l=2$, $l=6$, and $l=7$ cases. The $l=6$ case will be the prototype of a discontinuous orientational ordering transition that is, from the viewpoint of the Kim method \cite{Kim3}, non-trivial while the $l=7$ case will be the prototype of a non-trivial continuous orientational ordering transition.  Finally, we demonstrate how the Kim geometrical method (``Kim construction") works for these  prototypes. 
\subsection{Construction of Invariants.}
A square-integrable function defined on the surface of a sphere, such as the mass density $\rho(\theta, \phi)$, can be expanded in a spherical harmonics series in the form $\rho(\theta, \phi)=\sum_{l=0}^\infty\sum_{m=-l}^l c_{l,m}Y_l^m(\theta, \phi)$. If the scalar quantity of interest is to be real (as will be the case throughout this article) then
$c_{l,m} = (-1)^m c_{l,-m}^*$. 
Writing the expansion as 
\begin{equation}
\rho(\theta, \phi)=\sum_{l=0}^\infty\sum_{m=-l}^l\rho_{l,m}(\theta, \phi)=\sum_{l=0}^\infty\rho_l(\theta, \phi)
\label{eq:OP}
\end{equation}
then $\rho_{l,-m}$ is the complex conjugate of $\rho_{l,{m}}$ and $\rho_l$ is real. 

Turning to the Landau variational free energy for the density, if the expansion in powers of $c_{l,m}$ parameters is limited to terms no higher than fourth order and confined to a single value of $l$, then the most general form of the free energy $\mathcal{F}_l$ is
\begin{equation}
\mathcal{F}_l = \sum_{k=2}^{4} \sum_{n=1}^{n_l^{(k)}}f_{k,n}I^{(k,l)}_n \label{eq:singlel4}
\end{equation}
Here, $n_l^{(k)}$ is the number of independent $k^{\rm th}$ order invariant polynomials in the $c_{l,m}$ parameters for the $l$ value in question. Its value is determined by the Molien polynomial  \cite{Mukai}. Next, $I^{(k,l)}_n$ is the $n^{\rm th}$ invariant polynomial, a system-independent mathematical construct. Finally, the expansion coefficients $f_{k,n}$ depend on the thermodynamic parameters of the particular physical system in question. As discussed further in Appendix \ref{app:correlations}, the expansion coefficients $t_l/2=f_{2,1}$ of the $k=2$ quadratic invariant can be related to correlation functions of the system such as the linear susceptibility, the static structure factor, and the pair distribution function. For convenience, we will refer to $t_l$ as the ``reduced temperature" of the system.

\begin{figure}[htbp]
\begin{center}
\includegraphics[width=3in]{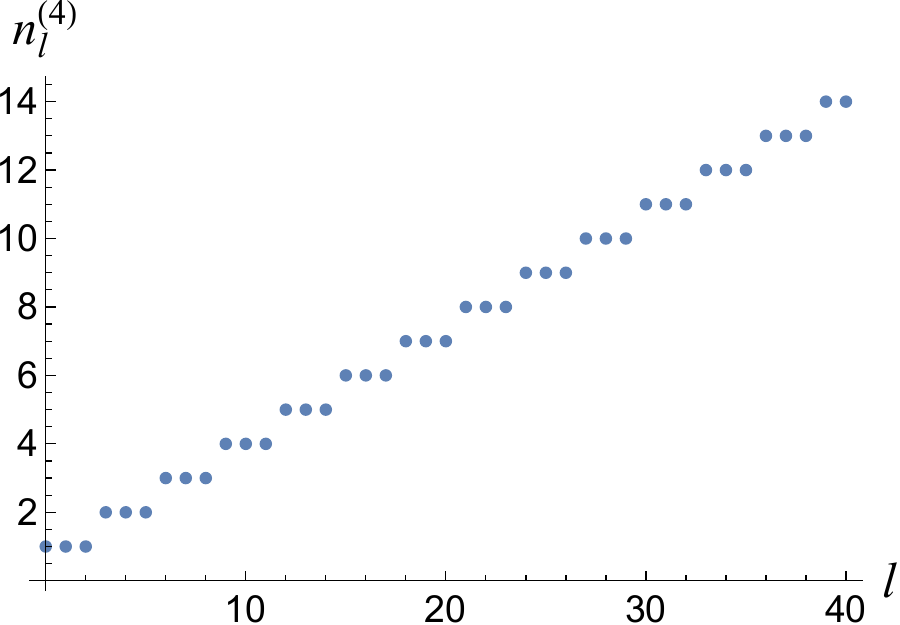}
\caption{The number of distinct quartic terms, $n^{(4)}_l$ plotted for $l$ ranging from 0 to 40. See Appendix \ref{sec:app2}, Eq. (\ref{eq:suppa3}).}
\label{fig:quarticnumber}
\end{center}
\end{figure}

We demonstrate in Appendix \ref{sec:app1}  that, for any value of $l$, there is only one quadratic invariant namely the integral $I^{(2,l)} = \int \rho_l( \theta, \phi)^2 \sin \theta \, d \theta d \phi$. There also is at most one cubic invariant $I^{(3,l)}=\int \rho_l( \theta, \phi)^3 \sin \theta \, d \theta d \phi$ for even values of $l$ and none for odd $l$ values. By contrast, using heuristic arguments we show in Appendix \ref{sec:app2} that the number $n^{(4)}_l$ of independent quartic invariants rises in a step-wise linear manner with $l$, as shown in Fig. \ref{fig:quarticnumber}
(note the triplet grouping).
Two of the quartic invariants are straightforward. They are obtained, respectively, from the integral of the fourth power of $\rho_l( \theta, \phi)$ and the square of the quadratic invariant (\ref{eq:supp6}).  We  term the first invariant the \textit{local} quartic invariant and the second the \textit{trivial} quartic invariant, which is non-local. For $l=0, 1$ and 2, those two quartic invariants are identical to within an overall multiplicative constant while for $l\geq3$ the two invariants differ in form (see Fig. \ref{fig:quarticnumber} and Appendix \ref{sec:app2}, Eq. (\ref{eq:suppa3}) ).

\begin{figure}[htbp]
\begin{center}
\includegraphics[width=3in]{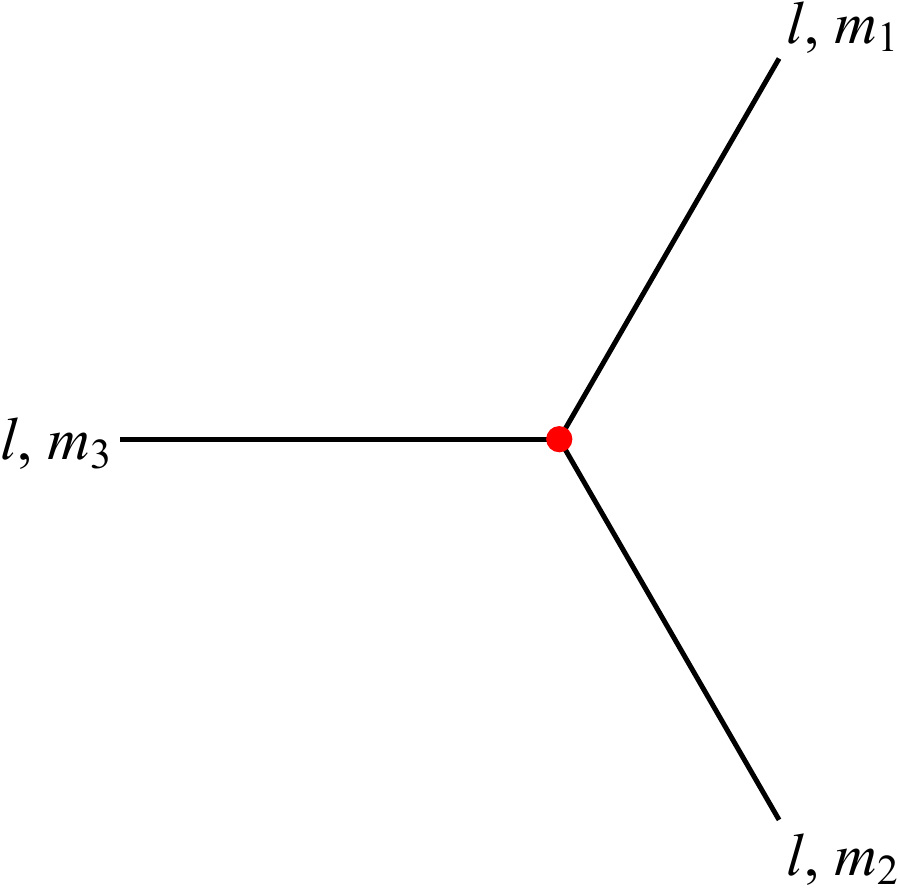}
\caption{Graphical representation of the quantity $ \mathcal{V}(l_1,m_1,l_2,m_2,l_3,m_3)$, as defined in (\ref{eq:singlel14}).}
\label{fig:vertex}
\end{center}
\end{figure}

\begin{figure}[htbp]
\begin{center}
\includegraphics[width=3in]{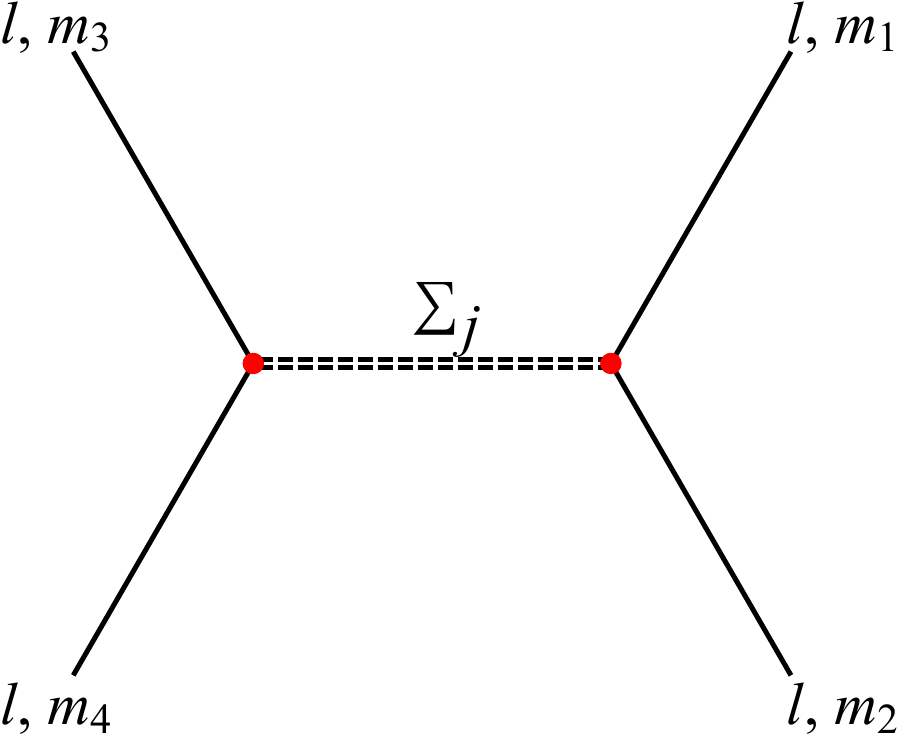}
\caption{Graphical representation of the product $\sum_j\mathcal{V}(l,m_1,l,m_2,j,-m_1-m_2)\mathcal{V}(l,m_3,l,m_4,j,-m_3-m_4)(-1)^{m_1+m_2}$ in (\ref{eq:singlel16}). The double dashed line expresses the fact that the overall index $j$ carried by that line is summed over. }
\label{fig:vertex1}
\end{center}
\end{figure}

One can systematically construct the fourth order invariants using the following quantity as a building block \cite{jaric}:
\begin{eqnarray}
\lefteqn{\mathcal{V}(l_1,m_1,l_2,m_2,l_3,m_3)} \nonumber \\ & = & \int Y_{l_1}^{m_1}(\theta, \phi) Y_{l_2}^{m_2}(\theta, \phi)  Y_{l_3}^{m_3}(\theta, \phi) \sin \theta \, d \theta d \phi  \nonumber \\ &  = & \sqrt{\frac{(2l_1+1)(2l_2+1)(2l_3+1)}{4 \pi}}\left(\begin{array}{ccc} l_1 & l_2 & l_3 \\ 0 & 0  & 0 \end{array} \right) \nonumber \\ && \times \left(\begin{array}{ccc} l_1 & l_2 & l_3 \\ m_1 & m_2  & m_3 \end{array} \right) \label{eq:singlel14}
\end{eqnarray}
The terms in brackets are Wigner $3j$ symbols \cite{Wigner}. For example, the cubic expression
\begin{equation}
\sum_{m_1,m_2,m_3=-l}^l\mathcal{V}(l,m_1,l,m_2,l,m_3) c_{l,m_1}c_{l,m_2}c_{l,m_3} \delta_{m_1+m_2+m_3} \label{eq:singlel15}
\end{equation}
is the invariant under rotation that is proportional to the local cubic invariant $\int \rho_l( \theta, \phi)^3 \sin \theta \, d \theta d \phi$. Next, the fourth order local invariant, arising from the integral of the fourth power of the density $\rho_l(\theta, \phi)$ over the surface of the sphere, can be expressed in terms of the Wigner $3j$ symbols as
\begin{eqnarray}
\lefteqn{\sum_{j=0}^{2l}\sum_{m_1,m_2,m_3,m_4 =-l}^l(-1)^{m_1+m_2} \mathcal{V}(l,m_1,l,m_2,j,-m_1-m_2)} \nonumber \\ & \times \mathcal{V}(l,m_3,l,m_4,j,-m_3-m_4)\nonumber c_{l,m_1}c_{l,m_2}c_{l,m_3}c_{l,m_4} \nonumber \\ & \times \delta_{m_1+m_2+m_3+m_4} \label{eq:singlel16}
\end{eqnarray}
Both (\ref{eq:singlel15}) and (\ref{eq:singlel16}) can be depicted graphically.
Figure \ref{fig:vertex} shows the graphical element for  $\mathcal{V}(l_1,m_1,l_2,m_2,l_3,m_3)$, while Fig. \ref{fig:vertex1} depicts the combination of $\mathcal{V}$'s in (\ref{eq:singlel16}).
Not all values of $j$ contribute to the summation over the internal line: for odd values of $j$ the expression evaluates to zero.

\begin{figure}[htbp]
\begin{center}
\includegraphics[width=3in]{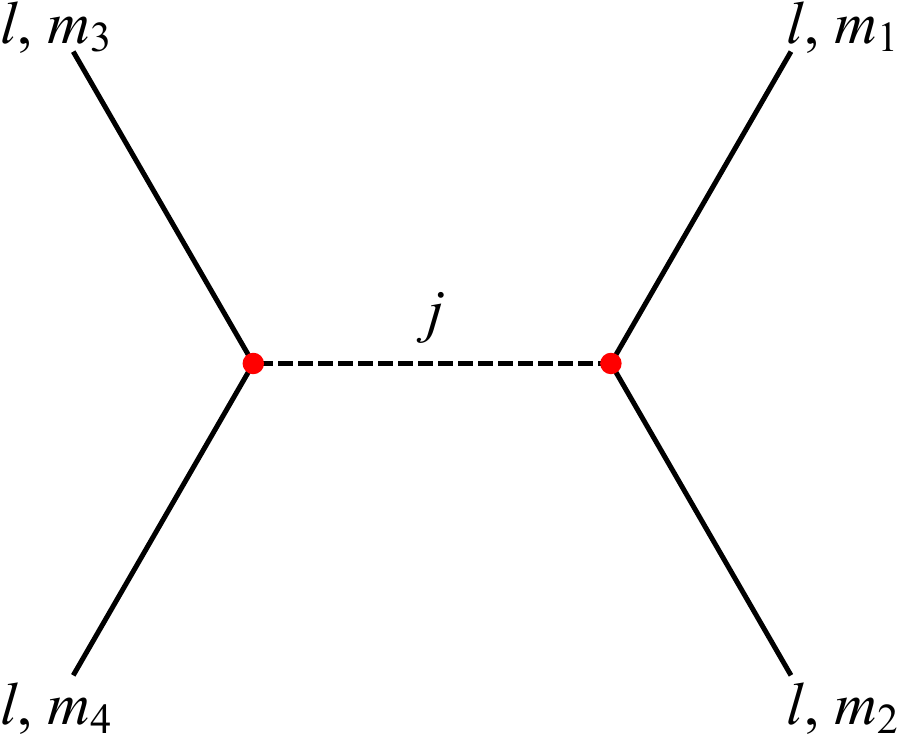}
\caption{Graphical representation of a quartic invariant generated from the third order vertex in Fig. \ref{fig:vertex} mediated by a single value of $j$. This diagram corresponds to the expression $\mathcal{V}(l,m_1,l,m_2,j,-m_1-m_2)\mathcal{V}(l,m_3,l,m_4,j,-m_3-m_4)(-1)^{m_1+m_2}$, the intermediate value of $j$ having been fixed.}
\label{fig:vertex2}
\end{center}
\end{figure}

Figure \ref{fig:vertex1} provides a clue how additional quartic invariants could be generated: the individual terms in the summation of the internal line over different $j$ each are \textit{separately} rotational invariants. The reason is that a rotation of the sphere in 3-space in general scrambles the $2l+1$ coefficients $c_{l,m}$ when the z axis is rotated. However, such a rotation cannot mix $c_{l,m}$ coefficients with $c_{l^{\prime},m^{\prime}}$ coefficients when $l^{\prime} \neq l$, since they belong to different irreducible representations. Similarly, rotations will not scramble the different $j$ terms in the summation. The new terms can be graphically represented as shown in Fig. \ref{fig:vertex2}. 
It is easy to check that the $j=0$ term corresponds to the trivial invariant. 

It might seem that this method provides a scheme to construct \textit{infinitely many} quartic invariants for all even $j$, though we know that these invariants can not be all independent. First, $j$ and the two $l$'s must satisfy the triangle inequalities: $0 \leq j \leq 2l$. For a given $l$, start from $j=0$ and recall that the trivial quartic invariant is separate from the local quartic invariant for $l\geq3$. Next, go to the $j=2$ case and check if this generates an independent invariant, which is the case for $l\geq 6$ (see Appendix \ref{sec:app2}). This suggests a pattern and one can indeed repeat this for any even $j$ until $j=2l$ after which no more invariants are generated. For all values of $l$ that we checked, this method produces the full number of independent invariants that is imposed by the Molien polynomial.  

The non-local invariants also can be viewed as being generated from the expression 
\begin{equation}
\int d \theta d \phi \int d \theta^{\prime} d \phi^{\prime} \left[ \rho(\theta, \phi)^2 \mathcal{K}_j( \theta, \phi, \theta^{\prime} \phi^{\prime}) \rho( \theta^{\prime}, \phi^{\prime})^2 \right] \label{eq:singlel18}
\end{equation} 
where 
\begin{equation}
\mathcal{K}_j ( \theta, \phi, \theta^{\prime} \phi^{\prime}) = \sum_{m=-j}^jY_j^m(\theta, \phi) Y_j^{m}(\theta^{\prime}, \phi^{\prime})^* \label{eq:singlel19}
\end{equation}
which, by inspection, is a rotational invariant for any $j$. The new invariants are produced by replacing the squared scalar densities with $\rho_l^2$. The graphical representation of this term is reminiscent of a Feynman diagram for the interaction between two particles mediated by the exchange of a mode with ``propagator" $\mathcal{K}_j$ so we will call these ``mediated invariants". In Appendix \ref{app:shape} we discuss how non-local invariants can be generated by coupling the density profile $\rho$ to a harmonic scalar field such as the \textit{shape profile} associated with variation of the sphere radius. The mediated invariants appear after the shape variables are integrated over. The coefficients of the mediated quartic invariants produced in this manner are always negative.

To recapitulate: the Landau free energy with invariants up to quartic order constructed from a single value of $l$ has only one quadratic invariant, obtained by integrating the square of the relevant density, $\rho_l(\theta, \phi)$, over the surface of the sphere. Next, it  has at most one cubic invariant, obtained in the same way from the cube of $\rho_l(\theta, \phi)$, or alternatively from Wigner $3j$ symbols via (\ref{eq:singlel15}); when $l$ is odd, the cubic invariant evaluates to zero. Quartic invariants will vary in number, depending on the value of $l$, but there is always the local invariant, obtained by integrating $\rho_l(\theta, \phi)^4$ over the surface of the sphere, and the trivial invariant, obtained by squaring the quadratic invariant. From $l=0$ to $l=2$ those two invariants are identical to within a multiplicative constant. For $l=6$ and above there are additional, mediated, quartic invariants, which can be obtained via the approach illustrated in Fig. \ref{fig:vertex2}, up to the required number shown in Fig. \ref{fig:quarticnumber}.

\subsection{Landau variational energies for $l=2$, $l=6$, and $l=7$.}
Having in hand a systematic construction method for the invariants, we are now in a position to construct Landau energies for specific cases.
\subsubsection{Landau energy for l=2}
 We start with $l=2$, which is realized by the familiar case of nematic ordering in liquid crystals~\cite{CL}. We will reformulate the standard treatment in a manner that brings out the connection with the Kim method. 
 Configuration space is five dimensional for $l=2$. Imposing the condition that the density is real leads to the relations
\begin{eqnarray}
c_{2,-2} & = & r_1 + i s_1 \label{eq:singlel6} \\
c_{2,-1} & = & r_2+is_2 \label{eq:singlel7} \\
c_{2,0} & = & \sqrt{2} r_3 \label{eq:singlel8} \\
c_{2,1} & = & -(r_2-is_2) \label{eq:singlel9} \\
c_{2,2} & = & r_1 -is_1 \label{eq:singlel10}
\end{eqnarray}
where the $r_l$'s and the $s_l$'s are real numbers. This parameterization is readily generalized to arbitrary values of $l$. In terms of these variables, the quadratic invariant has the form, 
\begin{eqnarray}
\langle \rho_2^2 \rangle & = & \int \rho_2(\theta, \phi)^2 \sin \theta \, d \theta \,d \phi \nonumber \\ &=&\sum_{m=-2}^2 c_{2,m}c_{2,-m}(-1)^m \nonumber \\ & = & 2(r_1^2+r_2^2+r_3^2+s_1^2 +s_2^2) \label{eq:singlel11}
\end{eqnarray}
Next, we construct a linear, five-dimensional vector space (``configuration space") from the five variables $r_i$ and $s_k$, with five-component vectors defined as
\begin{equation}
\left(\begin{array}{c} r_2 \\ r_2 \\  r_3 \\ s_1 \\ s_2 \end{array}\right)
\end{equation} 
The quadratic form on the right hand side of (\ref{eq:singlel11}), the square of the modulus of the five-component vectors, corresponds to the unique quadratic invariant under rotations in configuration space. As is well known, infinitesimal rotations in three-dimensional space (``Euclidean space") are generated by three 3-by-3 anti-symmetric matrices. There are three 5-by-5 matrices in configuration space that correspond to the three generators in Euclidean space:

$x$ axis:

\begin{equation}
\left(
\begin{array}{ccccc}
 0 & 0 & 0 & 0 & 1 \\
 0 & 0 & 0 & 1 & 0 \\
 0 & 0 & 0 & 0 & \sqrt{3} \\
 0 & -1 & 0 & 0 & 0 \\
 -1 & 0 & -\sqrt{3} & 0 & 0 \\
\end{array}
\right)
\end{equation}

$y$ axis:

\begin{equation}
\left(
\begin{array}{ccccc}
 0 & -1 & 0 & 0 & 0 \\
 1 & 0 & -\sqrt{3} & 0 & 0 \\
 0 & \sqrt{3} & 0 & 0 & 0 \\
 0 & 0 & 0 & 0 & -1 \\
 0 & 0 & 0 & 1 & 0 \\
\end{array}
\right)
\end{equation}

$z$ axis:

\begin{equation}
\left(
\begin{array}{ccccc}
 0 & 0 & 0 & 2 & 0 \\
 0 & 0 & 0 & 0 & 1 \\
 0 & 0 & 0 & 0 & 0 \\
 -2 & 0 & 0 & 0 & 0 \\
 0 & -1 & 0 & 0 & 0 \\
\end{array}
\right)
\end{equation}

Since these are anti-symmetric, it follows that an infinitesimal rotation in Euclidean space generates an infinitesimal rotation in configuration space. Similarly, finite rotations in Euclidean space generate finite rotations in configuration space. Next, in terms of the real expansion coefficients, the single cubic invariant is
\begin{eqnarray}
\lefteqn{\langle \rho_2^3 \rangle} \nonumber \\ & = & \int \rho_2(\theta, \phi)^3 \sin \theta \, d \theta \, d \phi \nonumber \\ & = & -\frac{6}{7} \sqrt{\frac{10}{\pi }} r_3 s_1^2+\frac{3}{7} \sqrt{\frac{10}{\pi }} r_3
   s_2^2-\frac{3}{7} \sqrt{\frac{30}{\pi }} r_1 s_2^2\nonumber \\ &&+\frac{6}{7} \sqrt{\frac{30}{\pi }}
   r_2 s_1 s_2 +\frac{2}{7} \sqrt{\frac{10}{\pi }} r_3^3 -\frac{6}{7} \sqrt{\frac{10}{\pi
   }} r_1^2 r_3 \nonumber \\ &&+\frac{3}{7} \sqrt{\frac{10}{\pi }} r_2^2 r_3+\frac{3}{7}
   \sqrt{\frac{30}{\pi }} r_1 r_2^2 \label{eq:singlel12}
\end{eqnarray}
This expression is \textit{not} invariant under general rotations in configuration space though---by construction---it still is an invariant under rotations in Euclidean space. 
Finally, the unique fourth-order term 
\begin{eqnarray}
\langle \rho_2^4\rangle  =  \frac{15 \left(r_1^2+r_2^2+r_3^2+s_1^2+s_2^2\right){}^2}{7 \pi } \label{eq:singlel13}
\end{eqnarray}
is invariant under rotations in both Euclidean and configuration space. 

Define spherical coordinates in configuration space with $A$ the modulus and with the four angular variables $\psi_k$'s, with $k=1,2,3,4$, determining direction in configuration space. In these coordinates, the $l=2$ Landau free energy has, up to fourth order, the general form
\begin{equation}
\mathcal{F}_2(A,\psi_k) = \frac{t_2}{2} A^2 + \frac{u}{3} A^3 Q_3(\{\psi_k\}) + \frac{v}{4} A^4 \label{eq:singlel25}
\end{equation} 
The system parameters $t_2$, $u$ and $v$ incorporate information about the physics of the particular system in question while $Q_3(\{\psi_k\})={\langle \rho_2^3 \rangle}/A^3$ is a combination of trigonometric functions of the four angles $\psi_k$ that is universal in the sense that it does not depend on the system parameters. 

Fix the set of angles $\{\psi_k\}$ and decrease $t_2$, starting from a large, positive value. For large and positive $t_2$, the only solution of the equation $\frac{\partial\mathcal{F}(A,\{\psi_k\})}{\partial A}=0$ is $A=0$, which is the symmetric state with $\mathcal{F}_2(A=0,\{\psi_k\})=0$. The transition temperature $t_c(\{\psi_k\})$ for the first-order phase transition is obtained by demanding that the pair of equations $\mathcal{F}_2(A,\{\psi_k\})=0$ and $\frac{\partial\mathcal{F}_2(A,\{\psi_k\})}{\partial A}=0$ has a non-trivial solution. Eliminating $A$ gives an expression for the transition temperature in terms of the direction in configuration space:
\begin{equation}
2t_c (\{\psi_k\})= \frac{u^2}{v} [Q_3(\{\psi_k\})]^2
\end{equation}
The equation marks the rotational symmetry breaking transition, along a particular direction in configuration space. Now, allow the set of angles $\{\psi_k\}$ to vary. Symmetry breaking takes place \textit{at the highest possible value} of $t_c (\{\psi_k\})$. The determination of the prevailing symmetry for orientational ordering in the $l=2$ sector is reduced to the purely mathematical question of determining the maximum of the modulus of the universal expression $Q_3(\{\psi_k\})$ in configuration space. Numerical minimization of  $|Q_3|$ is straightforward. Only $r_3$ is non-zero at the maximum of $|Q_3|$, which corresponds to the expected $l=2, m=0$ nematic state for the case of liquid crystals \footnote{Historically, maximizing the cubic invariant was the criterium proposed by Alexander and McTague \cite{AandM} in their pioneering study of melting viewed as an orientational phase transition.}.

\subsubsection{Landau energy for l=6}
Next we turn to $l=6$, the case explored by SNR. For $l=6$, configuration space expands to $12+1=13$ dimensions. There is, as always, only one quadratic and one cubic invariant, but now there are three quartic invariants: the trivial invariant, the local invariant, and the $j=2$ mediated invariant---or, equivalently, any three independent linear combinations of those three invariants.
It is instructive to start by first including only the local quartic invariant $Q_{4,1}$. Following the same steps as for $l=2$, the variational energy is
\begin{equation}
\mathcal{F}_6 = t_6A^2+\frac{u}{3} A^3 Q_3(\{\psi_i\}) + \frac{v}{4}A^4 Q_{4,1}(\{ \psi_i\}) \label{eq:osph48}
\end{equation}
where the set $\{ \psi_i\}$ refers to the 12 angular variables that collectively define a direction in the 13 dimensional configuration space. Next, apply the method we used to determine the transition temperature for $l=2$, i.e., set the derivative with respect to A to zero for a fixed set of angles $\{\psi_i\}$
\begin{eqnarray}
0 & = &  \frac{\partial \mathcal{F}_6}{\partial A}  \nonumber \\ & = & A\left(2t_6 + uAQ_3( \{ \psi_i \} ) + v A^2 Q_{4,1}( \{\psi_i \} ) \right) \label{eq:osph49}
\end{eqnarray}
and then demand that at the transition point the free energy itself must be zero
\begin{eqnarray}
0 & = &\mathcal{F}_6 \nonumber \\ & = & A^2 \left(t + \frac{u}{3} A Q_3 ( \{ \psi_i \}) + \frac{v}{4} A^2 Q_{4,1} (\{ \psi_i \} ) \right) \label{eq:osph50}
\end{eqnarray}
Solving the the simultaneous equations (\ref{eq:osph49}) and (\ref{eq:osph50}), for $t_6$ and non-zero $A$, we find
\begin{eqnarray}
t_6 & = & \frac{Q_3(\{ \psi_i \} )^2 u^2}{9 Q_{4,1}(\{ \psi_i \}) v} \label{eq:osph51} \\
A & = & -\frac{2Q_3(\{ \psi_i \} ) u}{3Q_{4,1}(\{ \psi_i \} ) v} \label{eq:osph52}
\end{eqnarray}
Now, as $t_6$ is lowered, ordering first  occurs for those values of $\{ \psi_i \}$ at which $t_6$ on the right hand side of (\ref{eq:osph51}) takes on the largest value. This means that we must seek the maximum value of the ratio $Q_3( \{ \psi_i \})^2/Q_4(\{ \psi_i \} )$ instead of the $l=2$ criterum of maximizing the modulus of $Q_3$. The numerical effort required to maximize $Q_3( \{ \psi_i \})^2/Q_4(\{ \psi_i \} )$ again is modest: the maximum corresponds to the $l=6$ icosahedral spherical harmonic, in agreement with SNR. The transition is again first order.

Now include the $j=2$ mediated quartic invariant and also the trivial non-local quartic invariant. The full $l=6$ Landau free energy can be expressed as 

\begin{equation}
{\mathcal{F}_6} = \frac{t_6}{2} A^2 + \frac{u}{3} A^3 Q_3+ \frac{1}{4} A^4 \big[aQ_{4,1}+bQ_{4,2}+c\big] \label{eq:singlel25}
\end{equation} 
where the subscript $4,1$ indicates the local quartic invariant and the subscript $4,2$ the $j=2$ mediated quartic invariant. Next, $a(=v)$, $b$ and $c$ are three system-dependent expansion coefficients. The coefficient $c$ accounts here for the trivial invariant. Thermodynamic stability requires that only coefficients $a, b, c$ are permitted such that for any set of angular variables the complete quartic term is positive. The transition temperature is 
\begin{eqnarray}
t_6 & = & \frac{u^2}{9}\frac{Q_3^2}{[aQ_{4,1}+bQ_{4,2}+c]}  \nonumber \\ \label{eq:osph51} 
\end{eqnarray}
(we suppressed here the dependence on the angular variables). The optimal direction in the 13 dimensional configuration space corresponds to the maximum of $Q_3^2/(aQ_{4,1}+bQ_{4,2}+c)$. Unlike the $l=2$ case, numerical maximization is no longer straightforward because the quantity to be maximized ($t_6$) now depends on the physical parameters $a$, $b$, and $c$.  

\subsubsection{Landau energy for l=7}
Unlike the $l=2$ and $l=6$ cases, $l=7$ has no (as yet known) physical realization. However, it represents an important contrast when it is compared to the $l=6$ case; there is no cubic invariant because $l$ is odd. This means that an $l=7$ orientational ordering transition will be continuous in Landau mean-field theory. Just as for $l=6$, there are three independent quartic invariants: the local invariant, the trivial invariant and the $j=2$ mediated invariant.  For $l=7$, configuration space has 15 dimensions with 14 angular variables $ \{ \psi_k \}$ plus the modulus $A$. The Landau variational free energy has, up to fourth order, the form:
\begin{equation}
\mathcal{F}_7 = \frac{t_7}{2} A^2 + \frac{1}{4} A^4 \left[aQ_{4,1}( \{ \psi_k \})+bQ_{4,2} (\{ \psi_k \})+c\right] \label{eq:singlel20}
\end{equation}
The four coefficients $a, b, c$ and $t_7$ are again system-dependent parameters while the expressions $Q_{4,1}( \{ \psi_k \})$ and $Q_{4,2}( \{ \psi_k \})$ are universal functions of the angular variables. 

For any set of angular variables, the critical point for the continuous symmetry breaking transition is now $t_7=0$, which provides no information about the selection of the angular variables. However, for $t_7< 0$ minimization of the the free energy with respect to $A$ leads to
\begin{equation}
t_7 + A  \left[aQ_{4,1}( \{ \psi_k \})+bQ_{4,2} (\{ \psi_k \})+c\right]  =0 \label{eq:singlel21}
\end{equation} 
The corresponding angle-dependent free energy is
\begin{equation}
\mathcal{F}_{\rm min}( \{ \psi_k \}) = - \frac{t_7^2}{4 \left[aQ_{4,1}( \{ \psi_k \})+bQ_{4,2} (\{ \psi_k \})+c\right] } \label{eq:singlel22}
\end{equation}
Minimization of this expression with respect to the angle variables determines the broken symmetry state. This means that for $l=7$ the broken symmetry must be determined by minimizing the positive quantity 
\begin{equation}
\Lambda=aQ_{4,1}( \{ \psi_k \})+bQ_{4,2} (\{ \psi_k \})+c  \label{eq:singlel22a}
\end{equation}
which is again dependent on the system parameters.

\subsection{Kim constructions for $l=6$ and $l=7$}

The Kim construction \cite{Kim3} can be used as a graphical method for performing the minimization of system-dependent quantities such as $\Lambda$ and ${Q_3^2}/{\Lambda}$ in a manner that  reveals system-independent information about competing broken symmetry states. Here, we apply the method to the cases $l=6$ and $l=7$. 
\subsubsection{l=7}

We will start with $l=7$, in which case we need to minimize $\Lambda$ as defined in (\ref{eq:singlel22a}). Construct a two-dimensional invariant vector space with linear combinations of the two independent non-trivial quartic invariants $Q_{4,1}$ and $Q_{4,2}$ as coordinate axes. While any independent pair of combinations of $Q_{4,1}$ and $Q_{4,2}$ can be used as coordinate pairs, we found, by trial and error, that $X=Q_{4,1}+Q_{4,2}$ and $Y=Q_{4,2}$ with $\Lambda(X,Y)=a(X-Y)+bY+c$ was a convenient choice for revealing the structure of the Kim regions in a more readily observable fashion. The set of points in the $X$-$Y$ plane for which $\Lambda(X,Y)=a(X-Y)+bY+c$ is constant is then a straight line. Let $\theta$ be the angle this line with the $X$ axis. It is convenient to absorb an overall factor $v=\sqrt{a^2+b^2}$ in $\Lambda$ and express the line as
\begin{equation}
\Lambda(X,Y)=(X-X_0)\cos\theta+(Y-Y_0)\sin\theta
\end{equation}
Note that a change of the values of the parameters $X_0$ and $Y_0$ amounts to an affine translation of the lines that leaves angles unchanged. 

\begin{figure}[htbp]
\begin{center}
\includegraphics[width=3in]{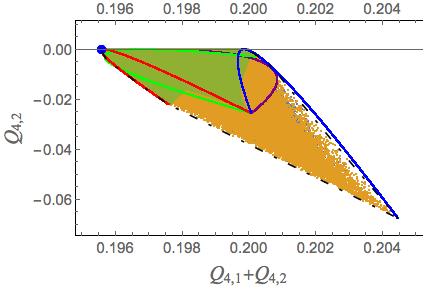}
\caption{Collection of allowed states for $l=7$ in the $X$-$Y$ plane where $X=Q_{4,1}+Q_{4,2}$ and $Y=Q_{4,2}$. Red curve: sixfold axis; Green: fivefold axis, Purple: fourfold axis; Blue: sevenfold axis. The long and short dashed line is the boundary of the dark orange stippled two-fold symmetry region; see Appendix \ref{app:boundaries}. The green region corresponds to three-fold symmetry. Finally, the blue dot on the upper left hand corner is a point of tetrahedral symmetry.}
\label{fig:Kimplotnewvars}
\end{center}
\end{figure}

Next, let the set of 14 angles $\{ \psi_k \}$ in the 15 dimensional configuration space adopt all mathematically allowed values. This generates the colored area in the $X-Y$ space shown in Fig. \ref{fig:Kimplotnewvars}. 
which is a first example of a Kim plot. Note that this volume still is independent of the physical system parameters. Additional symmetries can be imposed that generate subsets of the Kim plot. For example, the dark-orange colored area has two-fold symmetry, and the dark-green region corresponds to three-fold symmetry. As described in the caption, the curves in the Kim plot corresponds to the imposition of a four-fold, five-fold, six-fold, and seven-fold symmetry axis. Those curves meet at a point in the interior corresponding to $C_{\infty}$ symmetry. Imposition of tetrahedral symmetry leads to a single point in the plot (blue dot in the upper left-hand corner). This means that tetrahedral symmetry corresponds to unique values of $Q_{4,1}$ and $Q_{4,2}$. Within our numerical precision, the blue dot lies on a sharp corner of the perimeter of the plot.  Note that the $l=7$ Kim plot gives the impression of being the projection of a surface from a higher dimensional space (as indeed it is).

\begin{figure}[htbp]
\begin{center}
\includegraphics[width=3in]{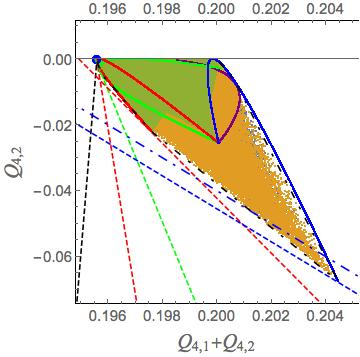}
\caption{Kim construction for $l=7$. The dashed lines are lines of constant $\Lambda$ that graze the Kim plot. They are drawn for increasing values of the system parameter $\theta$, starting with the dashed line for which $\theta=-0.01$. The point where it grazes the Kim plot has tetrahedral symmetry, indicated by the blue dot; Red dashed lines: $\theta=0.01$ and $\theta=0.11$: a six-fold axis; Green dashed line: $\theta=0.05$: a five-fold axis; Blue dashed line: $\theta=0.2$: a seven-fold axis.  The long and short black dashed curves are the boundary of the orange region in the plot, corresponding to two-fold symmetry. Finally, the blue long and short dashed line is parallel to the blue five-fold axis line and is associated with the same thermodynamic parameters. However, this line intrudes into the interior of the Kim plot and corresponds to a higher free energy.}
\label{fig:Kp1}
\end{center}
\end{figure}

The next step is the Kim construction. This involves, for l=7, drawing lines of constant $\Lambda(X,Y)$ in the Kim plot with different lines corresponding to a different sets of physical system parameters. These are the surfaces of constant free energy referred to in the Introduction. Examples are shown in Fig. \ref{fig:Kp1} for the case that $a$ is positive while $b$ and $c$ are negative. Recall that negative $b$ and $c$ can represent the physical case of coupling between density and an additional scalar field.
The dashed lines in Fig. \ref{fig:Kp1} are lines of constant $\Lambda$ for different values of $\theta$. The values of $X_0$ and $Y_0$ were---with the exception of the blue dash-dotted line---chosen so the constant $\Lambda$ line is tangent to the Kim plot. Changing the values of $X_0$ and $Y_0$ for fixed $\theta$ amounts to a parallel shift of the line. Suppose the shift is such that the line lies entirely in the white region, for example by sliding the blue dashed line to the left without changing its slope. The value of $\Lambda$ is reduced by this shift and this would lower the free energy  Eq. (\ref{eq:singlel22}). However, symmetry breaking is not possible in this case since there is no set of allowed invariants corresponding to the set of angles $\psi_i$ that are allowed along the line. We thus can disregard constant $\Lambda$ lines that lie outside the Kim plot.
Next, shift the blue dashed line to the right without changing its slope (so towards the blue dash-dotted line). The line enters  the interior of the Kim plot. While the states along the dash-dotted line in the interior of the Kim plot are mathematically allowed broken symmetry states, they are \textit{not} the minimum free energy states because the value of $\Lambda$ was increased in order to produce the rightward shift. This means that the free energy Eq. (\ref{eq:singlel22}) increased. We thus can also disregard lines of constant $\Lambda$ that enter the plot. In short: \textit{broken symmetry states that minimize the free energy are represented by straight lines in the Kim plot that graze the border of the plot without entering it.}, which is the essence of the method developed by Kim \cite{Kim3}.

\begin{figure}[htbp]
\begin{center}
\includegraphics[width=3in]{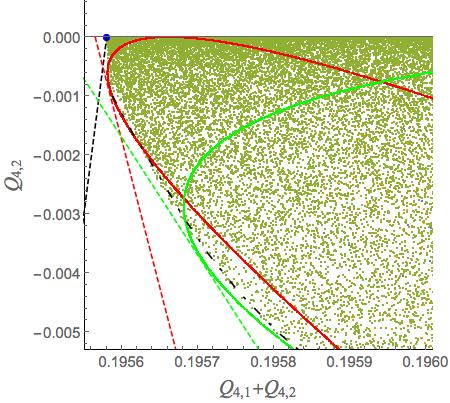}
\caption{Kim construction for the upper left hand portion of the Kim plot shown in Fig. \ref{fig:Kp1}. The colors of the straight lines and corresponds to those of Fig.\ref{fig:Kp1}). }
\label{fig:Kp2}
\end{center}
\end{figure}

We are now in a position to construct a phase diagram for $l=7$ symmetry breaking.  Restricting ourselves to constant energy surfaces in the form of straight lines that touch the Kim plot at one point, there is only one physical parameter that can be varied namely the angle $\theta$ of the lines. Start from the dashed black line that passes through the blue point corresponding to tetrahedral symmetry (with $\theta=-.01$). It is evident that tetrahedral symmetry has a large stability range: since lines passing through the blue point can be drawn over a range of angles. Continue to increase $\theta$. When the angle reaches $0.01$, the constant $\Lambda$ line (red dashed line) grazes the Kim plot at a point where a line of 6-fold symmetry states just touches the border of the Kim plot (see Fig. \ref{fig:Kp2}). Continuing on in this fashion, one finds that the system passes through states with five-fold symmetry, again six-fold symmetry, and then seven-fold symmetry. The stability range is small for six-fold and five-fold symmetries, while the seven-fold symmetry state has a larger stability interval. We observe that prominent asperities of the Kim plot correspond to states with large stability intervals.

\begin{figure}[htbp]
\begin{center}
\includegraphics[width=3in]{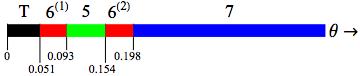}
\caption{Phase plot for l=7 for the case that the system parameter $a$ is positive while $b$ and $c$ are negative. For increasing $\theta$ and $t_7$ negative, the system passes from a tetrahedral state to a state with a single 6-fold symmetry axis, a 5-fold axis, again a six-fold axis, and finally a seven-fold axis. The loci of the transition points are independent of $t_7$. }
\label{fig:T71}
\end{center}
\end{figure}

A phase diagram for l=7 can be obtained by the following steps: (i) construct the Kim plot; (ii) draw a family of lines that graze the Kim plot; (iii) plot the symmetry of the point on the boundary of the Kim plot as a function of the angle of the straight lines. The final step is to determine the relation between the angle $\theta$ and the thermodynamic parameters that enter the Landau variational free energy. If we assume that $a$ is positive and that $b$ and $c$ are negative, then this last step excludes lines that graze the Kim plot along the solid blue border of the Kim plot and along the horizontal border that runs along the top of the plot at $Q_{4,2}=0$. A phase plot is shown in Fig.\ref{fig:T71}.
While the reduced transition temperature for states with different symmetry is the same, that does not mean that there are can be no transitions between states of different symmetry when the physical temperature is varied because the system parameter $\theta$ could depend on the physical temperature.

\begin{figure}[htbp]
\begin{center}
\includegraphics[width=3in]{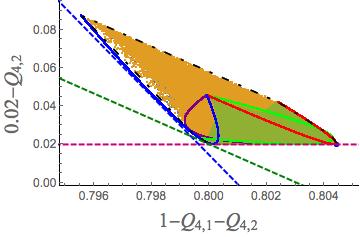}
\caption{Kim construction for the shifted Kim plot with newly-defined quartic invariants. The symmetries are seven-fold (blue dashed line), three-fold (green dashed line) and multiple (red dashed line).}
\label{fig:transplot2}
\end{center}
\end{figure}

\begin{figure}[htbp]
\begin{center}
\includegraphics[width=3in]{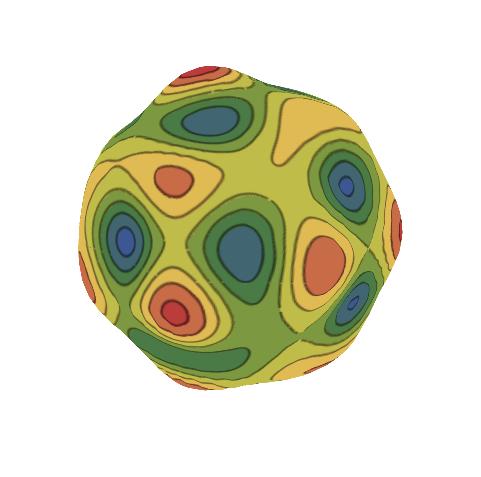}
\caption{Example of the outcome of an ``ordering" transition in the case $\theta=3 \pi/2$ that produces a state with no discernible symmetry . It corresponds to the red line exhibiting extended contact with the boundary of the Kim plot in Fig. \ref{fig:transplot2}.}
\label{fig:nosym}
\end{center}
\end{figure}

\begin{figure}[htbp]
\begin{center}
\includegraphics[width=3in]{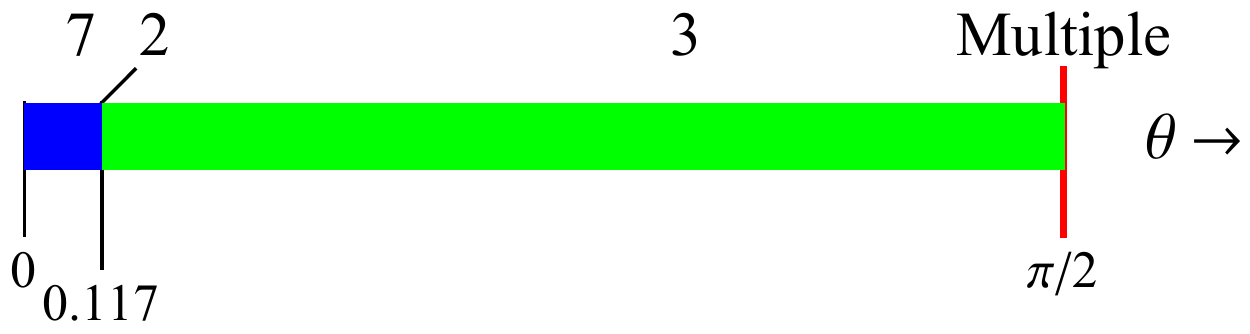}
\caption{Schematic phase diagram for l=7 for the case that the system parameter $a$ is negative while $b$ and $c$ are positive. For increasing $\theta$ and $t_7$ negative, the system passes from a state with a seven-fold axis (blue) to a state with a three-fold axis (green).  As indicated in the figure, there is also a very narrow window between the three-fold and seven fold states in which the minimum free energy possesses two-fold symmetry. If $\theta=\pi/2$ then a continuous degeneracy arises, which allows for a multiplicity of minimum free energy states.}
\label{fig:T72}
\end{center}
\end{figure}

\begin{figure}[htbp]
\begin{center}
\includegraphics[width=3in]{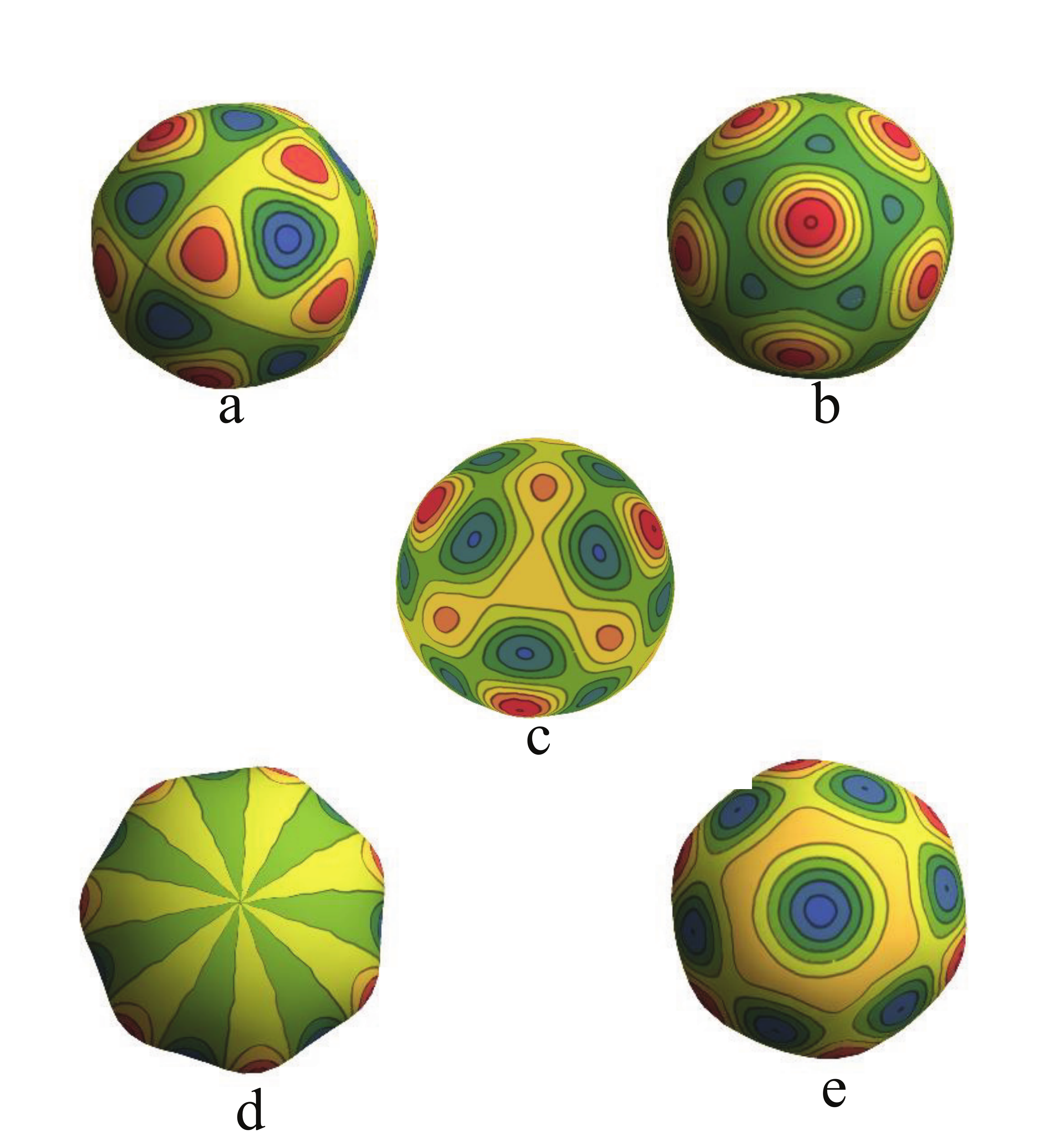}
\caption{Symmetric states encountered for the Kim plot $l=7$; { \bf a}: tetrahedral, { \bf b}: fivefold, { \bf c}: three-fold, { \bf d}: seven-fold, { \bf e}: six-fold.}
\label{fig:shapes}
\end{center}
\end{figure}

Next, assume that $a$ is negative while $b$ and $c$ are positive, with $a$ small enough so the overall sign of the quartic term remains positive. While this is (probably) an unphysical range, this case provides useful insights into the Kim construction. Fig. \ref{fig:transplot2} shows an example of the Kim plot and construction for that case. The net effect is an overall rigid-body translation and rotation of the Kim plot.
The Kim construction now is focused on the vertex of the Kim volume that previously was inaccessible in Fig. \ref{fig:Kp1}.  As noted in the caption to Fig. \ref{fig:transplot2}, there are now three regimes \cite{regimenote}. The ordering has a seven-fold axis in the regime associated with the dashed blue line and a three-fold axis in the regime indicated by the dashed green line. The fourth regime, indicated by the horizontal red line, allows for a variety of symmetries as a result of the degeneracy of state with respect to the local invariant $Q_{4,1}$. This is because of the structure of the mediated invariant $Q_{4,2}$, which is now the sole quartic invariant in the free energy. In this case, the ordered state in this case allows for a continuous, degenerate set of $\{ \psi_k \}$ angles. The ordering possibilities include a two-fold symmetry axis, a three-fold axis, a five-fold axis, a six-fold axis, seven-fold symmetry and tetrahedral ordering. Additionally, the ordering may have no discernible symmetry at all (an example is shown in Fig. \ref{fig:nosym}).
As shown in Appendix \ref{app:degeneracy}, the degeneracy arises from the structure of the quadratic invariant $Q_{4,2}$.  One might expect a physical realization of such a system to have the character of an \textit{orientational glass}. A phase-plot for negative $a$ as a function of $\theta$ is shown in Fig. \ref{fig:T72}.
Figure \ref{fig:shapes} shows realizations of the density associated with some of the other $l=7$ symmetric structures. 

The numerical effort involved in the Kim construction appears to be minimal as compared to a brute-force minimization of the Landau functional in a fifteen dimensional space. This is indeed the case \textit{if} one accurately knows the Kim plot. However, the boundary of the Kim plot of Fig. \ref{fig:Kp1} was obtained by random sampling and along part of the dash-dotted line of Fig. \ref{fig:Kp1} the boundary is quite sparse. This is due to the fact that the Kim plot is in this case actually the projection of a five dimensional volume onto a two dimensional plane. Consequently, random sampling can be expected to generate a far higher fraction of points in the interior of the volume than near its surface. This problem will only become worse for larger values of $l$. Because the precise location of the boundary of the Kim plot is crucial for predicting the possible symmetries of free energy minima, we developed a convenient method to precisely trace out the boundary of the Kim plot for any symmetry of interest, which is described in Appendix \ref{app:boundaries}. 
\newline

\subsubsection{$l=6$} \label{sec:leq6}

\begin{figure}[htbp]
\begin{center}
\includegraphics[width=3in]{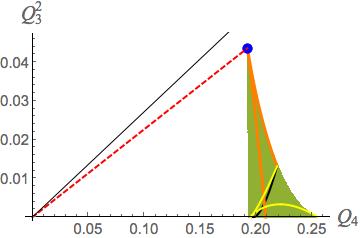}
\caption{Kim plot for l=6 with only the local invariant. Black curve: a single four-fold symmetry axis. Orange curve: single five-fold axis. Yellow curve: a single six-fold axis. The apex of the plot has icosahedral symmetry, as indicated by the blue point. The slope of the dashed red curve indicates corresponds to the highest possible ratio of $Q_3(\{ \psi_i \} )^2/Q_4(\{ \psi_i \} )$ and hence the highest value for $t_6$ at the transition.}
\label{fig:kla1}
\end{center}
\end{figure}

Our next example is the Kim construction for $l=6$. First consider the case that only the local quartic invariant is kept. To find the transition temperature $t_c$, we then only need to maximize the ratio  $Q_3( \{ \psi_i \})^2/Q_4(\{ \psi_i \} )$. It is instructive to do this by adapting the Kim method. Construct a two-dimensional invariant space with $Q_3( \{ \psi_i \})^2$ and $Q_4(\{ \psi_i \}$ as the coordinate axes and construct a Kim plot by random sampling of the two invariants over the allowed set of orientations $\{ \psi_i \}$. As shown in Fig. \ref{fig:kla1}, the resulting Kim plot has three sharp corners. 

The Kim construction that finds $t_c$ involves drawing straight lines starting at the origin in the $Q_4$-$Q_3^2$ plane with various slopes, corresponding to fixed values of $t_6 = Q_3( \{ \psi_i \})^2/Q_4(\{ \psi_i \} )$ for that orientation in the 13 dimensional configuration space. The highest transition temperature corresponds to the line with the highest slope that just grazes the Kim plot at its tip, which has icosahedral symmetry, so the Kim construction reproduces the results of SRN.
For lower values of $t_6$, we need to draw lines of constant free energy $\mathcal{F}_{\rm min}$ in the Kim plot. Such contours are constructed in Appendix \ref{sec:app3}, with an example shown in Fig. \ref{fig:Kp3a}.

Next, include the trivial and $j=2$ mediated invariants. Using the notation of the previous section, the full $l=6$ Landau free energy can be written as 
\begin{widetext}
\begin{eqnarray}
\mathcal{F}_6 &=& \frac{t_6}{2} A^2 + \frac{u}{3} A^3 Q_3 + \frac{v}{4} A^4 \big[(Q_{4,1}+0.95 Q_{4,2})-X_0)\cos\theta+(Q_{4,2}-Y_0)\sin\theta\big] \label{eq:singlel25}
\end{eqnarray} 
and the transition temperature that needs to be maximized is 
\begin{eqnarray}
t_6 & = & (u^2/9u) \frac{Q_3^2 }{[(Q_{4,1}+0.95 Q_{4,2})-X_0)\cos\theta+(Q_{4,2}-Y_0)\sin\theta]}  \nonumber \\ \label{eq:osph51} 
\end{eqnarray}
\end{widetext}
(where we did not explicitly display the dependence on the angle variables $\psi_i$). 

\begin{figure}[htbp]
\begin{center}
\includegraphics[width=3in]{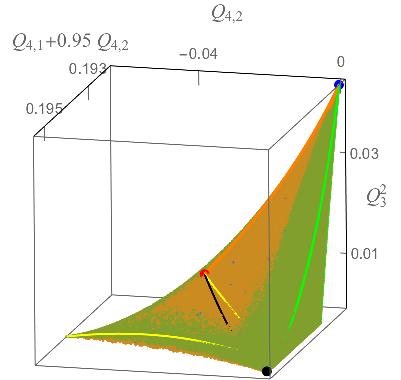}
\caption{The Kim plot for $l=6$ including all quartic invariants. For a description of the curves see accompanying text. } 
\label{fig:Kimplotleq61}
\end{center}
\end{figure}

\begin{figure}[htbp]
\begin{center}
\includegraphics[width=3in]{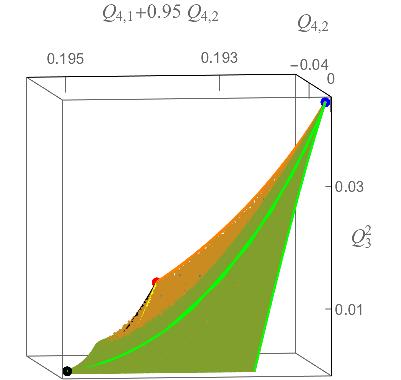}
\caption{Second perspective of the $l=6$ Kim plot.  }
\label{fig:Kimplotleq62}
\end{center}
\end{figure}

The Kim plot is now a three dimensional volume with a tent-like, concave surface spanned between four sharp corners. Redefine the coordinate axes as $X=Q_{4,1}+0.95 Q_{4,2}$, $Y=Q_{4,2}$, and $Z=Q_3^2$ (the small numerical shift in the definition of the X coordinate is for visual convenience.) Figures \ref{fig:Kimplotleq61} and \ref{fig:Kimplotleq62} show different perspectives of the three dimensional Kim plot in the space spanned by these three axes.
The orange curve is a ridge with five-fold symmetry; the green curve on the right hand side of the plot corresponds to tetrahedral symmetry; the blue dot, on the upper boundary of the plot, corresponds to icosahedral symmetry; the black dot at the lower right hand corner of the plot corresponds to octahedral symmetry, and the embedded purple dot at the lower end of the orange curve corresponds to $D_{\infty}$ (full rotational symmetry and mirror reflection about an axis). The black curve corresponds to a four-fold symmetry axis and the yellow curve to six-fold symmetry. A second perspective of the Kim plot is shown in Fig. \ref{fig:Kimplotleq61}. 
The invariant $Q_{4,2}$ is zero in the facing surface. The green tetrahedral symmetry curve has two branches that meet each other and the five-fold symmetry curve at the icosahedral symmetry point. The upper branch ends at the lower left hand corner of the plot, also a sharp point, at the black dot corresponding to the octahedral symmetry point noted earlier.

\begin{figure}[htbp]
\begin{center}
\includegraphics[width=3in]{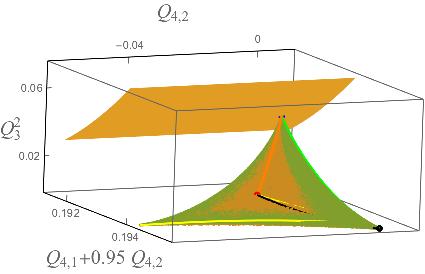}
\caption{The intersection of a constant free energy surface with the icosahedral symmetry point in the $l=6$ Kim plot.}
\label{fig:icosafigure}
\end{center}
\end{figure}

\begin{figure}[htbp]
\begin{center}
\includegraphics[width=3in]{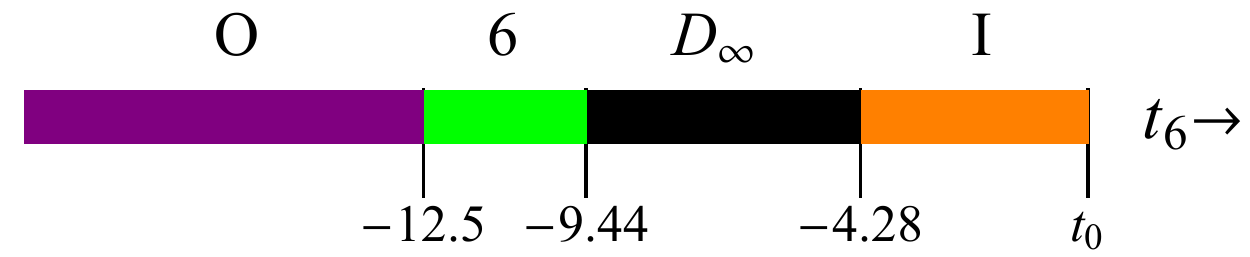}
\caption{Schematic phase diagram for l=6 for the case that the system parameter $a$ is negative while $b$ and $c$ are positive. The octahedral portion continues to arbitrarily large and negative $t_6$. The value of  $t_0$ is 0.0017.}
\label{fig:T6}
\end{center}
\end{figure}

Following the same steps as before, we first use the Kim method to locate the transition temperature $t_c$. This requires maximizing
\begin{eqnarray}
t_6=\frac{Q_3^2}{[(Q_{4,1}+0.95 Q_{4,2})-X_0)\cos\theta+(Q_{4,2}-Y_0)\sin\theta]}  \nonumber \\\label{eq:osph51} 
\end{eqnarray}
Surfaces of constant $t_6$ are planes in the three-dimensional invariant space. The surface with largest $t_6$ value that just grazes the ``Kim volume" determines the symmetry of the first broken symmetry state when $t_6$ is reduced.  For temperatures below $t_c$, we need to construct surface of constant free energy. As can be expected from the case of only local invariants, these surfaces are not flat planes.
Figure \ref{fig:icosafigure} shows an example of a constant free energy surface passing through the point of icosahedral symmetry.
The minimum free energy state \textit{always} has icosahedral symmetry for the physical case that the coefficient $a$ of the local quartic invariant is positive while the coefficients $b$ and $c$ of the non-local invariants are negative. 
We show in Appendix \ref{app:shape} that if the local quartic invariant has a negative coefficient and the non-local quartic invariants positive coefficients, then octahedral, six-fold and $D_{\infty}$ symmetries can be realized, together with icosahedral symmetry,  in agreement with the findings of Jari\'{c} \cite{jaric}.  A phase-plot is shown in Fig.\ref{fig:T6}

\section{Orientational ordering and the Kim construction for $l=15+16$.}

With the experience gained for $l=6$ and $l=7$, we now apply the Kim construction to the $l=15$ and $l=16$ sectors that are the focus of our physical interest. We will restrict ourselves to a variational free energy with only local invariants; even with this simplification, there still are six local cubic and quartic invariants in the combined $l=15+16$ sector. While we know of no simple method that would allow us to carry out complete Kim constructions in a six dimensional space, it is possible---as we will demonstrate---to combine Kim plots for restricted versions of the variational free energy with numerical minimization to arrive at a reasonably complete analysis. We start by examining the $l=15$ and $l=16$ sectors separately . 

\subsection{The $l=15$ sector.}
As for $l=7$, the $l=15$ ordering transition is continuous because of the absence of a cubic invariant. If only the single local quartic invariant is incuded, then the state with minimum free energy corresponds to the minimum of the ratio of quartic invariant $Q_4$ and the square of the quadratic invariant $Q_2$:
\begin{equation}
\mathcal{R}= \frac{ \langle \rho_{15}^4 \rangle}{ \langle \rho_{15}^2 \rangle^2}=\frac{Q_4}{Q_2^2}
\end{equation}
Numerical values of $\mathcal{R}$ are displayed in Table \ref{tab1}. In the table, $T$ refers to tetrahedral symmetry, $O$ to octahedral symmetry, and $I$ to icosahedral symmetry. The fact that odd $l$ spherical harmonics are odd under reflection precludes $D_n$ symmetry. 
The top entry, labeled ``All'', records the result of an unconstrained search for the minimum quartic invariant. The entry, labeled ``$C_5$" gives the value of $\mathcal{R}$ for a state with a five-fold symmetry axis. The two values are identical. Comparison with the results of  numerical minimization using the method of refs.\cite{sanjay,sanjay2} confirms that the global minimum in the $l=15$ sector has $C_5$ symmetry. 

\begin{table}[htbp]
\begin{center}
\begin{tabular}{|c|c|}
\hline
{ \bf Symmetry} & { \bf Magnitude of the quartic invariant} \\
\hline
\hline
All & 0.208797 \\
\hline
$C_5$ & 0.208797 \\
\hline
$C_3$ & 0.208809 \\
\hline
$C_2$ & 0.208857 \\
\hline
$C_7$ & 0.208904  \\
\hline
$C_4$ & 0.209495 \\
\hline
$T$ & 0.210448  \\
\hline
$C_6$ & 0.21052 \\
\hline
$I$ & 0.210534 \\
\hline
$C_{11}$ & 0.21121 \\
\hline
$C_{12}$ & 0.211582 \\
\hline
$C_{10}$ & 0.212131  \\
\hline
$C_{13}$ &  0.213575 \\
\hline
$C_9$ & 0.214088  \\
\hline
$C_8$ & 0.217175 \\
\hline
$C_{14}$ & 0.21796  \\
\hline
$O$ & 0.220681 \\
\hline
$C_{15}$ & 0.227049 \\
\hline
$C_{\infty}$ & 0.261008 \\
\hline
\end{tabular}
\label{tab1}
\caption{Ordered list of minimum quartic magnitudes, by imposed symmetry. Note that the magnitude for $C_5$ is the same as the magnitude for no imposed symmetry.}
\end{center}
\label{default}
\end{table}

The Hessian matrix---constructed by taking the second derivative of the quartic magnitude with respect to the $2l+1=31$ degrees of freedom of the density \cite{sanjay,sanjay2}---allows us to assess the stability of the various symmetry states. This matrix has three zero eigenvalues corresponding to the generators of global rotations in three dimensions. If all other eigenvalues are positive, then the symmetry state is locally stable with respect to infinitesimal distortions of the density. If any of the other eigenvalues is negative, then the quartic term can be reduced by introducing an additional density that distorts the symmetry. In this way, we find that the $C_5$ state is stable. The $C_3$ state also is stable and corresponds to a metastable free energy minimum. None of the other symmetries were stable. The table highlights an important point. The difference between the values of $\mathcal{R}$ for the $C_5$ and $C_3$ states appears only in the \textit{fifth} decimal: the two states are practically degenerate. The other symmetry states listed in the table are all unstable and have comparable values of $\mathcal{R}$. The free energy landscape of the $l=15$ sector is apparently quite flat with a only few shallow minima. 

\subsection{The $l=16$ sector.}

\begin{figure}[H]
\begin{center}
\includegraphics[width=3in]{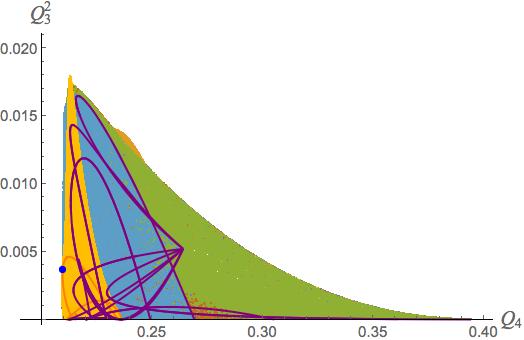}
\caption{The Kim plot for $l=16$ with only local invariants. The symmetries corresponding to the prominent colored regions are seven-fold (green region), eight-fold (blue region) and tetrahedral (orange region). The various dark purple curves correspond to symmetries ranging from nine-fold to sixteen-fold. Those curves meet at a point in the interior of the Kim region corresponding to $D_{\infty}$ symmetry. The orange curve in the lower left hand portion of the Kim region corresponds to octahedral symmetry. The blue dot is the point of icosahedral symmetry.}
\label{fig:KP1}
\end{center}
\end{figure}

\begin{figure}[H]
\begin{center}
\includegraphics[width=3in]{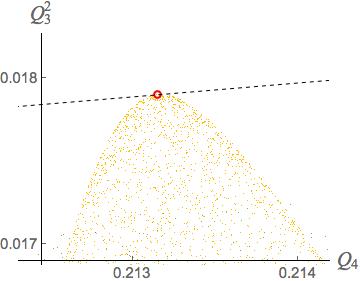}
\caption{The Kim plot in the vicinity of the tip at the top of the plot. The open red dot corresponds to the point on the border of the Kim plot that maximizes $Q_3^2/Q_4$. The stippled region has tetrahedral symmetry.}
\label{fig:KP2}
\end{center}
\end{figure}

For $l=16$, we follow the same steps as for $l=6$ with local invariants. There is a single cubic invariant ($Q_3$)---so the ordering transition must be first-order---and a single quartic invariant ($Q_4$). The state that appears at the point where the symmetry of the uniform state is broken corresponds to a maximum the ratio $Q_3^2/Q_4$. Figure \ref{fig:KP1} shows the corresponding Kim plot.
Just as for $l=6$, the Kim plot has a roughly triangular outline with a protruding tip, but here there is a crucial difference. For $l=6$ a unique state with icosahedral symmetry was located at the tip for $l=6$ with the result that only states with icosahedral symmetry emerged from the Kim construction. For $l=16$ this icosahedral state is missing from the tip.  By contrast, the tip has tetrahedral symmetry everywhere for $l=16$~\cite{tetrahedral_note}. An enlarged version of the tip region is shown in Fig. \ref{fig:KP2}. The tetrahedral point on the border of the Kim plot that maximizes $Q_3^2/Q_4$ is indicated by a red dot. 

\begin{figure}[htbp]
\begin{center}
\includegraphics[width=3in]{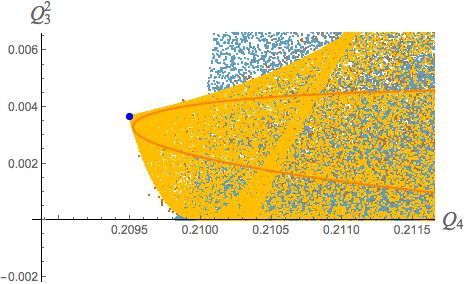}
\caption{The Kim plot in the vicinity of the icosahedral asperity. The orange curve corresponds to octahedral symmetry, which does not quite extend to the icosahedral point.}
\label{fig:KP3}
\end{center}
\end{figure}

While icosahedral symmetry has been demoted from the prominent position it had for $l=6$, it has not completely disappeared. There is a point with icosahedral symmetry on the boundary line of the Kim plot located at the tip of a small asperity in the lower left-hand side of the plot (see Fig.\ref{fig:KP3}). 

\begin{figure}[H]
\begin{center}
\includegraphics[width=3in]{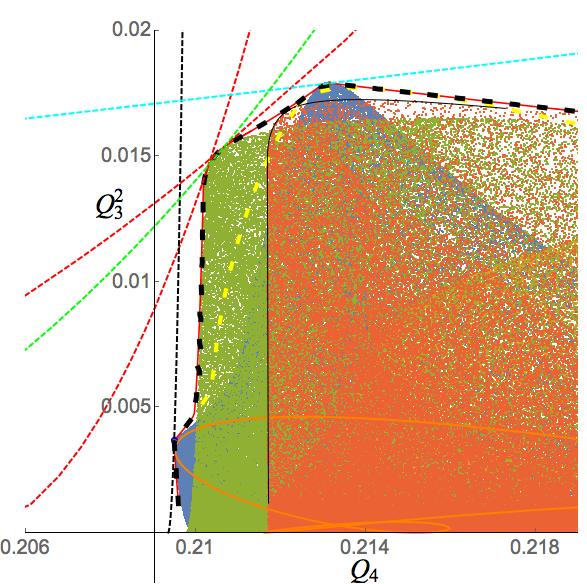}
\caption{Kim plot and construction for $l=16$ with local invariants. Curves outlining the boundary of the Kim plot are colored according to symmetry: five-fold (dashed yellow curve), three-fold (solid red curve), two-fold (dashed black curve), and eightfold (thin solid black curve). The dashed constant free energy curves of the Kim construction that graze the surface at a certain point on the boundary of the Kim plot are colored in correspondence to the symmetry of the point:  three-fold symmetry (dashed red curves), seven-fold symmetry (dashed green curve), icosahedral symmetry (dashed blue curve), and tetrahedral symmetry (dashed light blue curve).}
\label{fig:Kpencurvesleq16}
\end{center}
\end{figure}

We can now carry out the Kim construction for $l=16$. The result is shown in Fig.\ref{fig:Kpencurvesleq16}.
If the coefficient $t_{16}$ of the quadratic invariant is reduced starting in the isotropic state, then the first non-uniform state that appears has tetrahedral symmetry. It corresponds to the red dot in Fig. \ref{fig:KP2}. The value of the cubic invariant at the transition point (see Fig. \ref{fig:KP2}) is significant so it is a robust first-order transition. As the temperature is lowered, the tetrahedral state transits to a state with three-fold symmetry, then to a state with seven-fold symmetry, then again to a state with three-fold symmetry and then finally to a state with icosahedral symmetry. The phase plot is shown in Fig. \ref{fig:PD}.
 
\begin{figure}[htbp]
\begin{center}
\includegraphics[width=3in]{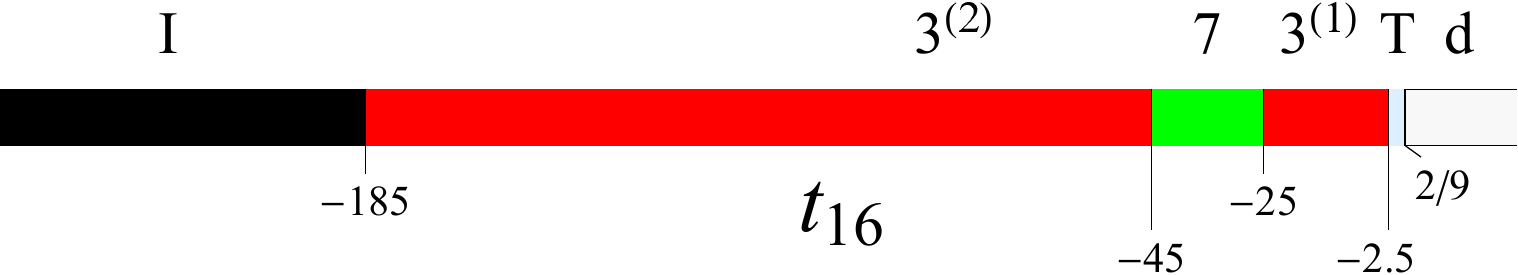}
\caption{Sequence of states produced by the Kim construction for $l=16$ as a function of the coefficient $t_{16}$ of the quadratic invariant. T: tetrahedral; $3^{(1,2)}$: three-fold axis; 7: seven-fold axis; I: icosahedral, d: disordered. The phase diagram was calculated for coefficients $u=v=1$; see Eq. (\ref{eq:l15161}).}
\label{fig:PD}
\end{center}
\end{figure}

\begin{figure}[htbp]
\begin{center}
\includegraphics[width=3in]{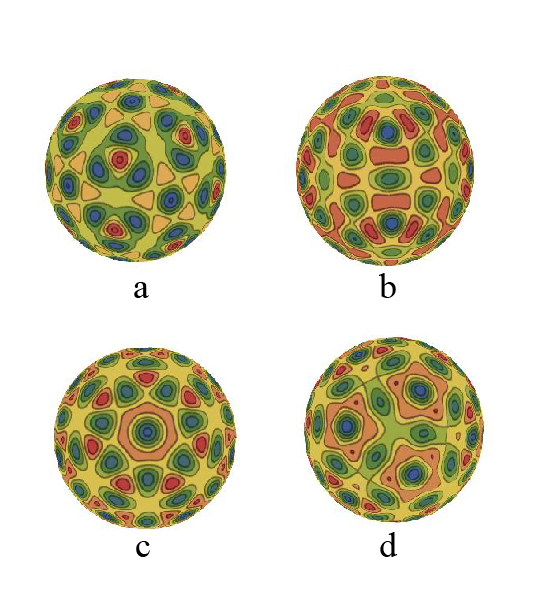}
\caption{The various symmetries that are possible for a purely local $l=16$ free energy as obtained from a numerical minimization of a variational free energy with one quadratic invariant, one cubic invariant and one quartic invariant. The symmetries are { \bf a}: icosahedral, { \bf b}: tetrahedral, viewed along a two-fold axis, { \bf c}: seven-fold, { \bf d}: three-fold. The last two symmetries are those of an antiprism.}
\label{fig:l16shapes}
\end{center}
\end{figure}

It is useful to complement the Kim construction for $l=16$ with direct free energy minimization to obtain explicit density profiles. Figure \ref{fig:l16shapes} shows examples that exhibit some of the symmetries obtained by numerical minimization of the local free energy for appropriate values of the coefficients of the quadratic, cubic and quartic invariants. We set here $u=v=1$, where $u$ is the cubic coefficient and $v$ the quartic coefficient; see for example Eq. (\ref{eq:l15161}).
In the earlier studies based on numerical free energy minimization \cite{sanjay,sanjay2}, we missed the $l=16$ icosahedral state in the phase plot. The utility of the Kim plot for the numerical minimization is evident: the symmetries will be realized by the Kim construction are obvious by inspection.

\subsection{The $l=15+16$ sector for fixed mixing ratio.}

\begin{figure}[htbp]
\begin{center}
\includegraphics[width=3in]{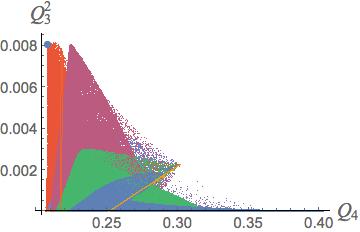}
\caption{The $l=15$ and $16$ Kim plot when $f_{16}=0.05$. The blue dot on the upper left boundary of the plot is the point of icosahedral symmetry. The red portion of the Kim region on whose boundary the dot sits corresponds to tetrahedral symmetry. The purple protrusion is a region of thirteen-fold symmetry. The prominent green region in the lower left hand portion of the plot corresponds to ten-fold symmetry. }
\label{fig:lowpartplot}
\end{center}
\end{figure}

\begin{figure}[htbp]
\begin{center}
\includegraphics[width=3in]{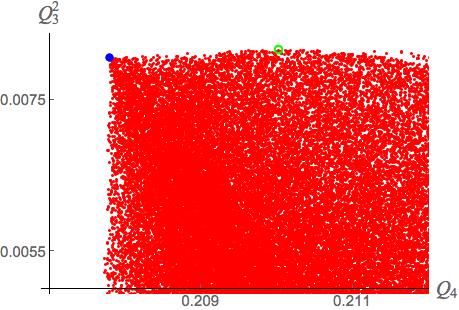}
\caption{Detailed portion of Fig. \ref{fig:lowpartplot} containing the icosahedral point (blue dot) and the point corresponding to the highest transition temperature (open green dot), along with the tetrahedral region (red stippled portion).  }
\label{fig:lowpartplotdetail1}
\end{center}
\end{figure}

Now we turn to the Kim construction in the enlarged $l=15+16$ space. As noted, the number of local invariants is significant. There are separate quadratic and quartic invariants for $l=15$ and for $l=16$, next there are the cubic invariant for $l=16$ as well as the mixed cubic invariant $\langle \rho_{16} \rho_{15}^2\rangle$ and finally the mixed quartic invariant $\langle \rho_{15}^2 \rho_{16}^2 \rangle$. To obtain insight, we used two different strategies. The first, discussed here, is to fix the relative contribution of $l=15$ and $l=16$. Define
\begin{eqnarray}
\rho_{15} & = & \cos\eta \, A \, q_{15}( \psi^{(15)}_i) \label{eq:l15163} \\
\rho_{16} & = & \sin\eta \, A \, q_{16}(\psi^{(16)}_i )\label{eq:l15164}
\end{eqnarray}
where $0 \leq \eta \leq \pi/2$ is a ``mixing angle" and where and $\psi^{(15)}_i$  and $\psi^{(16)}_i$ refer to the set of $2l$ internal angular variables that determine the precise forms of the two  densities. The two quadratic invariants are
\begin{eqnarray}
\langle \rho_{15}^2 \rangle & = &A^2 (\cos\eta)^2 \label{eq:l15165} \\
\langle  \rho_{16}^2 \rangle & = & A^2 (\sin\eta)^2 \label{eq:l15166}
\end{eqnarray}
where for any function $f( \theta, \phi)$,
\begin{equation}
\langle f \rangle = \int f(\theta, \phi) \, d \Omega \label{eq:l15162}
\end{equation}
Since $(\tan\eta)^2=\rho_{16}^2 / \rho_{15}^2$, the mixing angle is an invariant in its own right.  We will include it in the form of the relative participation $(\sin\eta)^2 = f_{16}$ of the $l=16$ density to the total density. If one fixes $f_{16}$ then one is effectively down to three invariants, a situation that can be managed by the methods described earlier. Figure \ref{fig:lowpartplot} shows the Kim region for the case that $f_{16}$ is fixed at $0.05$.  
This Kim plot now has \textit{two} vertical protrusions: the left protrusion has tetrahedral symmetry and the right protrusion thirteen-fold symmetry. Suprisingly, the icosahedral point has ``slid upwards" to a location close to the tip of the protrusion. Now, the tetrahedral state has to compete with the icosahedral state. Figure \ref{fig:lowpartplotdetail1} shows the portion of the plot in Fig. \ref{fig:lowpartplot} that contains the promoted icosahedral point (shown in blue) and also the point corresponding to the largest transition temperature in the Kim construction (the green point).  In an ensemble in which relative contributions from $l=15$ and $l=16$ are set at 0.95 and 0.05 respectively, the initial transition is to tetrahedral symmetry. As the quadratic coefficients are reduced, icosahedral symmetry takes over. The salient point is that \textit{the Kim plot of a dominant $l=15$ state with a small admixture of $l=16$ is qualitatively different from that of the pure $l=15$ Kim plot}. Icosahedral symmetry appears to be here intrinsically associated with a mixed $l=15+16$ state while it is largely unstable for pure $l=15$ and $l=16$ order parameter states. 

It again is useful to combine the Kim construction with the outcome of direct numerical minimization producing a representative phase diagram.. We allowed $\eta$ and $f_{16}$ to vary freely while for the variational free energy we used
\begin{eqnarray}
\lefteqn \nonumber \\ & F = &  \frac{t_{15}}{2} \langle \rho_{15}^2 \rangle + \frac{t_{16}}{2} \langle \rho_{16}^2  \rangle + \frac{u}{2} \langle ( \rho_{15} + \rho_{16})^3 \rangle \nonumber \\ &&+ \frac{v}{4} \langle ( \rho_{15}+ \rho_{16})^4 \rangle \nonumber \\ & = &   \frac{t_{15}}{2} \langle \rho_{15}^2 \rangle + \frac{t_{16}}{2} \langle \rho_{16}^2 \rangle + \frac{u}{3} \left( \langle \rho_{16}^3 \rangle + 3 \langle \rho_{16} \rho_{15}^2 \rangle \right) \nonumber \\ &&+ \frac{v}{4} \left( \langle \rho_{15}^4 \rangle + 4 \langle \rho_{15}^2 \rho_{16}^2 \rangle + \langle \rho_{16}^4 \rangle \right)   \label{eq:l15161}
\end{eqnarray} 
In arriving at the last line, we used the symmetry properties of $\rho_{15}(\theta, \phi)$ and $\rho_{16}(\theta, \phi)$ under reflection. Because of the orthogonality of the $l=15$ and $l=16$ densities, the square of the total density, $\langle \rho^2 \rangle$, is  $\langle \rho_{15}^2 \rangle + \langle \rho_{16}^2 \rangle$. Next, because of the second cubic term, a non-zero $l=15$ density necessarily entrains an $l=16$ density but the the reverse is not true. This means that  pure $l=16$ states are possible but pure $l=15$ states are not. Note that this is not the most general local variational energy: the cubic and quartic contributions were expressed in terms of the total density $( \rho_{15} + \rho_{16})$ but the $l=15$ and $l=16$ quadratic invariants have separate prefactors $t_{15}$ and $t_{16}$ for the corresponding density. 

\begin{figure}[htbp]
\begin{center}
\includegraphics[width=3in]{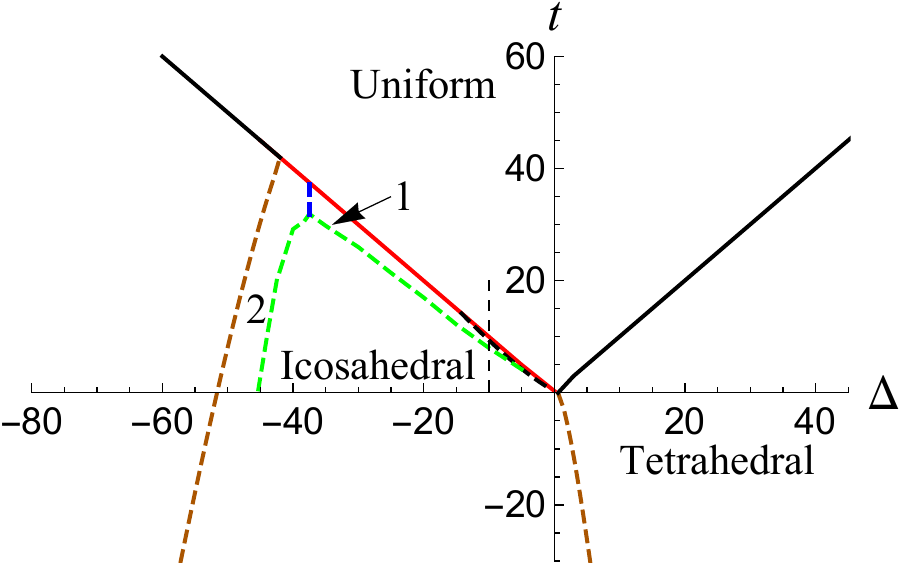}
\caption{Phase diagram obtained by numerically minimizing the free energy (\ref{eq:l15161}) for $u=v=10$. The green dashed curve separates the region in which icosahedral order is a global free energy minimum (below the curve) from the region in which it is a local, but not global, free energy minimum. The curve corresponds to a line of first order phase transitions. The thin vertical dashed line at $\Delta =-10$ indicates the parameter range plotted in Fig. \ref{fig:participation1}. The brown dashed lines indicate stability limits of icosahedral symmetry, corresponding to spinodal lines. Finally, the blue dashed line between the green curve and the line of transitions from uniform to ordered phases separates the two regions of non-icosahedral order: tetrahedral (region 1) and non-tetrahedral (region 2).}
\label{fig:phasediagram7}
\end{center}
\end{figure}

The result of numerical free energy minimization for $u=v=10$ is shown in Fig. \ref{fig:phasediagram7} \cite{figurenote}. The $t$ and $\Delta$ axes are defined by the relations $t_{15} = t + \Delta$ and $t_{16} = t-\Delta$.
The phase diagram has four principal regions. The top region is the uniform state. It is bordered by a wedge of solid black lines that separates it from the phases with orientational order. In the region below the wedge to the right, the order is pure $l=16$ with tetrahedral symmetry. In the third region, labeled ``Icosahedral,'' a mixed state with icosahedral symmetry is at least locally stable but it is only the global free energy minimum below the green dashed curve. Between the green dashed curve and the two nearly vertical solid black lines, other symmetries have lower free energy. The two dashed brown lines play the role of \textit{spinodals} for the icosahedral state. The red line along the top of the icosahedral region marks either continuous or weakly first-order transitions from the isotropic state to a tetrahedral state that quickly transforms to an icosahedral state as the temperature is lowered further. Even when the parameter $\Delta=(t_{15}-t_{16})/2$ is as low as $-50$, the tiny amount of residual $l=16$ density suffices to destabilize the $C_5$ minimum free energy state of pure $l=15$. This is consistent with our earlier observations concerning the  fragility of the free energy minimum of the pure $l=15$ sector. Finally, the fourth region to the left of the icosahedral region has complex symmetries that are neither icosahedral nor does it have the $C_5$ symmetry of pure $l=15$, again a consequence of the near degeneracy of the $l=15$ sector. 

The numerical results can be compared with the Kim construction for fixed mixing ratio. The Kim construction predicted that the icosahedral state should be stable over a large range of parameters but that the symmetry breaking transition of the uniform state should produce a tetrahedral state with a short stability interval. The utility of the Kim construction is obvious:  it would be easy to miss the tetrahedral sliver in a numerical minimization while the tetrahedral state is obvious by inspection of the Kim plot. In summary, numerical minimization of the variational free energy in the enlarged $l=15+16$ space of invariants confirms that the icosahedral state is globally stable over a significant parmeter range. The region of icosahedral stability is separated from the uniform region by a narrow interval of tetrahedral dominance.  

\begin{figure}[htbp]
\begin{center}
\includegraphics[width=2in]{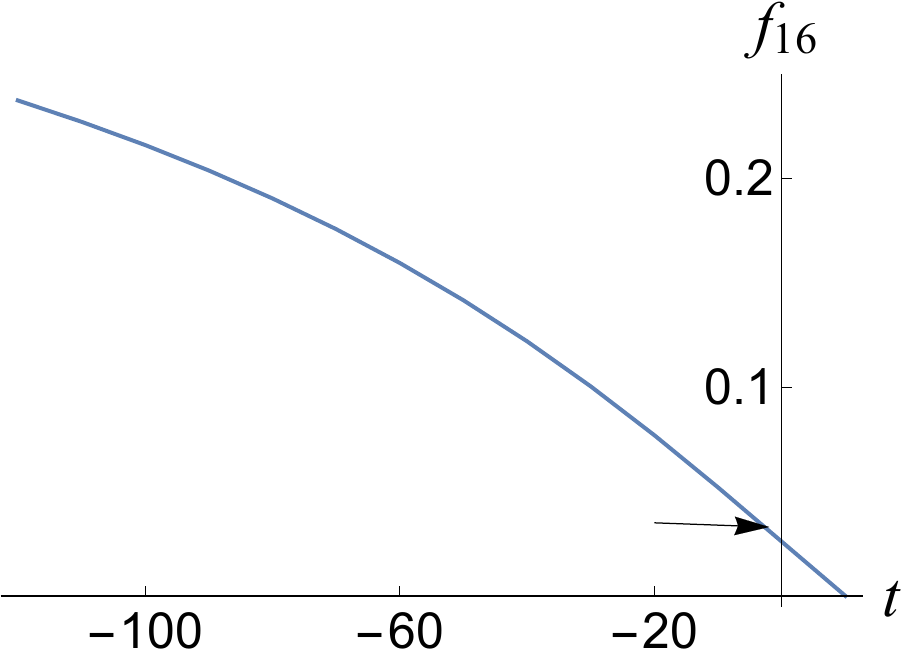}
\caption{The fractional participation of $l=16$ in the total density, labeled by $f_{16}$, is plotted as a function of $t$ for fixed  $\Delta=-10$, along the vertical dashed line in Fig. \ref{fig:phasediagram7} (and its extension to negative values of $t$). For the very rightmost values of $t$ the density is tetrahedral. The arrow identifies the point at which the minimum free energy symmetry changes from tetrahedral to icosahedral. The curve hits the horizontal axis at the onset of ordering.}
\label{fig:participation1}
\end{center}
\end{figure}

Finally, how good is the assumption that $f_{16}$ is constant? Figure \ref{fig:participation1} shows a plot of dependence of $f_{16}$ on $t$ for fixed $\Delta=-10$.  
For large and negative $t$, $f_{16}$ varies only modestly so the assumption of fixed $f_{16}$  is reasonable. However, the assumption of fixed $f_{16}$ definitely fails near the onset of orientational ordering with $f_{16}$ going linearly to zero at the transition point. This can be understood from the form of the coupling term $ u\langle \rho_{16} \rho_{15}^2 \rangle$ between the $l=15$ and $l=16$ states. Combined with the term that is quadratic in $ \rho_{16}$, it follows that, just below the ordering temperature, the average  $\langle \rho_{16} \rangle$ is proportional to $\frac{u}{t_{16}}\langle\rho_{15}\rangle^2$. Since there is no cubic invariant in the $l=15$ sector it follows that $\langle\rho_{15}^2\rangle$ goes to zero linearly at the transition point so the participation ratio  $\langle \rho_{16} \rangle^2/(\langle\rho_{15}\rangle^2+\langle\rho_{16}\rangle^2)$ also goes linearly to zero. Close to the ordering transition, we need a different approach that will be discussed in the next section.
 
From the viewpoint of Landau theory, these results are disconcerting. While it is reasonable that $l=15$ ordering will entrain a certain amount of $l=16$ density as a secondary order parameter---because of the mixed cubic invariant $\langle \rho_{16} \rho_{15}^2\rangle$---it is anomalous that the mixing between primary and secondary order parameters ($l=15$, respectively, $l=16$) destabilizes the primary $l=15$ order parameter (with $C_5$ five-fold symmetry) and that it stabilizes an icosahedral state new state that is unstable in the single $l$ subspaces. It would seem that canonical Landau theory, based on a dominant order parameter that transforms according to a single irreducible representation of the high temperature symmetry group, does not produce the actual minimum free energy state. In the next section, we will see why this conclusion has to be modified. 

\subsection{Diagrammatic perturbation theory and the Kim construction.}

In this final section we use the Kim construction to investigate the competition between icosahedral and tetrahedral symmetry close to the ordering transition where $f_{16}$ is very small. This second method is based on \textit{perturbation theory}. It starts from the assumption that the contribution from $l=16$ contribution is sufficiently small so the $l=16$ density can be described by a quadratic Hamiltonian. Expanding the variational free energy to second order in $\rho_{16}({\bf{\Omega}})$ gives 
\begin{eqnarray}
\lefteqn \nonumber \\ & \Delta H_{16} \simeq & \int d{\bf{\Omega}}\left( \frac{t_{16}}{2}\rho_{16}({\bf{\Omega}})^2  + u\rho_{16}({\bf{\Omega}}) \rho_{15}({\bf{\Omega}})^2\right)~      \label{eq:l15161}
\end{eqnarray} 
with $t_{16}$ positive. The next step is to integrate out $\rho_{16}({\bf{\Omega}})$ to arrive at a renormalized variational free energy for $\rho_{15}({\bf{\Omega}})$. The mathematical steps of integrating-out the $l=16$ component are very similar to the steps that are taken if one integrates out shape fluctuations (see Appendix \ref{app:shape}). Just as in that case, the integration generates a negative, non-local quartic contribution to the $l=15$ variational free energy:
\begin{equation}
\begin{split}
- \frac{u^2}{2t_{16}}\sum_{m=-16}^{16}\int d{\bf{\Omega}}\int d{\bf{\Omega}}^{\prime} \rho_{15}({\bf{\Omega}})^2Y_{16}^m({\bf{\Omega}}) Y_{16}^{m}({\bf{\Omega}}')^* \rho_{15}({\bf{\Omega}}')^2\label{eq:singlel18}
\end{split}
\end{equation}
Using the notation introduced in Section II, this term can be represented by the graph shown in Fig. \ref{fig:newterms}.
\begin{figure}[htbp]
\begin{center}
\includegraphics[width=2in]{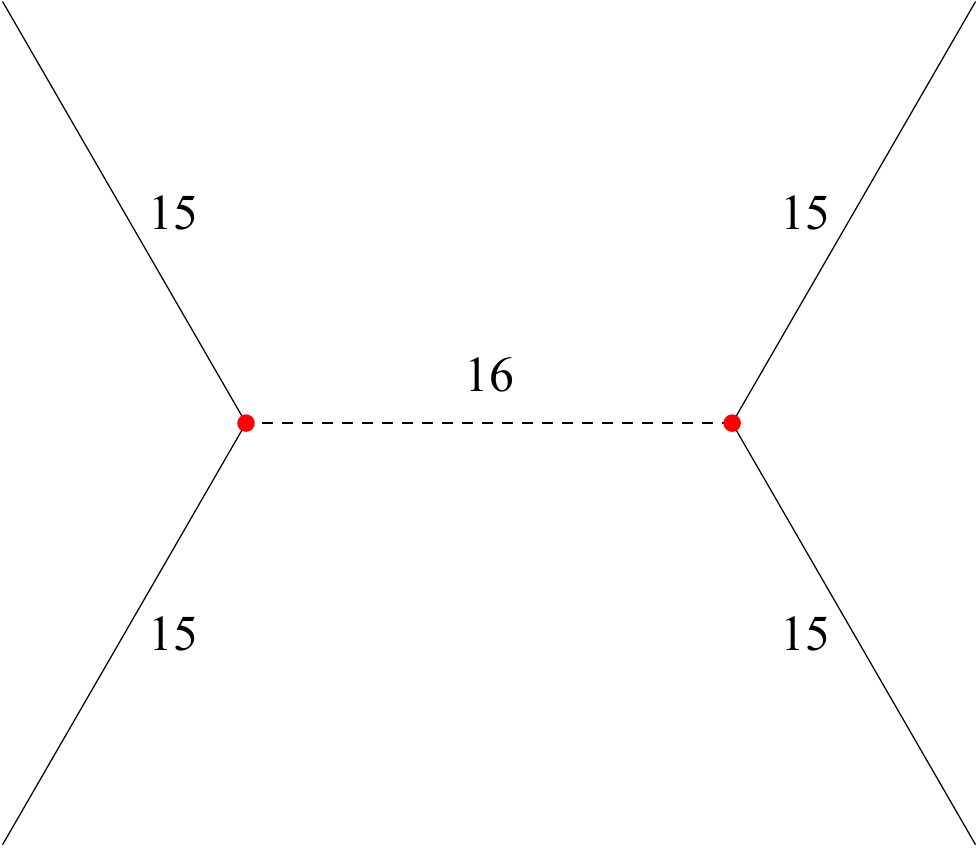}
\caption{Diagrammatic representation of the non-local quartic term generated in the $l=15$ free energy by integrating out the $l=16$ density. The dashed line represents the $l=16$ propagator with a weight $1/t_{16}$ and the two red dots represent the three point vertex with a weight $u$.}
\label{fig:newterms}
\end{center}
\end{figure}
Two three-point vertices are connected by an $l=16$ ``propagator". By connecting two of the external lines in the graph, a fluctuation correction to the quadratic term of the $l=15$ variational energy could be generated but we only will include tree diagram contributions in this section. So, even though we started from a local variational free energy, the step of integrating out the $l=16$ component generates non-local invariants \cite{perturbation_note}.

\begin{figure}[htbp]
\begin{center}
\includegraphics[width=3in]{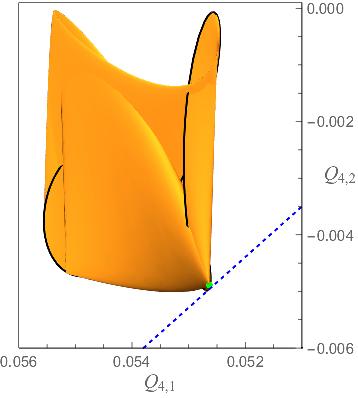}
\caption{The Kim plot for $l=15$ with the $l=16$ contribution integrated out to lowest non-trivial order. The term $Q_{4,1}$ is the local quartic term for $l=15$ and the term $Q_{4,2}$ is the new non-local quartic term. Only states with tetrahedral symmetry are included, along with the state with icosahedral symmetry represented by the green dot. The black line denotes edges of areas with tetrahedral symmetry. The dashed blue line depicts a constant free energy surface that grazes the boundary of the tetrahedral symmetry region and also the boundary of the full Kim plot. It passes close to the green icosahedral symmetry point, but it does not impinge on it, as shown in Fig. \ref{fig:newkimplot2}. }
\label{fig:newkimplot1}
\end{center}
\end{figure}

Next, construct a two-dimensional Kim plot with the normalized local invariant and the new non-local quartic invariant as axes (see Figs. \ref{fig:newkimplot1} and \ref{fig:newkimplot2}).
Only states with tetrahedral symmetry are shown plus a single point with icosahedral symmetry. The tetrahedral area is folded on itself. As the 31 expansion coefficients of the $l=15$ spherical harmonics are varied over the range of allowed values, the same pair of values for the invariants $Q_{4,1}$ and $Q_{4,2}$ can be associated with different sets of expansion coefficients, which leads to fold lines.The external edges of the tetrahedral surface are indicated by red lines. The position of the icosahedral point close to the outer edge suggests that a Kim construction could be performed that would reproduce the transition from a tetrahedral to an icosahedral state, but that is not the case. Figure \ref{fig:newkimplot2}, an enlarged version of the plot near the icosahedral point, shows why: 
\begin{figure}[htbp]
\begin{center}
\includegraphics[width=3in]{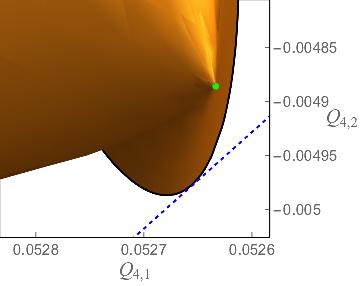}
\caption{The portion of the Kim plot shown in Fig. \ref{fig:newkimplot1} containing the icosahedral point. The icosahedral point is located in the interior of the Kim plot. As shown in Fig. \ref{fig:newkimplot1}, the constant free energy surface indicated by the dashed blue line grazes the surface of the tetrahedral symmetry region. }
\label{fig:newkimplot2}
\end{center}
\end{figure}
The icosahedral point is definitely located in the \textit{interior} of the Kim plot, which means that in actuality the icosahedral state should not show up.

Because the icosahedral point is very close to the boundary, it makes sense to include higher-order terms in perturbation theory. We restart from a Hamiltonian for the $l=16$ degrees of freedom that now includes both ``three-point" and ``four-point" interaction terms between $l=16$ and $l=15$:
\begin{equation}
\begin{split}
 \Delta H_{16} \simeq  \int d{\bf{\Omega}}&\bigg( \frac{t_{16}}{2}\rho_{16}({\bf{\Omega}})^2  +  u\rho_{16}({\bf{\Omega}}) \rho_{15}({\bf{\Omega}})^2 \\ & +4v\rho_{15}({\bf{\Omega}}) ^2 \rho_{16}({\bf{\Omega}}) ^2\bigg)~      
\end{split}
\end{equation} 
Integrating out the $l=16$ density using perturbation theory generates a sixth-order, positive non-local invariant contribution to the $l=15$ variational free energy with prefactor $u^2v/2t_{16}^2$. It is represented by the diagram shown in Fig. \ref{fig:newterms}.
\begin{figure}[htbp]
\begin{center}
\includegraphics[width=2in]{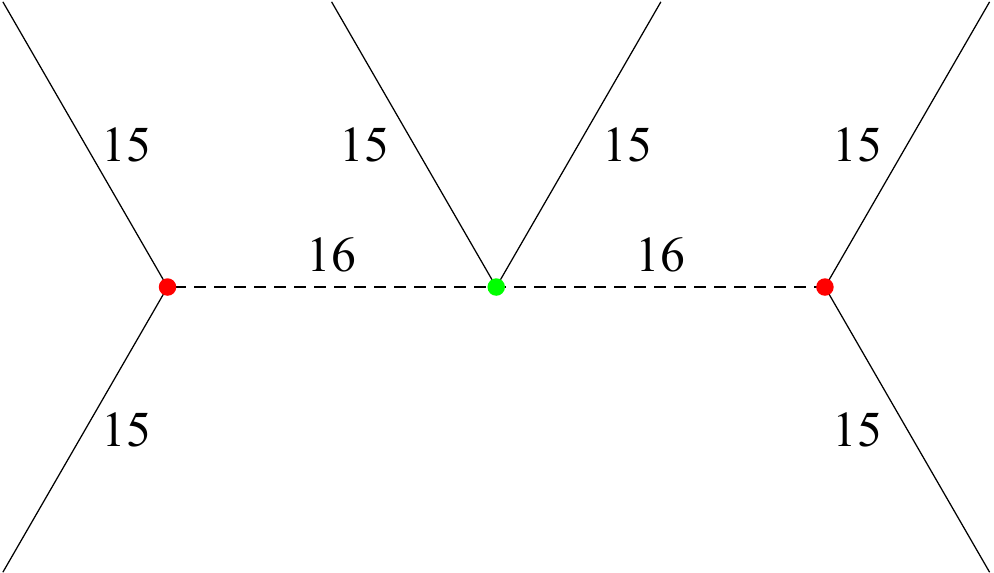}
\caption{Diagrammatic representation of a sixth-order non-local contribution to the $l=15$ variational energy generated by integrating out the $l=16$ contributions perturbatively to second order in $1/t_{16}$. The red dots represent the three-point vertex with weight $u$ and the green dot a four-point vertex with weight v}
\label{fig:newterms}
\end{center}
\end{figure}
We will denote this new invariant by $Q_6$. We now can construct a three dimensional Kim plot with the two earlier quartic invariants plus the new sixth order invariant as coordinate axes. 
\begin{figure}[htbp]
\begin{center}
\includegraphics[width=3.5in]{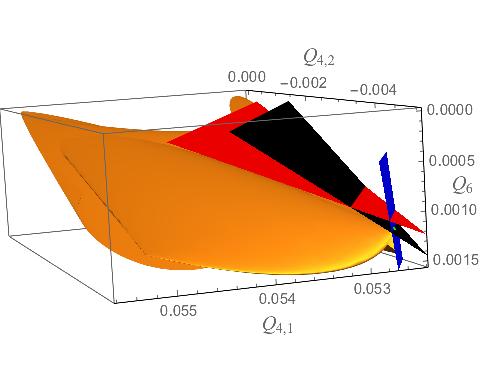}
\caption{The Kim plot in the three dimensional space spanned by the two quartic terms and the sixth order term, along with constant free energy surfaces that grazes the Kim region at the point corresponding to icosahedral symmetry (blue and black surfaces) and a constant free energy surface that grazes the plot along the tetrahedral surface (red surface).  }
\label{fig:newkimplot3}
\end{center}
\end{figure}
The icosahedral point now does lie on the boundary of the Kim plot and, as shown, it is accessible as a free energy minimum \cite{sixthordernote}.  In Fig. \ref{fig:newkimplot3} the coefficients of the two quartic terms were arbitrarily set equal to each other, so the results shown in that figure must be viewed as qualitative. At very low values of $|t_{15}|$, the constant free energy surface, which is  nearly perpendicular to the $Q_6$ plane grazes the boundary of the Kim plot along the edge of the tetrahedral symmetry surface. For sufficiently large negative coefficient $t_{15}$ the constant free energy surface grazes the subspace of allowed invariants is through the point of icosahedral symmetry. However, as that coefficient grows in absolute value the constant free energy surface once again grazes the Kim plot along a point of tetrahedral symmetry. Figure \ref{fig:PD} illustrates the sequence of states when the free energy is of the form
\begin{equation}
\mathcal{F} \frac{t_{15}}{2} Q_2 + \frac{v}{4} (Q_{4,1} + Q_{4,2}) + \frac{w}{6} Q_6 \label{eq:pertfreen}
\end{equation}
with $v=w=1$.

\begin{figure}[htbp]
\begin{center}
\includegraphics[width=2.5in]{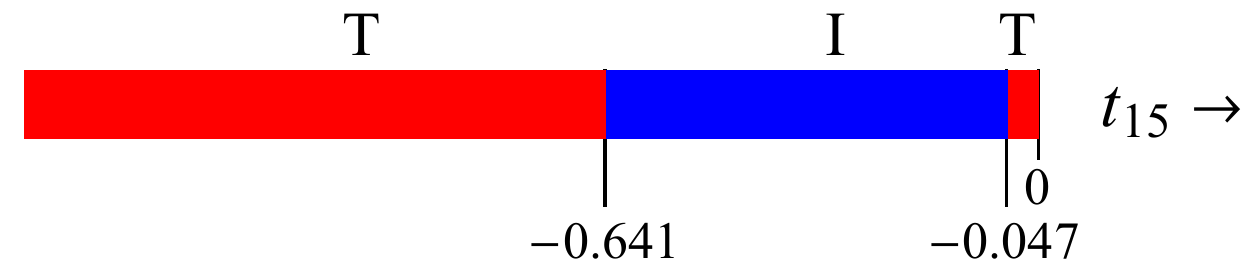}
\caption{Sequence of states produced by the Kim construction for $l=15$ as a function of the coefficient $t_{15}$ of the quadratic invariant including non-local invariants that are mediated by $l=16$. T: tetrahedral; I: icosahedral. The phase plot was calculated for coefficients $v=w=1$ in the free energy (\ref{eq:pertfreen}). The tetrahedral symmetry regime continues to $|t_{15}|$ well in excess of 100.}
\label{fig:PD}
\end{center}
\end{figure}

This phase plot can be compared with that of Fig. \ref{fig:phasediagram7} for the case that the mixing ratio between the $l=16$ to $l=15$ states is held fixed. In both cases, the initial symmetry-breaking transition of the uniform state leads to a state with tetrahedral symmetry. A stable icosahedral state appears as the reduced temperature is lowered further. In both cases, an $l=16$ component is essential for the stability of the icosahedral state. The key difference is that now the phase plot formally is obtained inside the $l=15$ sector with the $l=16$ components absorbed by the introduction of non-local invariants.

\section{Conclusion.}

The Introduction posed the question whether an order-parameter theory can be constructed for the transition from an isotropic state to an icosahedral state if the primary order parameter is an $l=15$ or an $l=16$ icosahedral spherical harmonic. We have addressed this question using the Kim construction method that allows one to obtain the general structure of a phase diagrams without having to take recourse to numerical minimization of a variational free energy for certain specific values of the physical system parameters. We found that the answer to the question is no if the variational free energy is constructed from the local invariants of the $l=15$ or of the $l=16$ sectors. In the $l=15$ sector, the icosahedral state with local invariants was found to be completely unstable. In the $l=16$ sector, a stable icosahedral state did appear but only well below the orientational ordering transition. On the other hand, in the enlarged $l=15+16$ space stable icosahedral states are present over a large range of system parameters, at least for fixed mixing ratio, confirming earlier numerical results~\cite{sanjay,sanjay2}. A slice of states with tetrahedral symmetry interposes between the isotropic and icosahedral states. In the Kim construction method, the competition between the icosahedral and tetrahedral states is very clear: the Kim plot of mixed $l=15+16$ states has an asperity with icosahedral symmetry that competes with a rounded peak with tetrahedral symmetry, itself a characterstic of the $l=16$ Kim plot.

These results appear to be in glaring contradiction with the basic tenet of Landau theory that continuous phase transitions can be described by an order parameter that transforms according to a single irreducible representation of the symmetry group of the high-temperature isotropic phase. This contradiction disappeared when non-local invariants were included in the variational free energy: there is a stable icosahedral state in the $l=15$ sector. By combining the Kim method with a diagrammatic perturbation expansion, we showed that the required non-local invariants of the $l=15$ sector appear when a purely local free energy functional is confined to the $l=15$ sector by integrating out the $l\neq15$ components. Because the local variational free energy in the $l=15$ sector is quasi-degenerate, the minimum free energy state is very sensitive to the presence of even weak non-local $l=15$ invariants. The non-local invariants necessary for the stabilization of the icosahedral state are mediated by $l=16$. 

More generally, if one starts from a variational free energy expression with only local invariants of the density, and if the coefficient $t_{l^*}$ of the quadratic invariant of one of the irreducible representations is significantly smaller than that of the other irreducible representations then, at the point where $t_{l^*}$ is close to zero, the other irreducible resprentations can be integrated out diagrammatically, keeping such terms only to quadratic order. This procedure generates non-local invariants up to the maximum number permitted by the Molien polynomial.

The great advantage of the Kim geometrical method over brute-force numerical minimization is that it replaces a hit-and-miss choice of specific parameters by global geometrical analysis. The Kim construction method has already been known for decades but it is the availability of modern visualization methods that makes it such a useful tool for the study of ordering transitions. The Kim method does become cumbersome if one is forced to carry out the geometrical constructions in an invariant space with more than three dimensions. Because of the proliferation of invariants for larger $l$, this would seem to be a serious objection because, as we have just shown, there are instances in which the non-local invariants really must be included. However, if the original free energy functional is local---with at most two invariants---and if the coefficient $t_{l^*}$ of the quadratic invariant of the dominant irreducible representation is significantly smaller then that of the other $l$ then only a limited number of non-local invariants, say with $l=l{^*}\pm1$ may need to be included. A Kim geometrical analysis may remain practical for $l$ larger than 15-16 but this will need to be verified in future work. It is interesting to note in this context that icosahedral spherical harmonics come in the form of neighboring even/odd pairs of the form of $l,l+1$, which suggests that this strategy may work for icosahedral ordering for general $l$ as it did for the l=15/16 pair. 

We have found that the basic tenets of Landau theory formally can be saved at the cost of introducing the non-local invariants permitted by the Molien polynomial. In actuality, the description of the broken symmetry states as involving multiple irreducible representations is the more economical. The difficulties with ``single-l" canonical Landau theory are expected to multiply for shell structures that would require even larger $l$ values. An icosahedral state in the form of, say, a large icosahedral "buckyball" is composed of twenty rounded equilateral triangular facets where particles have six-fold coordination. The same is true for large viral capsids that are constructed by the Caspar-Klug method~\cite{caspar}. These structures are only very poorly represented by any icosahedral spherical harmonic. It would appear that spherical harmonics are not the best basis set for such cases but it is not clear what would be a better choice.  

\begin{figure}
	\centering
	\includegraphics[width=3.0in]{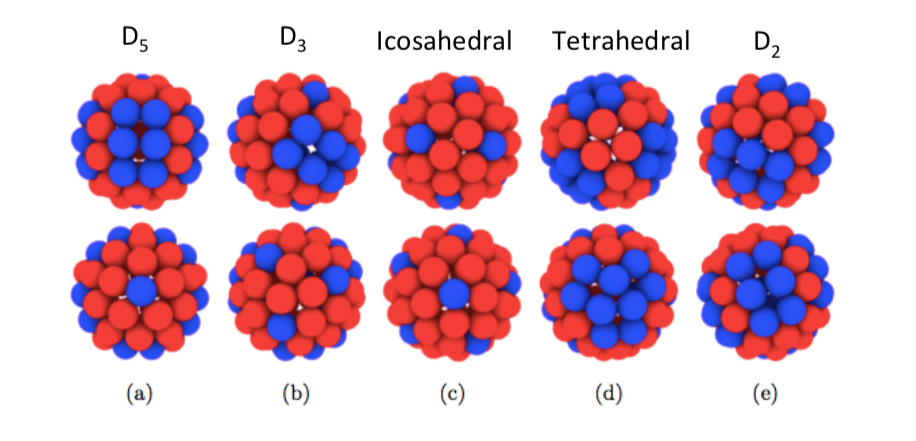}
	\caption{The five Lennard-Jones packings for N = 72 with the lowest potential energy. (a) D5h packing with energy per particle 3.0564 in units of the LJ binding energy, (b) D3 packing with 3.0559, (c) icosahedral packing with 3.0548, (d) tetrahedral packing with 3.04636 and (e) Baseball symmetry ($D_2$) with 3.04630. The color indicates the coordination numbers five (blue) or six (red); from Ref. \cite{paquay}.}
	\label{P}
\end{figure}

Other interesting questions await resolution. In the Introduction we discussed that an $l=16$-like icosahedral state is generated by numerical simulations of 72 point particles that were interacting via the LJ pair interaction. The stability range of the icosahedral state is quite small when system parameters such as temperature and interaction range are varied. A variety of other symmetries appear as well (see Fig. \ref{P}). 
The icosahedral state competes with states that have different symmetries: $D_{5h}$, $D_3$, tetrahedral, and $D_2$. The method that we explored has to account for these other states.
We found that in the pure $l=16$ sector, states with five-fold, three-fold, two-fold, and eightfold symmetry compete with icosahedral symmetry. This is very encouraging but the mirror planes that characterize the $D_5$, $D_3$, and $D_2$ symmetry groups are missing for reasons that are not clear. Higher-order invariants may have to be included to explain this. Note though that in the $l=16$ one cannot introduce non-local invariants mediated by $l=15$. 

Chirality is believed to play an important role~\cite{lorman2} for the case of viral capsids. The high-temperature symmetry group is $SO(3)$ in this case. Chiral invariants have to be included as part of the expansion. The lowest-order chiral invariant is fourth order in the density~\cite{sanjay2}, and it would be interesting to see how this invariant will affect the $l=15+16$ phase diagrams.

We conclude by noting that symmetry arguments play an important role for the design of synthetic molecular shells as exemplified by the work in the Yeates group~\cite{todd}. It would be interesting to apply the Kim construction method to analyse the symmetry-based strategies developed for shell design. 

\section{Acknowledgements}

 Sanjay Dharmavaram, Amit Singh, Vladimir Lorman and Alexander Grosberg contributed to the paper with discussions. A special debt of gratitude is owed to Alec Stein, whose recognition of the utility of the Kim approach and whose ideas on the nature of invariants of the rotation group form the foundation of the work reported in this article. We thank the NSF for support under DMR Grant No. 1006128 and the Aspen Center for Physics for hosting a workshop on the physics of viral assembly. This paper is dedicated to the memory of our friends Vladimir Lorman and Marko Jar\'{i}c.
\clearpage
\begin{appendix}

\section{The quadratic invariant} \label{sec:app1}

Given the form of $\rho_l(\theta, \phi)$ in (\ref{eq:OP}), in order to be unaffected with respect to rotations about the $z$ axis, the most general form of a quadratic invariant must be
\begin{equation}
\sum_{m=0}^{l} B_m c_{l,m}c_{l,-m}
\end{equation}
The next question is what restriction rotational invariance places on the coefficients $B_m$. We can arrive at that restriction by noting that the generators of rotations about the $x$ or $y$ axis are  combinations of angular momentum raising and lowering operators. The raising operator $a^{\dagger}$ has the following action on the coefficients $c_{l,m}$.
\begin{equation}
a^{\dagger}c_{l,m} \propto \sqrt{l(l+1)-m(m+1)}c_{l,m+1} \label{eq:supp1}
\end{equation}
Consider the two consecutive terms in (\ref{eq:supp1})
\begin{equation}
B_mc_{l,m}c_{l,-m} + B_{m+1}c_{l,m+1}c_{l,-m-1} \label{eq:supp2}
\end{equation}
If we act on these two terms with the operator $\mathds{1} + \delta a^{\dagger}$, then two of the $O(\delta)$ terms generated are proportional to
\begin{equation}
\delta \sqrt{l(l+1) -m(m+1)} c_{m+1}c_{m}(B_m+B_{m+1})
\end{equation}
In order for this to vanish, we must have
\begin{equation}
 B_m=-B_{m+1} \label{eq:supp3}
 \end{equation}
 The equality above holds for all $m > 0$. In the case $m=0$, the same procedure yields
 \begin{equation}
B_1=-2B_0 \label{eq:supp4}
\end{equation}
Thus, the quadratic invariant must have the form
\begin{eqnarray}
\lefteqn{K \left(2 \sum_{m=1}^l (-1)^l c_{l,m}c_{l,-m} + c_{l,0}^2\right)} \nonumber \\ & =& K \sum_{m=-l}^l (-1)^m c_{l,m}c_{l,-m} \label{eq:supp5}
\end{eqnarray}
From the orthonormality of the spherical harmonics and their symmetry properties, this expression is equivalent to 
\begin{equation}
K\int \Phi_l(\theta, \phi)^2 \sin \theta \, d \theta d \phi \label{eq:supp6}
\end{equation}

It is possible to carry out the same analysis by requiring that  the invariant is unchanged under the action of the lowering operator, $a$. However, given that this is just the Hermitian conjugate of the raising operator, the analysis is fundamentally identical to the one above, leading to exactly the same conclusion. 

\section{Higher order invariants} \label{sec:app2}

The way in which one determines the number of invariants of a particular order is to compare the number of terms that can contribute to an invariant with the number of restrictions on those terms arising from application of the raising operator. At a given order $n$ and angular quantum number $l$, the invariant is the sum of terms going as 
\begin{equation}
B_{m_1, \ldots, m_n}c_{l,m_1} c_{l,m_2} \cdots c_{l,m_n} \delta_{m_1 + \cdots +m_n} \label{eq:suppa1}
\end{equation}
The number of such terms is the number of distinct ways of finding $n$ integers between $-l$ and $l$ that sum to zero. This can be expressed in terms of the number distinct of ways representing the integer $n(l+1)$ as a sum of $n$ positive and non-zero integers less than or equal to $2l+1$. The  restrictions are a set of requirements on terms of the form
\begin{equation}
B^{\prime}_{m_1, \ldots, m_n}c_{l,m_1} c_{l,m_2} \cdots c_{l,m_n} \delta_{m_1 + \cdots +m_n-1} \label{eq:suppa2}
\end{equation}
The number of such terms is the number of distinct ways of finding $n$ integers between $-l$ and $l$ that sum to one. This can be expressed in terms of the number of a way of expressing the integer $n(l+1)+1$  as a sum of $n$ positive and non-zero integers less than or equal to $2l+1$. The total number of distinct $n^{\rm th}$ order invariants is just the difference between the two numbers above. 

Figure \ref{fig:quarticnumber} shows that difference in the instance of fourth order invariants, for values of $l$ ranging from 0 to 40. In the case of third order invariants, the difference is always zero or one: zero for odd values of $l$ and one for even values of $l$. This is because the only third order invariant is the integral of the density cubed, and given the symmetry properties of spherical harmonics, such an integral is guaranteed to vanish for odd $l$. 
Based on Fig. \ref{fig:quarticnumber}, which also follows from the Molien series \cite{Mukai}, it is reasonable to conjecture that that $n_l^{(4)}$ is given by
\begin{equation}
n_l^{(4)} = \lfloor \frac{l}{3} \rfloor +1 \label{eq:suppa3}
\end{equation}
where the first term on the right hand side of (\ref{eq:suppa3}) is the largest integer less than or equal to $l/3$. 
 
\section{Tracing out boundaries in the Kim plot} \label{app:boundaries}

\begin{figure}[htbp]
\begin{center}
\includegraphics[width=3in]{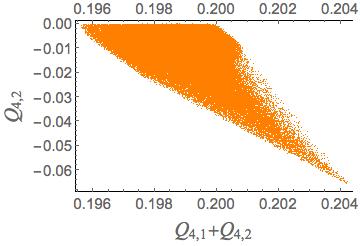}
\caption{The portion of the Kim plot in Fig. \ref{fig:Kimplotnewvars} that corresponds to two-fold symmetry. This region is bounded from above by $Q_{4,2}=0$. The points in the region become sparse in the vicinity of the other boundaries.}
\label{fig:tworegion}
\end{center}
\end{figure}

Figure \ref{fig:tworegion} shows the portion of the Kim plot in Fig. \ref{fig:Kimplotnewvars} corresponding to two-fold symmetry.
Although the upper boundary of this region at $Q_{4,2}=0$ is well-defined, the other boundaries are somewhat diffuse, especially towards the bottom of the plot, as the points generated by random sampling of the parameters defining the density are sparse in the immediate vicinity of some of the plot's edges. This can be understood heuristically as the consequence of projecting a high dimensional region---seven dimensional in this case---onto two dimensions. Consider, for example projecting a collection of uniformly distributed points in a  seven dimensional sphere onto a two dimensional flat plane. The number of points directly above the plane in the vicinity of the circular surface of the projected sphere will be considerably smaller than the number directly above the center of the circular region into which the points fall. 

\begin{figure}[htbp]
\begin{center}
\includegraphics[width=3in]{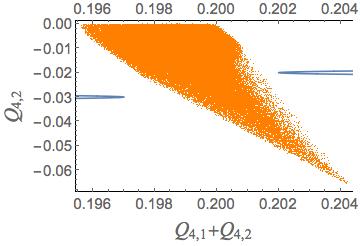}
\caption{The two-fold symmetry region and two stylus curves. }
\label{fig:stylus1}
\end{center}
\end{figure}

As an alternative to generating more points, which for large $l$ becomes computationally demanding as well as memory intensive, we adapt the Kim method to trace out the boundary. Recall that minimizing the free energy entails finding the point at which a curve of constant free energy---in the instance of $l=7$ a straight line---impinges tangentially on the Kim plot. If we were to take all possible orientations for this constant energy surface we would trace out the convex hull of the Kim plot. Given a different constant free energy surface, one can perform a more detailed probe of the boundary. To this end, we devise a new surface which, for lack of a better term,  we call a ``stylus.''.  It is of the general form 
\begin{equation}
x \cos \phi + y \sin \phi + K( -x \sin \phi + y \cos \phi - D)^2 = C \label{eq:stylus1}
\end{equation}
where $K$,  $C$ and $D$ are constants. For K sufficiently large this is a very steep parabola. Figure \ref{fig:stylus1} shows the two-fold symmetry region and two of the stylus surfaces, for $\phi=0$ and  $K = \pm 5,000$. In practice, we used $K=10,000$. 
We locate the two bounding surfaces by varying $D$, thus scanning vertically, and determining the minimum value, effectively the quantity $C$, of the left hand stylus function and the maximum value of the right hand one. This process yields the two boundary curves shown in Figs. \ref{fig:Kimplotnewvars} and \ref{fig:Kp1}, as well as the translated versions of those Kim plots.

\section{Kim plot with a cubic invariant: details} \label{sec:app3}

The Kim method can also be applied to cases in which third order invariants arise, as when $l$ is even. Take the instance of a single $l$ system with only one relevant quartic term, say the local one. The free energy is, then,
\begin{equation}
\mathcal{F}[\rho_l] = \frac{t_l}{2} A^2 + \frac{u}{3} A^3 Q_3\{\psi_k\} + \frac{v}{4} A^4 Q_4\{ \psi_k \} \label{eq:singlel25}
\end{equation} 
The outcome of minimizing (\ref{eq:singlel25}) with respect to $A$ and discarding the possibility of $A=0$ is
\begin{widetext}
\begin{eqnarray}
\mathcal{F}_{\rm min} &=&-\frac{u^4Q_3^4}{v^3Q_4^3}\frac{1+6(tvQ_4/u^2Q_3^2)^2  -6 t vQ_4/u^2Q_3^2+ \left(1-4 t vQ_4/u^2Q_3^2\right)^{3/2}}{24}  \nonumber \\ & \equiv & -\frac{u^4Q_3^4}{v^3Q_4^3}\frac{1+6W^2 - 6W + \left( 1-4W\right)^{3/2}}{24} \label{eq:singlel26}
\end{eqnarray}
\end{widetext}
From the equations above, we find
\begin{eqnarray}
Q_3^2&=&-\frac{t^3}{u^2}\frac{1}{W^3}\frac{1+6W^2 - 6W + \left( 1-4W\right)^{3/2}}{24 \mathcal{F_{\rm min}}} \label{eq:singlel27} \\
Q_4& = & -\frac{t^2}{v}\frac{1}{W^2}\frac{1+6W^2 - 6W + \left( 1-4W\right)^{3/2}}{24 \mathcal{F_{\rm min}}} \nonumber \\ & = & \frac{u^2}{vt}WQ_3^2 \label{eq:singlel28}
\end{eqnarray}
Both $Q_3^2$ and $Q_4$ are positive, the latter to ensure thermodynamic stability. Furthermore, $\mathcal{F}_{\rm min}$ will be negative. Given this we see from (\ref{eq:singlel27}) and (\ref{eq:singlel28})  that the parameter $W$ will have the same sign as $t$. 

The above equations allow us to plot curves of constant free energy in the space spanned by $Q_4$ and $Q_3^2$. One important point is that the approach can be generalized to more than one $Q_4$. If, for instance, there are two quartic invariants, then we replace $Q_4$ with $a Q_{4,1} + b Q_{4,2}$. The curve defined by (\ref{eq:singlel26}) and (\ref{eq:singlel27}) becomes a surface in which the $Q_4$ axis is replaced by lines of constant $aQ_{4,1} + b Q_{4,2}$. This is readily extended to the case of more quartic invariants. 

In light of the last line of (\ref{eq:singlel26}) and the fact that $Q_4$, $Q_3^2$, $v$ and $u^2$ are positive, it is clear that the signs of the quantity $W$ and the parameter $t$ must be the same. Given this and the fact that the contribution of the $W$-dependent expressions to the right hand sides of (\ref{eq:singlel26}) and the first line of (\ref{eq:singlel27}) is zero when $W=2/9$, we can distinguish between two regimes in those equations. The first is $- \infty < W < 0$, which applies when $t<0$. The second is is $0<W<2/9$, appropriate to $t>0$. Outside of those regimes, the right hand sides of (\ref{eq:singlel26}) and (\ref{eq:singlel27})  either apply to the case $\mathcal{F}_{\rm min} >0$, which is not of interest, or possess imaginary parts. 

\begin{figure}[htbp]
\begin{center}
\includegraphics[width=3in]{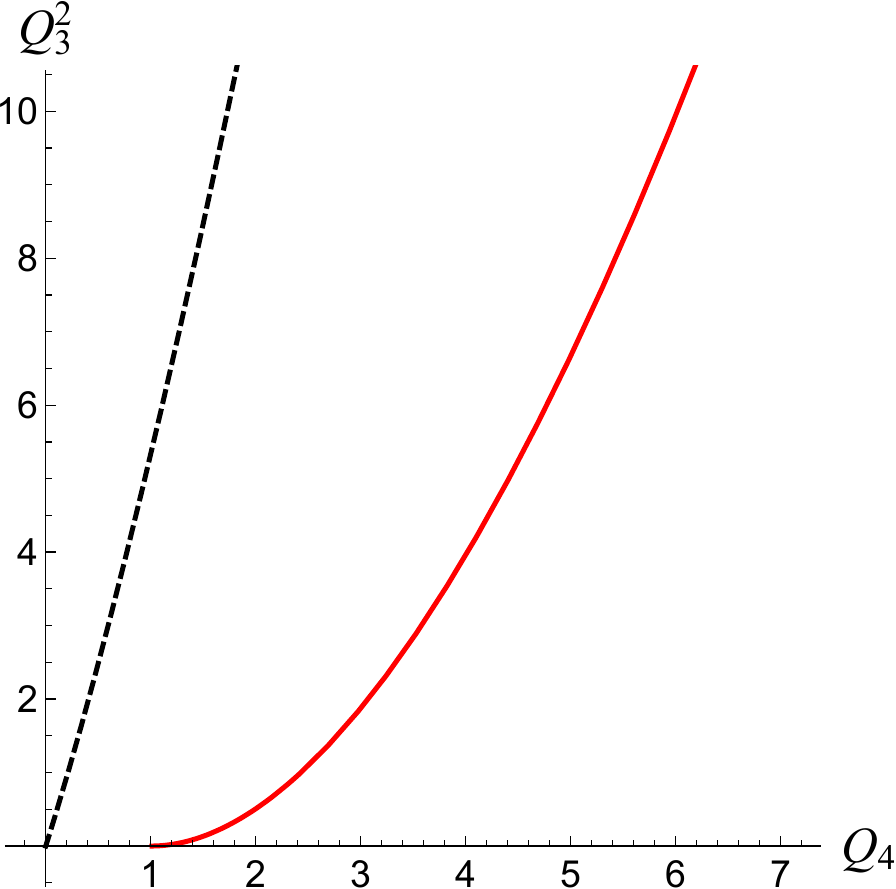}
\caption{Plots of $Q_3^2$ versus $Q_4$ for values of $t$ that are positive (dashed black curve) and negative (solid red curve). The free energy $\mathcal{F}_{\rm min}$ has been set equal to $-1/4$, and the parameters $u$, $v$ and $|t|$ have been set equal to 1.   }
\label{fig:Kp3a}
\end{center}
\end{figure}

\begin{figure}[htbp]
\begin{center}
\includegraphics[width=3in]{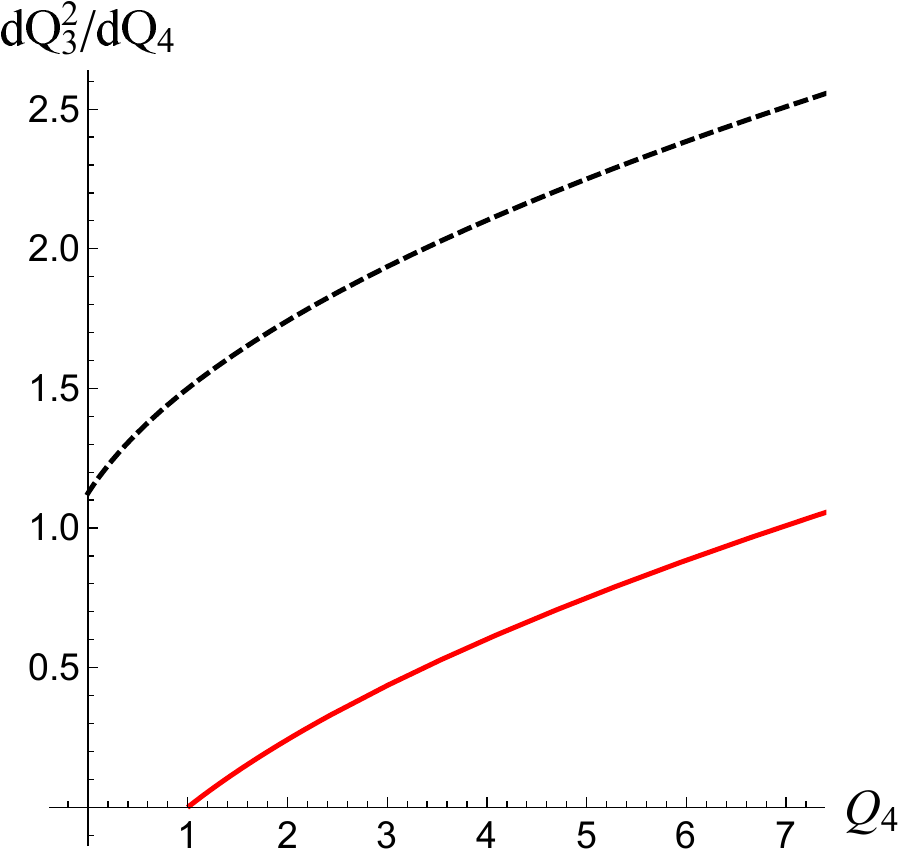}
\caption{The slopes of the curves in Fig. \ref{fig:Kp3a}, plotted against $Q_4$. The parameters have been set to the same values as in that figure.}
\label{fig:Kp3b}
\end{center}
\end{figure}

The relationships between $Q_3^3$ and $Q^4$ are illustrated in Figs. \ref{fig:Kp3a} and \ref{fig:Kp3b}, in which all terms aside from $W$ have been set equal to convenient values. 

From these plots---and a bit of analysis---we see that for large amplitudes of the two invariants $Q_3^2 \propto Q_4^{3/2}$. Furthermore as is evident from the plot, the dependence is monotonic with increasing positive slope. Additionally, inspection reveals that there are three independently adjustable quantities in the two relationships, which can be chosen to be $W$, $\mathcal{F}_{\rm min} v/t^2$ and $\mathcal{F}_{\rm min}u^2/t^3$. This means that we can in principle choose two of those quantities to ensure that a $Q_3^2$ versus $Q_4$ curve passes through a given point in the Kim plot. The third quantity can then be chosen so as to adjust the slope of that curve. Given the values of those three quantities
\begin{eqnarray}
Q_4 & = & x \label{eq:Q4val} \\
Q_3^2 & = & y \label{eq:Q3val} \\
\frac{dQ_3^2}{dQ_4} & = & s \label{eq:slopeval}
\end{eqnarray}
with the additional conditions
\begin{eqnarray}
x & >&0 \label{eq:xcond} \\
y &>& 0 \label{eq:ycond} \\
s &>&0 \label{eq:scond}
\end{eqnarray}
we find
\begin{equation}
W = \frac{2 s x (3 y-2 s x)}{9 y^2} \label{eq:Wsol}
\end{equation}
This result is of interest in the range $s>y/x$.  The relationships yield
\begin{eqnarray}
|\mathcal{F}_{\rm min} |u^2/t^3 & = &\frac{9 y (y-s x)}{2 (2 s x-3 y)^3} \label{eq:asol2} \\
|\mathcal{F}_{\rm min} | v/t^2 & = & \frac{s (s x-y)}{(2 s x-3 y)^2} \label{eq:bsol2}
\end{eqnarray}
Given these equations, it is relatively straightforward to construct the desired constant free energy curve. 

\begin{figure}[htbp]
\begin{center}
\includegraphics[width=3in]{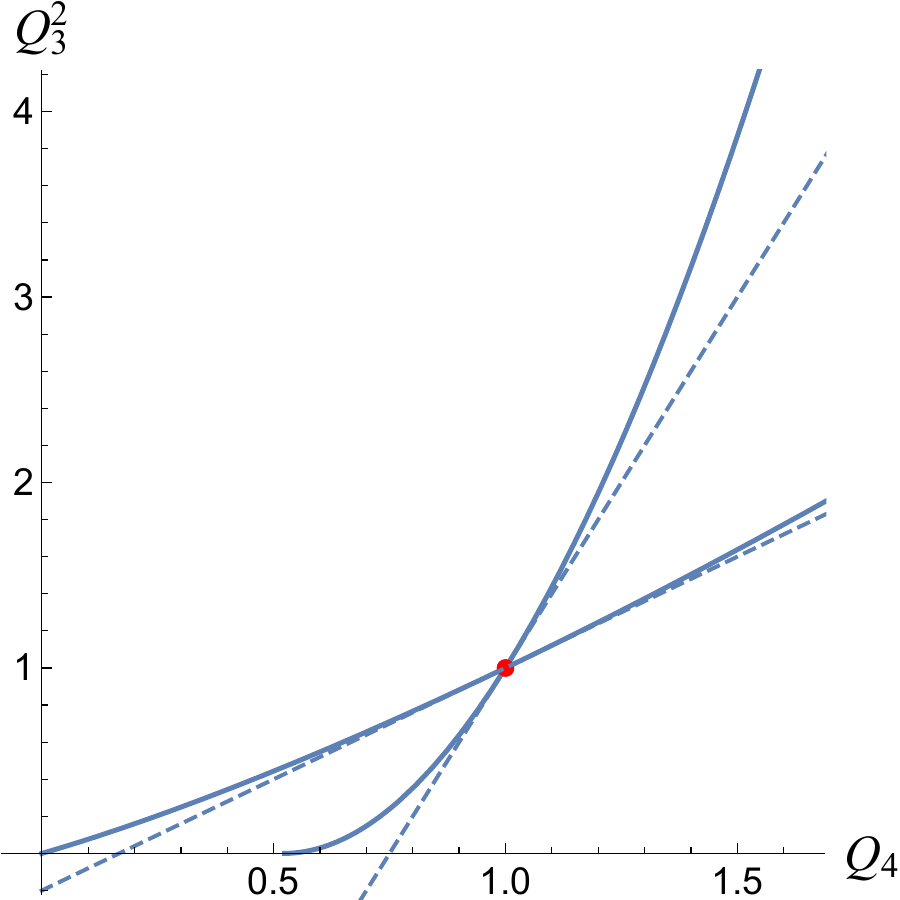}
\caption{Two constant free energy curves passing through the point $Q_4=1$, $Q_3^2 =1$, one with a slope of 1.2, corresponding to a positive value of $t$, the other with a slope of 4, corresponding to a negative value of $t$. The dashed lines indicate the slopes.}
\label{fig:twoqplot}
\end{center}
\end{figure}

One final point: the regime $y/x<s<3y/2x$ corresponds to positive values of $t$, and the regime $s>3y/2x$ corresponds to negative values of $t$. Figure \ref{fig:twoqplot}  displays the results of such a fit, in which two constant free energy curves have been produced, both going through the point $Q_4 =1$, $Q_3^2=1$, one with a slope of 1.2 and the other with a slope of 4.

\section{Landau functional for point particles on a spherical surface.} \label{app:correlations}
Here we construct a mean-field Landau free energy functional for the freezing of a system of point particles confined to a spherical surface. The aim is to provide physical insight into the various local and non-local invariants that were introduced in the main text on mathematical grounds.

Assume that $N$ point particles are restricted to the surface of a sphere with radius $R$. They interact via a radial pair potential with a range $\sigma$ that is small compared to the sphere radius. We will express energy in units of $k_BT$ and length in units of the radius of the sphere, which thus becomes a unit sphere. The arc distance between two points on the sphere surface then reduces to the angle $\theta$ in radians subtended by the two particle and the center of the sphere.  Assume that a monolayer of particles covers the sphere relatively uniformly so with $N$ of the order of $1/\sigma^{2}$. Let the temperature be quite close to the melting temperature. In the liquid phase, the density is uniform and equal to $\rho_0=N/4\pi$. The amplitude of the density modulation $\rho({\bf\Omega})$ in the solid phase will be assumed to be small compared to $\rho_0$. 

Let $F[\rho]$ be a functional that gives the free energy for an imposed density profile $\rho({\bf{\Omega}}$.  In mean-field theory, $F[\rho]$ is an analytical functional so $F[\rho]$ can be expressed as a functional Taylor expansion in $\rho({\bf{\Omega}})/\rho_0$. In general, near a continuous or weakly first-order transition with $|\rho({\bf{\Omega}})| \ll \rho_0$ a limited number of terms in this expansion suffices. 

\subsection{Linear and Quadratic Invariants.}

Let $F^{(2)}[\rho]$ denote the contributions to $F[\rho]$ from the first and second terms of the functional Taylor expansion. The general form of $F^{(2)}[\rho]$ for a particle system covering a sphere is a direct generalization of the form of $F^{(2)}[\rho]$ for an infinite system on a planar surface~\cite{CL}:
\begin{equation}
\begin{split}
F^{(2)}[\rho]=& -\int \mu({{\bf{\Omega}}})\rho({{\bf{\Omega}}}) \, d{\bf{\Omega}}\\&+\frac{1}{2}\int \rho({{ \bf {\Omega}}})K^{(2)}({{\bf{\Omega}}},{{\bf{\Omega}}}^{\prime})\rho({{\bf{\Omega}}}^{\prime}) \, d{{\bf{\Omega}}}d{{\bf{\Omega}}^{\prime}} 
  \label{K}
\end{split}
\end{equation}
The physical meaning of $\mu({{\bf{\Omega}}})$ is that of an externally imposed, position-dependent, chemical potential. It is the thermodynamic conjugate of the density $\rho({\bf{\Omega}})$. Because of the rotational symmetry of the uniform liquid phase, the kernel $K^{(2)}({{\bf{\Omega}}},{{\bf{\Omega}}}^{\prime})$ can only depend on the angle difference ${{\bf{\Omega}}}-{{\bf{\Omega}}}^{\prime}$. Define $K^{(2)}({{\bf{\Omega}}},{{\bf{\Omega}}}^{\prime})=\chi^{-1}({{\bf{\Omega}}}-{{\bf{\Omega}}}^{\prime})$. The physical meaning of $\chi^{-1}({{\bf{\Omega}}}-{{\bf{\Omega}}}^{\prime})$ follows from minimizing $F^{(2)}[\rho]$. Setting the functional derivative $\delta F^{(2)}[\rho]/\delta\rho$ to zero gives
\begin{equation}
\begin{split}
\rho({\bf{\Omega}})=&\int\chi({{\bf{\Omega}}}-{{\bf{\Omega}}}^{\prime})\mu({{\bf{\Omega}}}^{\prime}) d{{\bf{\Omega}}^{\prime}}
  \label{eq:chi}
\end{split}
\end{equation}
where $\chi({{\bf{\Omega}}}-{{\bf{\Omega}}}^{\prime})$ is the functional inverse of $\chi^{-1}({{\bf{\Omega}}}-{{\bf{\Omega}}}^{\prime})$ so with $\int\chi({{\bf{\Omega}}}-{{\bf{\Omega}}}^{\prime \prime})\chi^{-1}({\bf{\Omega^{\prime \prime}}}-{{\bf{\Omega}}}^{\prime}) d{{\bf{\Omega}}^{\prime \prime}}=\delta({{\bf{\Omega}}}-{{\bf{\Omega}}^{\prime}})$. It follows from Eq. (\ref{eq:chi}) that $\chi({{\bf{\Omega}}}-{{\bf{\Omega}}}^{\prime})$ can be identified as the \textit{susceptibility} of the system when exposed to the external perturbation $\mu({{\bf{\Omega}}}^{})$. 
If we expand the inverse susceptibility in a spherical harmonics series
\begin{equation}
\begin{split}
\chi^{-1}({{\bf{\Omega}}}-{{\bf{\Omega}}}^{\prime})=\sum_{l=0}^\infty \chi^{-1}_{l}\sum_{m=-l}^{l}Y_{l}^{m}({{\bf{\Omega}}})Y_{l_1}^{m}({{\bf{\Omega}}}^{\prime}) 
  \label{eq:lBEnergy}
\end{split}
\end{equation}
then the expansion coefficient $\chi^{-1}_l$ is the inverse of the expansion coefficient $\chi_l$ of the linear response susceptibility. Using Eq.\ref{eq:lBEnergy} in Eq.1, one obtains
\begin{equation}
\begin{split}
F^{(2)}[\rho]=\sum_{l=1}^\infty \sum_{m=-l}^{l}\left(\frac{1}{2}\chi^{-1}_{l} |c_{l,m}|^2-(c^*_{l,m}\mu_{l,m}+c.c.)\right)
  \label{F2}
\end{split}
\end{equation}
with ${\rho}(\Omega)=\sum_{l=1}^\infty\limits\sum_{m=-l}^l\limits c_{l,m}Y_{l,m}(\Omega)$ (as in section II). The $\mu_{l,m}$ are here the expansion coefficients of the chemical potential. Because density is a conserved quantity, there is no $l=0$ term. The first term has the same form as the quadratic invariant of Section II. The expansion coefficient $t_l$ of the $l^{\rm th}$ irreducible representation in the sum over quadratic invariants of Section II is thus the inverse of the expansion coefficient of the linear response function. 
An alternative interpretation of $t_l$ is obtained by treating $F^{(2)}[\rho]$ as the quadratic Hamiltonian for thermal density fluctuations. This leads to
\begin{equation}
\begin{split}
\langle|c_{l,m}|^2\rangle=1/t_l
  \label{eq:lBEnergy}
\end{split}
\end{equation}
We thus can also identify $1/t_l$ as the equivalent of the \textit{static structure factor} of the particles on the spherical surface. 
Finally, a third meaning for $t_l$ is obtained by relating the static structure factor to the pair-distribution function (PDF)~\cite{CL}. The PDF is defined as:
\begin{equation}
\begin{split}
g({\bf \Omega})=\frac{4\pi}{N}\sum_{i\neq0} \langle \delta({\bf \Omega}-{\bf \Omega}_i)\rangle
\end{split}
\end{equation}
With this definition, the PDF of the set of particles on a spherical surface transforms in the large $R$ limit to the PDF of a two-dimensional system of particles on a flat surface with the same area density and temperature. Place the fixed $i=0$ particle at the North pole of the sphere (i.e., $ \theta_0=0$). By rotational symmetry, the $g(\theta,\phi)$ in the liquid state can not depend on the azimuthal angle $\phi$. Expand $g(\theta)$ in spherical harmonics. The expansion coefficients
\begin{equation}
\begin{split}
g_{l}=2\pi\sqrt{\frac{(2l+1)}{4\pi}}\int_{-1}^{1} g(\theta) P_l(\cos\theta)d\cos\theta
\end{split}
\end{equation}
are equal to expectation values of spherical harmonics:
\begin{equation}
\begin{split}
g_{l}=&\frac{4\pi}{N}\sum_{i\neq0} \langle Y_{l,0}^*({\bf \Omega}_i)\rangle
\end{split}
\end{equation}
Following the same steps as for bulk systems~\cite{CL}, it can be shown that the PDF expansion coefficients can be related to the static structure factor by
\begin{equation}
\begin{split}
{t_l}^{-1}=\frac{N}{4\pi}\left(1+\frac{N}{4\pi}\sqrt{\frac{4\pi}{2l+1}}g_l\right)
\end{split}
\end{equation}

\subsection{Large $R$ limit.}

In the limit that the sphere radius is very large compared to the mean inter-particle spacing, the physics of a distribution of particles on a spherical surface must reduce to that of a distribution of particles on a two-dimensional (2D) surface. In that limit, the PDF $g(\theta)$ of N particles on a sphere approaches $g_2(r)$, the PDF of a planar array of particles with the same density and temperature, if one sets $r=R\theta$. The PDF $g_2(r)$ has been computed numerically for a two dimensional system of point particles interacting with a Lennard-Jones (LJ) potential just above the freezing temperature~\cite{ran}. $g_2(r)$ has a ``correlation hole" for $r/\sigma\lesssim1$ with $\sigma$ the zero of the LJ interaction, followed by an extended sequence of maxima and minima whose amplitude decays as $\exp(-r/\xi)$. Here, $\xi$ is the \textit{correlation length} $\xi$ of the fluid. 

\begin{figure}[htbp]
\begin{center}
\includegraphics[width=3in]{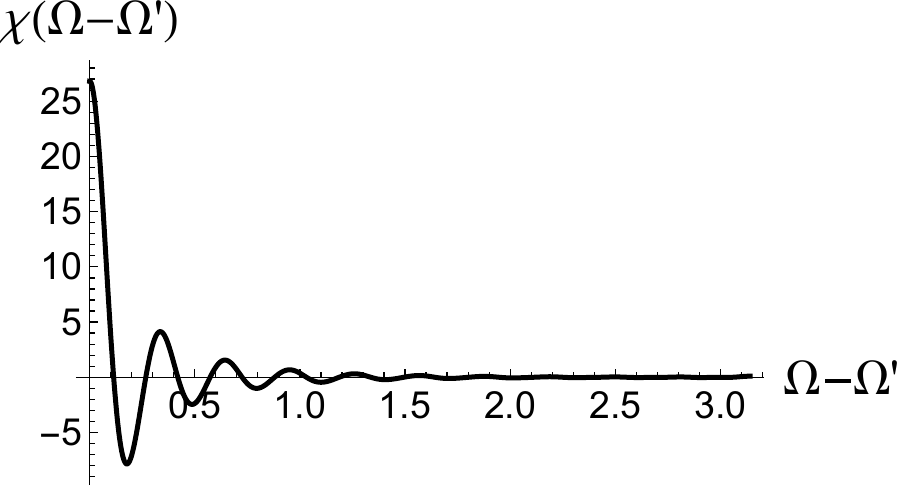}
\caption{The susceptibility $\chi(|{\bf \Omega} - \bf{\Omega}^{\prime}|)$ resulting from (\ref{eq:chi}), with $t=5$, $R=1$ and $l^*=20$.}
\label{fig:chitheta}
\end{center}
\end{figure}

The function $g_2(r)$ is related to the Fourier transform of the static structure factor $S(k)$ of the 2D fluid by:
\begin{equation}
\begin{split}
&g_2(r)=1+2\pi\int_0^{\infty}J_0(kr)(S(k)-1)dk
  \label{g2}
\end{split}
\end{equation}
For the LJ system near the melting point, $S(k)$ has a sharp maximum at a wavenumber $k^*$ with $k*\simeq 6/\sigma$ while it approaches one for large $k$. 
Now go to the large $R$ limit and use $g(\theta)\simeq g_2(R\theta)$ 
\begin{equation}
\begin{split}
&g(\theta)\simeq1+2\pi\int_0^{\infty}J_0(kR\theta)(S(k)-1)dk
  \label{g2}
\end{split}
\end{equation}
Insert this in the expression for the expansion coefficient $g_l$:
\begin{equation}
\begin{split}
&g_{l}\propto\int_{-1}^{1} g(\theta) P_l(\cos(\theta))d\cos(\theta)\\&\propto\int_{-1}^{1} \left(\int_0^{\infty}J_0(kR\theta)(S(k)-1)dk\right) P_l(\cos(\theta))d\cos(\theta)
\end{split}
\end{equation}
In the large $l$ limit, which will be justified afterwards, we can use the large $l$ approximation $J_0(l\theta) \simeq P_l(\cos{\theta})+\mathcal{O}(1/l)$. Next, convert the integral over $k$ into a summation over $k=m/R$ with m an integer, which is justified in the large R limit. The result is:
\begin{equation}
\begin{split}
&g_{l}\propto \int_{-1}^{1} \left(\sum_mP_m(\cos\theta)(S(m/R)-1)\right) P_l(\cos(\theta))d\cos(\theta)
\end{split}
\end{equation}
It finally follows from the orthonormality of Legendre polynomials that  
\begin{equation}
\begin{split}
{g_l}\propto (S(l/R)-1)
  \label{est}
\end{split}
\end{equation}
The justification of the large $l$ limit applies in particular to the sharp maximum of the structure factor at $k=k^*$. The function $g_l$ has a correspondingly sharp maximum around  $l^*\simeq Rk^*\simeq R/\sigma >> 1$.  In that same range of $l$ values
\begin{equation}
\begin{split}
{t_l}\propto \frac{1}{S(l/R)}
\end{split}
\end{equation}
assuming $S(l/R)>>1$. The primary maximum of the structure factor for bulk systems is usually fitted to a Lorentzian form $S(k)\propto 1/(t+(k-k^*)^2)$, where $t$ is of the order $1/\xi^2$ in order to produce the correct width for the first peak in the structure factor. The parameter $t$ is proportional to $T-T_c$ with $T_c$ the transition temperature of the ordering transition in d=2. It follows that
\begin{equation}
t_l \sim t + [(l-l^*)/R]^2 \label{eq:chil}
\end{equation} \label{LB}
It follows from this expression that---ignoring the higher order invariants---the ordering transition of particles on a spherical surface takes place at the same temperature as that of the bulk system. The density modulation that appears at the transition point is dominated by spherical harmonics with index $l\simeq k^*R$ proportional to the radius of the sphere. It also follows from this expression that the range $\Delta l$ of values of $l$ that contributes to the density modulation is of the order of $R\sqrt{t}\simeq R/\xi$.
In the limit of small $t$, so close to the transition point, the susceptibility function $\chi(|{ \bf \Omega}- { \bf \Omega^{\prime}}|)$ on the sphere's surface has the form of the graph in Fig. \ref{fig:chitheta}.
As $t \rightarrow 0$, the range of the oscillations grow in magnitude until it reaches the sphere size. The range $\Delta l$ of values of $l$ that contribute to the transition is of the order of one at that point.
We conclude by noting that the form Eq. (\ref{LB}) for $t_l$ is similar to the Landau-Brazovskii variational free energy~\cite{CL}, which is used extensively to describe ordering transitions of bulk systems.

\subsection{Cubic and Quartic Invariants; Shape Fluctuations.} \label{app:shape}

We now turn to the higher-order terms of the functional Taylor expansion of $F[\rho]$. First consider local terms. One can define $n^{\rm th}$ order terms defined in analogy to Eq. (\ref{K}) for $n=2$. Such terms would be characterized by a kernel $K^{(n)}({{\bf{\Omega}}}_1,{{\bf{\Omega}}}_2,..{{\bf{\Omega}}}_n)$. Following the discussion for $n=2$, the range of the kernels will be of the order of the correlation length $\xi$ of the liquid. The simplest terms of this form are obtained by assuming that the kernels are products of delta functions. This produces the local invariants 
\begin{equation}
\begin{split}
\langle \rho^n \rangle = \int \rho^n({{\bf{\Omega}}})d{\bf{\Omega}}
  \label{newK}
\end{split}
\end{equation}
of the main text. Physically, the local terms follows form free energy density $f(\rho)$ of a liquid by a Taylor expansion in powers of the deviation $(\rho-\rho_0)$ of the density from the mean density. The expansion coefficients of the terms are derivatives of the free energy with respect to the density.  Other higher-order terms can be constructed by combining powers of $\rho({\bf \Omega})$ with derivatives such as $\nabla\rho({\bf \Omega})$ or $\triangle\rho({\bf \Omega})$, ......, but for our purposes these are effectively local invariants.

The non-local non-linear terms that were discussed in the main text appear only if the particle density couples to another scalar variable. Here, we will discuss the case where the density couples to radial displacement associated with shape fluctuations of the sphere on which the particles are located. Assume the shape fluctuations are described by the Helfrich Hamiltonian~\cite{milner}:
\begin{equation}
\begin{split}
\mathcal{H}_H=\frac{\kappa}{2}\int dA(H-H_0)^2
  \label{Helfrich}
\end{split}
\end{equation}
where the integral is over the surface of the sphere, H is the mean curvature of the deformed surface and $H_0=2/R$ the mean curvature of the undeformed surface. Express the radius of the deformed surface as $R(\Omega)=R(1+u(\Omega))$ with $u$ a dimensionless variable small compared to one. Expand in spherical harmonics:
 
\begin{equation}
\begin{split}
R(\Omega)=R\left(1+\sum_{l,m}u_{l,m}Y^m_l(\Omega)\right)
  \label{Helfrich}
\end{split}
\end{equation}

Expanding the Helfrich Hamiltonian to second order in $u$ and using the orthonormality of spherical harmonics gives
\begin{equation}
\begin{split}
\mathcal{H}_H=\frac{1}{2}\sum_{l>1,m}\kappa_l|u_{l,m}|^2
  \label{Helfrich}
\end{split}
\end{equation}
where $\kappa_l/\kappa=[(l+2)(l+1)l(l-1)-2l(l+1) + 4 ]$ ~\cite{milner}. If the shape fluctuations are decoupled from the density fluctuations then
\begin{equation}
\begin{split}
\langle|u_{l,m}|^2\rangle=1/\kappa_l
  \label{eq:lBEnergy}
\end{split}
\end{equation}
which can be rewritten as
\begin{equation}
\begin{split}
\langle u({\bf{\Omega}})u({\bf{\Omega}}')\rangle=\sum_l {\kappa_l}^{-1}\sum_{m=-l}^lY_l^m({\bf{\Omega}}) Y_l^{m}({\bf{\Omega}}')^* \label{eq:e22}
\end{split}
\end{equation}

The density and radial displacement variables are however coupled, which can be seen by considering the effect of a uniform decrease of the radius by an amount $Ru$ with u negative. The resulting decrease in surface area amount to placing the particle system under an external pressure. This will have two effects: a reduction in density and an increase in melting temperature. The two effects can be included phenomenologically by a coupling energy $\Delta\mathcal{H}$
\begin{equation}
\begin{split}
\Delta\mathcal{H}=\int dA\left(a_1 u\rho+a_2 u\rho^2\right)
  \label{Helfrich}
\end{split}
\end{equation}
Next, perform a Boltzmann average over thermal shape fluctuations $u(({\bf{\Omega}})$ while keeping the density profile $\rho({\bf{\Omega}})$ fixed with the aim of identifying contributions to the Landau variational free energy $F([\rho({\bf{\Omega}})])$. The Boltzmann average is perform treating the coupling energy $\Delta\mathcal{H}$ perturbatively. To zeroth order in perturbation theory, so without coupling, the correction to the variational free energy is independent of $\rho({\bf{\Omega}})$ and of no interest. The first-order term $\langle\Delta\mathcal{H}\rangle$ is zero. The second order term $-\frac{1}{2}\langle\left(\Delta\mathcal{H}\right)^2\rangle$ produces a sum of three expressions with the general form
\begin{equation}
I_{m,n}([\rho({\bf{\Omega}})])=\int d{\bf{\Omega}}\int d{\bf{\Omega}}^{\prime} \rho({\bf{\Omega}})^m\langle u({\bf{\Omega}})u({\bf{\Omega}}')\rangle \rho({\bf{\Omega}})^n
\end{equation} 
with m and n integers. The first expression has the form $-(a_1^2/2)I_{1,1}([\rho({\bf{\Omega}})])$. It can be absorbed into the unperturbed functional by the redifining $t_l$ as $t_l-a_1^2/2\kappa_l$. The second term has the form $-a_1a_2 I_{1,2}([\rho({\bf{\Omega}})]$. This produces a non-local cubic invariant. Since there can be only one independent cubic invariant, this term can for our purposes be absorbed into the local cubic invariant. Finally, the third term that appears in second-order perturbation theory is $-(a_2^2/2)I_{2,2}([\rho({\bf{\Omega}})])$. In explicit form, this contributes a non-local quartic term to $F([\rho({\bf{\Omega}})])$. 
\begin{equation}
-\frac{a_2^2}{2}\int d{\bf{\Omega}}\int d{\bf{\Omega}}^{\prime} \rho({\bf{\Omega}})^2\langle u({\bf{\Omega}})u({\bf{\Omega}}')\rangle \rho({\bf{\Omega}'})^2 
\end{equation}

When combined with Eq. (\ref{eq:e22}) , one obtains Eq. (\ref{eq:singlel18}) of the main text. Specifically,
\begin{equation}
\begin{split}
-\int d{\bf{\Omega}}\int d{\bf{\Omega}}^{\prime} \rho({\bf{\Omega}})^2\sum_l \frac{a_2^2}{2\kappa_l}\sum_{m=-l}^lY_l^m({\bf{\Omega}}) Y_l^{m}({\bf{\Omega}}')^* \rho({\bf{\Omega}}')^2
\end{split}
\end{equation}

\section{On the accessibility of points in the Kim plot for $l=6$}

\begin{figure}[htbp]
\begin{center}
\includegraphics[width=3in]{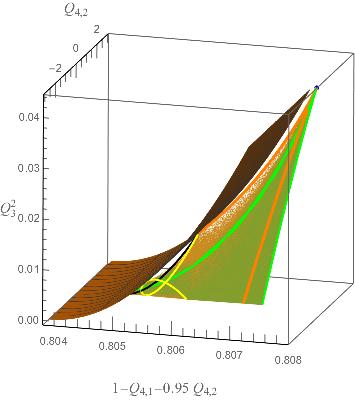}
\caption{Kim plot with a constant free energy surface that grazes the boundary of the Kim region at the point on its boundary corresponding to $D_{\infty}$ symmetry. Note the invariant axis in which the combination $Q_{4,1}+0.95 Q_{4,2}$ has been replaced by $1-Q_{4,1}-0.95Q_{4,2}$. }
\label{fig:infinitysolution}
\end{center}
\end{figure}

In order to render the minimum free energy solutions for six-fold, octahedral and $D_{\infty}$ symmetries accessible, it is necessary to make a change in the quartic invariants utilized. As noted in the text of the article, one way to do this is to replace the combination $Q_{4,1}+0.95 Q_{4,2}$ by $1-Q_{4,1} -0.95Q_{4,2}$. This corresponds to adding in the trivial quartic invariant, which is independent of the angles $\psi_i$, and changing the sign of the coefficient of the original quartic invariant. The relative amplitude of the trivial quartic invariant ensures that for positive overall multiplicative factor the free energy remains stable. Given this change, constant free energy surfaces can be constructed that graze the Kim region at points corresponding to those symmetries. Figure \ref{fig:infinitysolution} shows such a free energy surface in the case $D_{\infty}$. 
Constant free energy surfaces that graze the Kim region at the point of octahedral symmetry and along the the yellow, six-fold symmetry curve can be similarly constructed.

\begin{figure}[htbp]
\begin{center}
\includegraphics[width=3in]{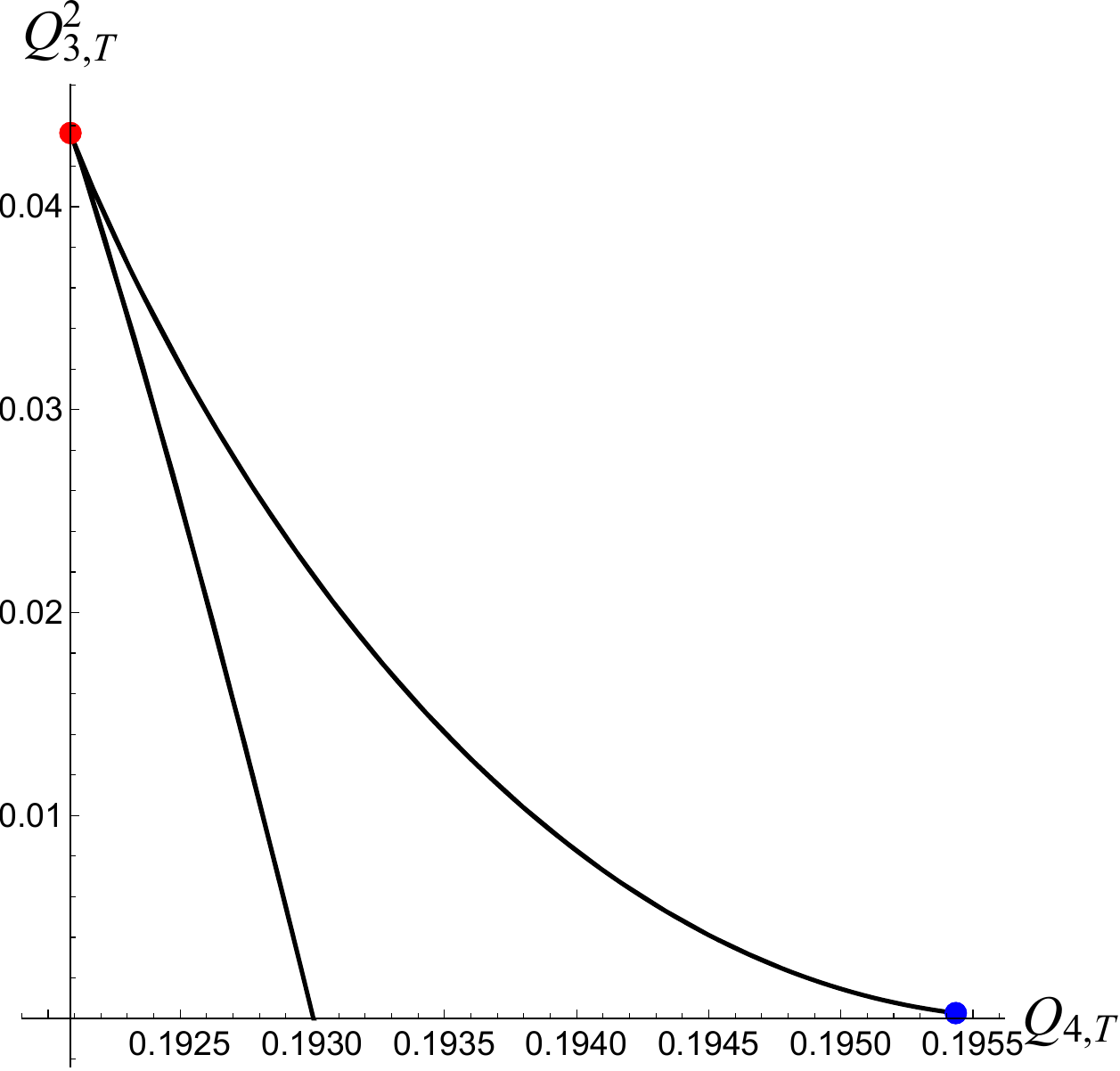}
\caption{The curve for tetrahedral ordering in the Kim plot. The two small dots correspond to icosahedral ordering (red dot) and octahedral ordering (blue dot).}
\label{fig:Kimcurve}
\end{center}
\end{figure}

\begin{figure}[htbp]
\begin{center}
\includegraphics[width=3in]{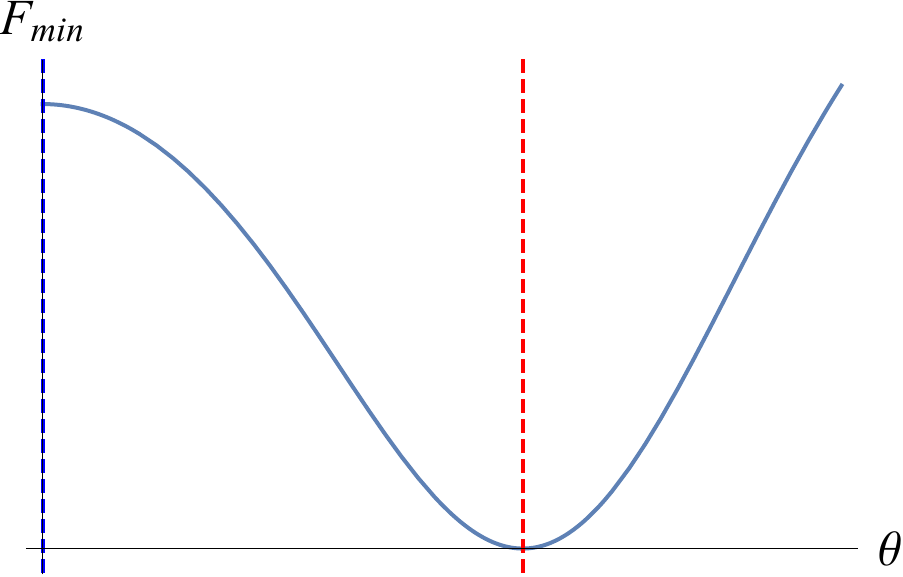}
\caption{The minimum free energy curve as a function of $\theta$. The two vertical dashed lines indicate the values of $\theta$ for which the symmetry is octahedral (blue line) and icosahedral (red line).}
\label{fig:freecurve1}
\end{center}
\end{figure}

One remaining question is whether there is a minimum free energy solution in the case of tetrahedral symmetry. The curve corresponding to this symmetry, shown in green in Fig. \ref{fig:infinitysolution}, clearly lies on the surface of the Kim region, so there is reason to address the question of the possibility of tetrahedral---but not octahedral or icosahedral---symmetry. In order to do this, it is useful to  look at the properties of the tetrahedral ordering subspace in the $l=6$ Kim plot. It turns out that tetrahedral order can be described in terms of two variables---an amplitude $a$ and an angle $\theta$. Given the standard parameterization of an $l=6$ density, as an extension of the $l=2$ case we discuss in the article,
\begin{equation}
\{r_1 ,r_2, r_3, r_4, r_5, r_6, r_7, s_1, s_2, s_3, s_4, s_5, s_6\} \label{eq:to1}
\end{equation}
a general tetrahedral state is generated when we replace the above by 
\begin{eqnarray}
\lefteqn{\{\frac{1}{9} \sqrt{\frac{77}{6}} a \cos (\theta ),0,0,-\frac{1}{9}
   \sqrt{\frac{35}{3}} a \cos (\theta ),0,0,}\nonumber \\ && \frac{4}{9} a \cos (\theta ),\frac{1}{3}
   \sqrt{\frac{5}{6}} a \sin (\theta ),0,0,\frac{1}{3} \sqrt{\frac{11}{3}} a \sin (\theta
   ),0,0 \}  \nonumber \\ \label{eq:to2}
   \end{eqnarray}
With this replacement, we can find the possibilities for the quadratic, the cubic and the two quartic invariants
\begin{eqnarray}
Q_2 & = & a^2 \label{eq:to3} \\
Q_3 & = & \frac{20 \sqrt{\frac{26}{\pi }} a^3 (8 \cos (3 \theta )-9 \cos (\theta ))}{3553} \label{eq:to4} \\
Q_{4,1} & = & \frac{39 a^4 (80 \cos (2 \theta )+64 \cos (4 \theta )+12721)}{817190 \pi } \label{eq:to5} \\
Q_{4,2} & = & 0 \label{eq:to6}
\end{eqnarray}
From this we note that the curve of the normalized invariants $Q_{4,1}/a^4$ and $Q_3^2/a^6$ for tetrahedral symmetry lies in a two dimensional subspace of the original $l=6$ Kim plot. Calling these normalized invariants $Q_{4,T}$ and $Q_{3,T}^2$, we have the curve shown in Fig. \ref{fig:Kimcurve}.
As noted in the caption, there are two special ordering points on the curve: icosahedral, at the peak on which the red dot sits and octahedral, at the lower right hand terminus where there is a blue dot. The latter point lies near to, but not on, the $Q_{3,T}^2=0$ axis. The actual location of that point is at\begin{equation}
\{Q_{4,T}, Q_{3,T}^2 \} = \left\{\frac{100347}{163438 \pi },\frac{10400}{12623809 \pi }\right\} \label{eq:to7}
\end{equation}
The values of the parameter $\theta$ that yield the two points are
\begin{eqnarray}
\mbox{Icosahedral} & : & \theta = \arctan(\sqrt{21/11}) \label{eq:to8} \\
\mbox{Octahedral} & : & \theta = 0 \label{eq:to9}
\end{eqnarray}
Given all this, we are in a position to look for free energy minima along the tetrahedral ordering curve. We can safely say that if a point on the curve does not correspond to a free energy minimum, then it will never correspond to a global free energy minimum. On the other hand, if it does, then the possibility that it is also a global free energy minimum exists. We proceed by making use of the expression for the free energy minimized with respect to overall amplitude amplitude, expressed as a function of invariants in a system with a cubic invariant, utilized in the Kim method as applied to such systems; see Appendix D. This expression is 
\begin{eqnarray}
\lefteqn{F_{\rm min}} \nonumber \\ &=&-\frac{u^4Q_3^4}{v^3Q_4^3}\big[1+6(tvQ_4/u^2Q_3^2)^2  -6 t vQ_4/u^2Q_3^2 \nonumber \\ && + \left(1-4 t vQ_4/u^2Q_3^2\right)^{3/2}\big]/24  \nonumber \\ & \equiv & -\frac{u^4Q_3^4}{v^3Q_4^3}\frac{1+6W^2 - 6W + \left( 1-4W\right)^{3/2}}{24} \label{eq:to10}
\end{eqnarray}
We simply replace $Q_4$ with $Q_{4,T}$ and $Q_3^2$ with $Q_{3,T}^2$. For the set of invariants discussed in Section II.C.2, all free energy curves exhibit the feature of the curve shown in Fig. \ref{fig:freecurve1}, which is to say a minimum at the angle, $\theta$, corresponding to icosahedral ordering.
This means that the only possible ordering with tetrahedral symmetry is icosahedral. 

\begin{figure}[htbp]
\begin{center}
\includegraphics[width=3in]{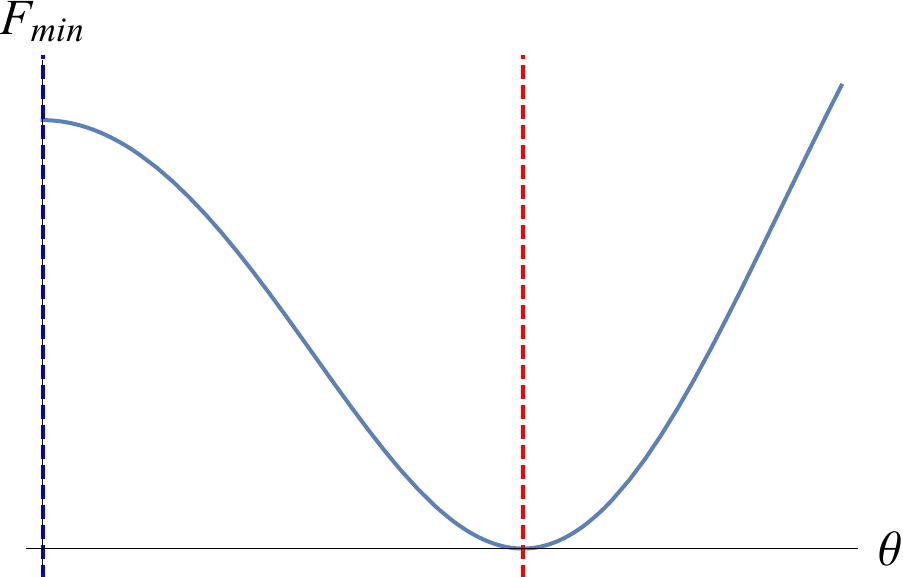}
\caption{The minimum free energy curve with the new quartic invariant $Q_{4,T}^{\prime}$, as given in (\ref{eq:to11}). Once again, icosahedral ordering is indicated by the red dashed line and octahedral ordering by the blue dashed line. }
\label{fig:freecurve2}
\end{center}
\end{figure}

\begin{figure}[htbp]
\begin{center}
\includegraphics[width=3in]{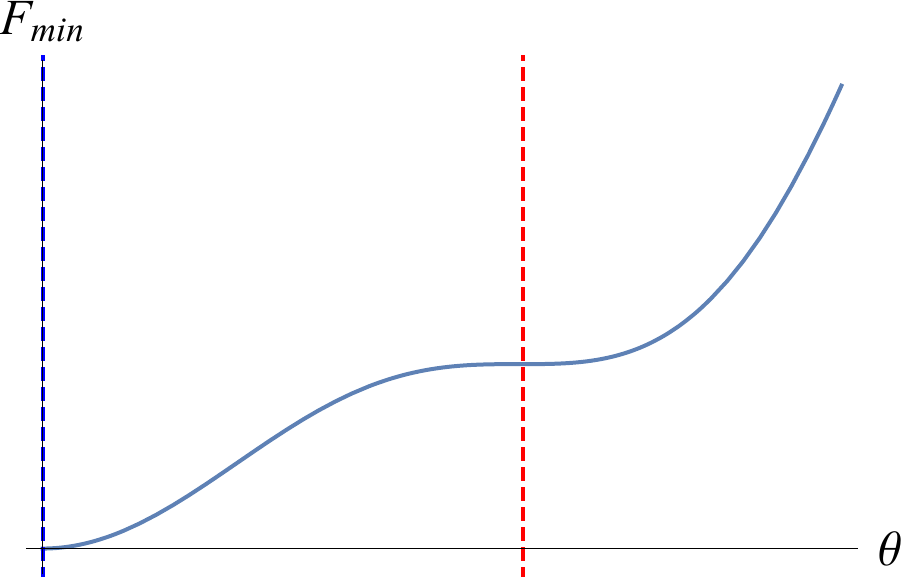}
\caption{The minimum free energy curve with the new quartic invariant $Q_{4,T}^{\prime}$ in a different parameter range. Now octahedral ordering is preferred.}
\label{fig:freecurve3}
\end{center}
\end{figure}

To fully explore the possibilities for tetrahedral ordering, we replace $Q_{4,T}$ with $Q_{4,T}^{\prime}$, where
\begin{equation}
Q_{4,T}^{\prime} = 1-Q_{4,T} \label{eq:to11}
\end{equation}
In this case,  two possibilities emerge, as shown in Figs. \ref{fig:freecurve2} and \ref{fig:freecurve3}.
A comprehensive survey leads to the conclusion that  the only possibilities for ordering are icosahedral and octahedral. We can thus rule out the lower tetrahedral symmetry as a possible global free energy minimum.

\section{Constant free energy surface for a sixth order free energy}
We start with the mean field free energy
\begin{equation}
\mathcal{F} = -\frac{R}{2}x^2+\frac{V}{4}x^4+\frac{W}{6}x^6
\label{eq:wb1}
\end{equation}
where
\begin{eqnarray}
R&=&t_{15}Q_2 \label{eq:def1} \\
V & = & vQ_{4,1} \label{eq:def2} \\
W & = & w Q_{6,1} \label{eq:def3}
\end{eqnarray}

Taking the first derivative with respect to $x$ and then dividing by $x$ we end up with the equation of state
\begin{equation}
-R+Vx^2+Wx^4 =0 \label{eq:wb2}
\end{equation}
The relevant solution is 
\begin{equation}
x = \pm \frac{\sqrt{\frac{\sqrt{4 R W+V^2}}{W}-\frac{V}{W}}}{\sqrt{2}} \label{eq:wb3}
\end{equation}
If we plug this solution into the free energy expression (\ref{eq:wb1}), we end up with the expression for the free energy minimum
\begin{equation}
\mathcal{F}_{\rm min} = \frac{V^3}{24 W^2}\left[1+6X -(1+4X)^{3/2} \right] \label{eq:wb4}
\end{equation}
where
\begin{equation}
X=\frac{RW}{V^2} \label{eq:wb5}
\end{equation}
Making use of (\ref{eq:wb4}) and (\ref{eq:wb5}) we obtain the following expressions for the coefficients $v$ and $w$
\begin{eqnarray}
V & = & \left(\frac{R}{X} \right)^2 \frac{(1+4X)^{3/2} - 1-6X}{24|\mathcal{F}_{\rm min}|} \label{eq:wb6} \\
W & = & \left( \frac{R}{X} \right)^3 \left( \frac{(1+4X)^{3/2} - 1-6X}{24|\mathcal{F}_{\rm min}|} \right)^2 \label{eq:wb7}
\end{eqnarray}
We can verify by direct substitution that (\ref{eq:wb6}) and (\ref{eq:wb7}) are consistent with (\ref{eq:wb5}). Furthermore, given that $X$ as defined by (\ref{eq:wb5}) is necessarily positive---$w$ must be greater than zero to guarantee free energy stability, and we are assuming that the quadratic term is negative and the quartic term positive as we expect to be in that regime---we can easily demontrate that $\mathcal{F}_{\rm min}$ as given by (\ref{eq:wb4}) will be negative.  

\begin{figure}[htbp]
\begin{center}
\includegraphics[width=3in]{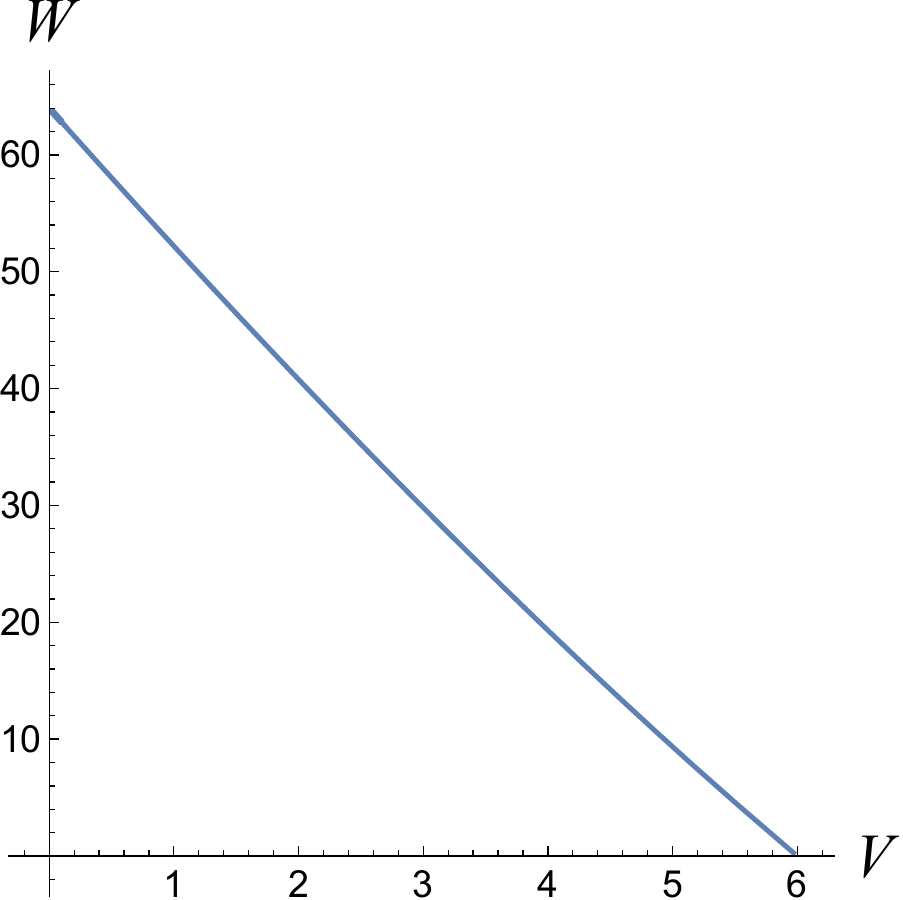}
\caption{Plot of the quantities $v$ and $w$, as given by Eqs. (\ref{eq:wb8}) and (\ref{eq:wb9}), with $a=b=1$. }
\label{fig:parametricplot}
\end{center}
\end{figure}

It is instructive to plot the two variables $v$ and $w$ against each other as functions of the new composite quantity $W$. Rewriting the above equations as \begin{eqnarray}
V & = &  \frac{(1+4X)^{3/2} - 1 - 6X}{aX^2} \label{eq:wb8} \\
W & = &  \frac{\left( (1+4X)^{3/2} - 1 - 6X\right)^2}{bX^3} \label{eq:wb9}
\end{eqnarray}
with both $a$ and $b$ positive, we start by setting $a=b=1$. Then, the parametric plot of $w$ versus $v$ is as shown in Fig. \ref{fig:parametricplot}.

\begin{figure}[htbp]
\begin{center}
\includegraphics[width=3in]{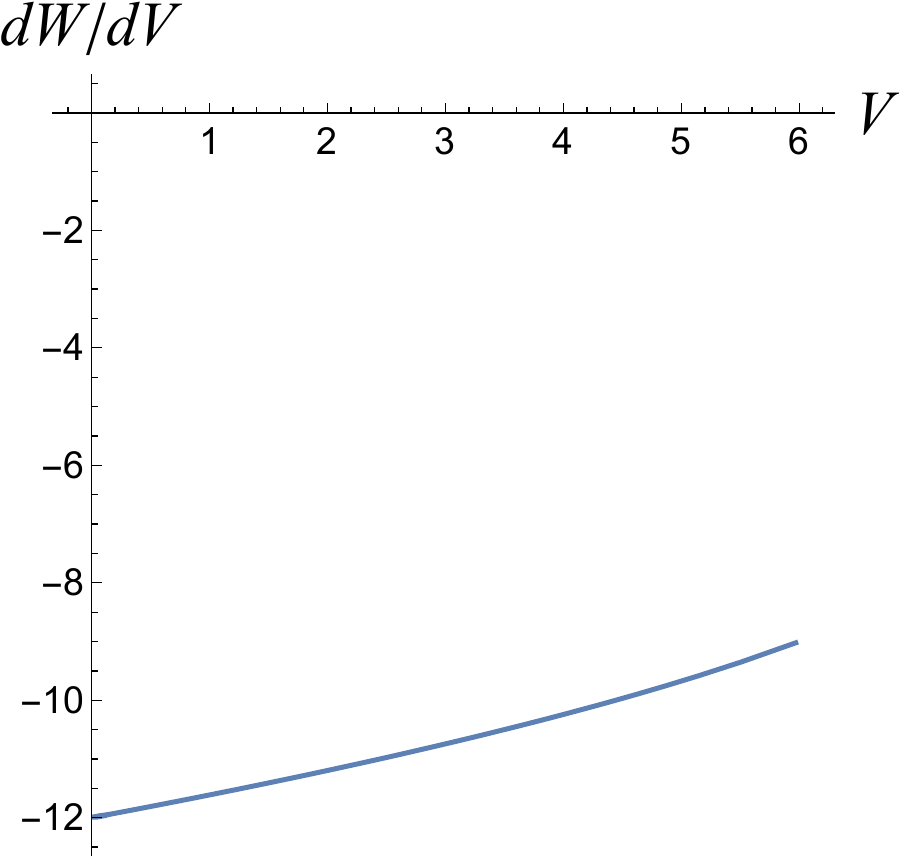}
\caption{The slope of the plot in Fig. \ref{fig:parametricplot}.}
\label{fig:parametricslope}
\end{center}
\end{figure}

It is also useful to plot the slope of the plot in Fig. \ref{fig:parametricplot}. This graph is shown in Fig. \ref{fig:parametricslope}.

A feature of the plot in Fig. \ref{fig:parametricplot} is that it spans the range of the two free energy coefficients, $V$ and $W$. It is not a straight line, as is clear from Fig. \ref{fig:parametricslope}. However, it has the general form of a constant free energy surface when there are two quartic invariants. 

Now, let's see how those surfaces behave as the quartic coefficient $r$ varies. In particular, let's look at what happens when $r$ approaches 0. From (\ref{eq:wb6}) and (\ref{eq:wb7}) we see that the for fixed values of $X$ and $\mathcal{F}_{\rm min}$ both $v$ and $w$ decrease in absolute value. However, $w$ decreases faster than $v$. Thus, the constant free energy surface in Fig. \ref{fig:parametricplot} becomes more nearly horizontal; the sixth order term becomes less and less important, which is to be expected. In light of what we already know, the ordering is going to be tetrahedral. This makes sense, as in the regime contemplated here the fourth order coefficient is going to be more important than the sixth order one as one approaches the transition point ($r=0$) from below. 

To explore the phase diagram we vary the ratio of $b$ to $a^2$ in Eqs. (\ref{eq:wb8}) and (\ref{eq:wb9}). At the transition point, $b/a^2=0$. As we move away from that point the ratio grows.

\section{Degeneracy of mediated quartic invariants} \label{app:degeneracy}

As it turns out, some of the mediated quartic invariants are degenerate, in that they take on zero values in a subspace of the configuration space spanned by the real parameters defined in for $l=2$ in Section II.B.1 of the text. This can be understood in terms of the structure of those invariants. Consider such invariants as depicted in Fig. 8 of the text. The explicit form for the invariant is
\begin{eqnarray}
\lefteqn{\sum_m \Bigg[ \sum_{m_1,m_2,m_3,m_4}\mathcal{V}(l,m_1,l,m_2,j,m) } \nonumber \\ && \mathcal{V}(l,m_1,l,m_2,j,-m) \delta_{m_1+m_2+m} \delta_{m_3+m_4-m} \Bigg] \label{eq:deg1}
\end{eqnarray}
Setting $j=2$, let's consider individually each contribution for a particular value of $m$. Breaking the product in the brackets into its two components, we find for each of those components
\begin{equation}
\sum_{m_1,m_2} \mathcal{V}(l,m_1,l,m_2,j,m) \delta _{m_1+m_2\pm m} \label{eq:deg2}
\end{equation}
Those two components are a complex conjugate pair. In the case $l=7$ and $m=\pm 2$ we find for one of them
\begin{eqnarray}
\lefteqn{\bigg(\frac{14}{221} \sqrt{\frac{30}{\pi }} r_7^2-\frac{45}{221} \sqrt{\frac{10}{\pi }} r_5
   r_7-\frac{1}{17} \sqrt{\frac{210}{13 \pi }} r_1 r_3} \nonumber \\ &&-\frac{6}{17} \sqrt{\frac{15}{13
   \pi }} r_2 r_4-\frac{6}{221} \sqrt{\frac{330}{\pi }} r_3 r_5 -\frac{10}{221}
   \sqrt{\frac{165}{\pi }} r_4 r_6 \nonumber \\ &&-\frac{36}{221} \sqrt{\frac{35}{\pi }} r_6
   r_8-\frac{14}{221} \sqrt{\frac{30}{\pi }} s_7^2-\frac{1}{17} \sqrt{\frac{210}{13 \pi
   }} s_1 s_3 \nonumber \\ && -\frac{6}{17} \sqrt{\frac{15}{13 \pi }} s_2 s_4 -\frac{6}{221}
   \sqrt{\frac{330}{\pi }} s_3 s_5-\frac{10}{221} \sqrt{\frac{165}{\pi }} s_4
   s_6\nonumber \\ &&-\frac{45}{221} \sqrt{\frac{10}{\pi }} s_5 s_7 \bigg) \nonumber \\ && + i \bigg( \frac{1}{17} \sqrt{\frac{210}{13 \pi }} r_3 s_1+\frac{6}{17} \sqrt{\frac{15}{13 \pi }}
   r_4 s_2-\frac{1}{17} \sqrt{\frac{210}{13 \pi }} r_1 s_3  \nonumber \\ && +\frac{6}{221}
   \sqrt{\frac{330}{\pi }} r_5 s_3-\frac{6}{17} \sqrt{\frac{15}{13 \pi }} r_2
   s_4+\frac{10}{221} \sqrt{\frac{165}{\pi }} r_6 s_4 \nonumber \\ && -\frac{6}{221} \sqrt{\frac{330}{\pi
   }} r_3 s_5+\frac{45}{221} \sqrt{\frac{10}{\pi }} r_7 s_5-\frac{10}{221}
   \sqrt{\frac{165}{\pi }} r_4 s_6 \nonumber \\ &&+\frac{36}{221} \sqrt{\frac{35}{\pi }} r_8
   s_6-\frac{45}{221} \sqrt{\frac{10}{\pi }} r_5 s_7-\frac{28}{221} \sqrt{\frac{30}{\pi
   }} r_7 s_7 \bigg)
\end{eqnarray}
and for the other the complex conjugate of the expression above. Their product is thus a sum of squares of quadratic forms. The complete expression yields five such squares, two each for intermediate $m$'s equal to $\pm 2$ and $\pm 1$, and one for intermediate $m$ equal to zero. If we require that the mediated invariant be zero, then we have five constraints. Given that the $l=7$ parameter space is fifteen dimensional, this reduces that space to ten dimensions. If we specify the overall modulus we are now down to nine dimensions. Given the degeneracy of the space to the three generators of overall rotations in real space, we are left with a six dimensional subspace in which the $j=2$ mediated invariant is equal to zero.

As an interesting side-note, one finds for $l=5$, for which there are two independent quartic invariants, the local and the trivial one, that the $l=2$ mediated invariant is a linear combination of those two, as it must be. In light of the fact that the trivial invariant is degenerate with respect to all rotations in configuration space, this means that the local invariant is itself degenerate, in a two dimensional subspace in this instance. As far as we know $l=5$ is the only case in which the local quartic invariant possesses such an additional degeneracy.

\end{appendix}

\pagebreak

\begin{thebibliography}{30}
\bibitem [{Kim()}]{Kimgeneralization}%
  \BibitemOpen
  \href@noop {} {}\bibinfo {note} {To be precise, we generalize the approach of
  Kim \cite{Kim3}, which applies to a free energy, or effective Hamiltonian,
  with quadratic and quartic, but no cubic, contributions.}\BibitemShut {Stop}%
\bibitem [{reg()}]{regimenote}%
  \BibitemOpen
  \href@noop {} {}\bibinfo {note} {Actually, there is a fourth regime. See Fig.
  \ref{fig:T72}.}\BibitemShut {Stop}%
\bibitem [{tet()}]{tetrahedral_note}%
  \BibitemOpen
  \href@noop {} {}\bibinfo {note} {That the initial ordering in the case $l=16$
  should be tetrahedral was noted by Matthews \cite{Matthews}.}\BibitemShut
  {Stop}%
\bibitem [{fig()}]{figurenote}%
  \BibitemOpen
  \href@noop {} {}\bibinfo {note} {This figure is an elaborated version of a
  phase diagram for the same system in \cite{sanjay2}.}\BibitemShut {Stop}%
\end{thebibliography}
\end{document}